\documentclass[12pt]{article}
\pdfoutput=1

\usepackage{amssymb}
\usepackage{epsfig}
\usepackage{amsmath}
\usepackage{multirow}

\usepackage{graphicx}
\usepackage{color}
\definecolor{rossos}{cmyk}{0,1,1,0.55}

\ifx\pdfoutput\undefined
\usepackage[dvips,bookmarks=false]{hyperref}	
\else
\usepackage{hyperref}	
\fi
\def\hhref#1{\href{http://arxiv.org/abs/#1}{#1}} 

\def\lsim{\mathrel{\rlap{\lower3pt\hbox{\hskip0pt$\sim$}}
   \raise1pt\hbox{$<$}}}         
\def\gsim{\mathrel{\rlap{\lower4pt\hbox{\hskip1pt$\sim$}}
   \raise1pt\hbox{$>$}}}         

\newcommand{\eps}{\epsilon}

\newcommand{\Tr}{\text{Tr}}
\newcommand{\calG}{{\cal G}}
\newcommand{\calH}{{\cal H}}
\newcommand{\calHp}{{\cal H}^\prime}

\newcommand{\trace}{\text{Tr}}
\newcommand{\acap}{ {\hat{a}} }
\newcommand{\capp}[1]{ {\hat{#1}} }

\newcommand{\ampl}{\mathcal{A}}

\newcommand{\be}{\begin{equation}}
\newcommand{\ee}{\end{equation}}
\newcommand{\bea}{\begin{eqnarray}}
\newcommand{\eea}{\end{eqnarray}}
\newcommand{\bit}{\begin{itemize}}
\newcommand{\eit}{\end{itemize}}

\begin{document}

\setcounter{page}{0}
\thispagestyle{empty}

\begin{flushright}
CERN-PH-TH/2009-036 
\end{flushright}

\vskip 50pt

\begin{center}
{\bf \Huge {
On the effect of resonances \\[0.4cm]
in composite Higgs phenomenology
}}
\end{center}

\vskip 30pt

\begin{center}
{\large Roberto Contino$^{\,a}$, David Marzocca$^{\, a,c}$, \\[0.5cm]
Duccio Pappadopulo$^{\, b}$,  and Riccardo Rattazzi$^{\, b}$ }
\end{center}

\vskip 20pt

\begin{center}
\centerline{$^{a}$ {\small \it Dipartimento di Fisica, Sapienza Universit\`a di Roma and INFN, Roma, Italy}}
\vskip 5pt
\centerline{$^{b}$ {\small \it Institut de Th\'eorie des Ph\'enom\`enes Physiques, EPFL, Lausanne, Switzerland}}
\vskip 5pt
\centerline{$^{c}${\small \it International School for Advanced Studies (SISSA) and INFN, Trieste, Italy}}
\vskip 5pt
\end{center}

\vskip 13pt 

\begin{abstract}
\vskip 3pt
\noindent
We consider a generic composite Higgs model based on the coset $SO(5)/SO(4)$ and study its phenomenology beyond the leading low-energy 
effective lagrangian approximation. Our basic goal is to introduce in a controllable and simple way the lowest-lying, possibly narrow, resonances that 
may exist is such models. We do so by proposing a criterion that we call {\it partial UV completion}. We characterize the simplest cases, corresponding 
respectively to a scalar in either singlet or tensor representation of $SO(4)$ and to vectors in the adjoint of $SO(4)$. 
We study the impact of these resonances on the signals associated to high-energy vector boson scattering, pointing out for each resonance the 
characteristic patterns of depletion and enhancement with respect to the leading-order chiral lagrangian. En route we derive the $O(p^4)$ general 
chiral lagrangian and discuss its peculiar accidental and approximate symmetries. 
\end{abstract}

\vskip 13pt
\newpage

\section{Introduction}
\label{sec:introduction}
It is rather plausible that a new strong dynamics be hiding behind  electroweak symmetry breaking (EWSB).
Unfortunately the modeling of the generic collider signals of such scenario is limited by our scarce ability to  control strongly coupled quantum 
field theory.  In particular, while in the low-energy regime effective chiral lagrangians provide in principle a reliable and universal description of 
the dynamics, it is in the physics of the massive states that strong coupling and model dependence represent a real limitation. Moreover, the 
separation of energy scales is often not enough in practice, so that the corrections to the simple chiral lagrangian results can be important. 
The study of $WW$ scattering with realistic cuts (to beat the Standard Model (SM) background) is an example of that. 
Modeling the effects of heavy resonances, 
or at least assess their qualitative effects is thus a very relevant phenomenological problem.
Some progress has come in recent years from extra dimensional constructions. The resulting models can be viewed as {\it deformations} of 
strongly coupled field theories where a parameter, basically the inverse volume of compactification, controls a weak coupling expansion. 
The couplings of the resonances are thus controlled to some extent.
However, the studies in the literature are often made complicated by the presence of more structure and more parameters than
one would hope to need in order to parametrize the physics of the low-lying resonances. In order to best serve  the needs of the experimental 
community it would be preferable to introduce as simple as possible a description without sacrificing too much theoretical consistency. 
That is the main goal of the present paper.
More precisely, we will be focussing on the scenario where the Higgs boson is a pseudo Nambu-Goldstone (NG) boson resulting from the 
coset $SO(5)/SO(4)$ \cite{Kaplan:1983fs,Agashe:2004rs}, and study the impact of different resonances on the scattering of NG bosons (longitudinal vector bosons plus the Higgs boson). 
Our study is very much in line with the recent effort to construct ``simplified models'' to describe the broad LHC signatures of classes of models \cite{Alves:2011wf}. 
Among previous studies that are related to ours,  we should mention Refs.~\cite{Donoghue:1989qw,Donoghue:1988ed,Bagger:1993zf,chanowitz} 
and the more  recent~\cite{Barbieri:2008cc,Barbieri:2009tx,Hernandez:2010iu,Falkowski:2011ua}, where the impact of resonances in $WW$ scattering for 
the Higgsless $SO(4)/SO(3)$ case was studied. 
Closer perhaps to our approach is Ref.~\cite{Panico:2011pw}, where two- and three-site deconstructed models for $SO(5)/SO(4)$ were introduced.
In fact, our approach is even more minimal, as we are not attempting to parametrize the physics that gives rise to the Higgs potential, but just focussing 
on the scattering of NG bosons. Moreover, as (preliminary) part to our study of resonances we will construct the subleading $O(p^4)$ effective 
chiral lagrangian for the $SO(5)/SO(4)$ coset. That will allow us to compare to the specific $O(p^4)$ effects mediated by the exchange of each specific 
resonance. We will also describe in details the peculiar pattern of accidental and approximate discrete symmetries that arise in general in the models 
based on $SO(5)/SO(4)$.

\section{The $SO(5)/SO(4)$ chiral lagrangian at $O(p^4)$}

A systematic method to construct the chiral lagrangian 
for a generic Lie group $\calG$ spontaneously broken to a subgroup $\calH$
has been introduced by Callan, Coleman, Wess and Zumino (CCWZ)~\cite{CCWZ}.~\footnote{
The results of Ref.~\cite{CCWZ} apply to any compact, connected, semisimple Lie group $\calG$.}
Their main result can be summarized as follows: the most general 
lagrangian invariant under global $\calG$ transformations can be built following the
rules for a \textit{local} symmetry~$\calH$, where $\calH$ is the linearly-realized subgroup of $\calG$,
by means of suitably defined covariant variables.
Here we apply the CCWZ formalism to construct the $SO(5)/SO(4)$ chiral lagrangian at order $O(p^4)$.

Let us start by considering a generic global symmetry $\calG\to \calH$.
We denote the Nambu-Goldstone fields by $\Pi(x) = \Pi^{\hat a}(x) T^{\hat a}$ and define
\begin{gather}
\label{eq:Udef}
U(\Pi) = e^{i \Pi(x)} \\[0.2cm]
\label{eq:DefOfdE}
-i\, U^\dagger  \partial_\mu U  = d_\mu^{\hat a} T^{\hat a} + E_\mu^a T^a \equiv d_\mu + E_\mu\, .
\end{gather}
Here and in the following $T^{\hat a}$, $T^{a}$  are respectively the broken and unbroken  generators
normalized as $\Tr(T^A T^B) = \delta^{AB}$ (the explicit expressions for the $SO(5)$ generators can be found in Appendix~\ref{app:SO5generators}).
The action of $g\in \calG$ on the NG fields is given by 
\begin{equation}
g\, U(\Pi) \equiv U(g(\Pi))\, h(\Pi,g))\, ,
\end{equation}
where $h(\Pi(x),g)$ is an element of $\calH$ which depends on the space-time point through $\Pi(x)$, 
and $U(g(\Pi)) = \exp(i\, T^{\hat a} \xi^{\hat a}(\Pi, g))$. 
Hence, under  global $\calG$ transformations
\begin{equation}
\label{eq:Utransf}
U(\Pi) \to g\, U(\Pi)\, h^\dagger(\Pi,g))\, ,
\end{equation}
which implies
\begin{align}
d_\mu(\Pi) &\to h(\Pi, g)\, d_\mu(\Pi) \, h^\dagger(\Pi, g) \\[0.2cm]
E_\mu(\Pi) &\to h(\Pi, g)\, E_\mu(\Pi) \, h^\dagger(\Pi, g) - i \, h(\Pi, g)  \partial_\mu h^\dagger(\Pi, g) \, .
\end{align}
The above equations show that  $d_\mu(\Pi)$, $E_\mu(\Pi)$ transform under a local symmetry $\calH$, and that in particular $E_\mu(\Pi)$
transforms like a gauge field.  It is thus possible to define a covariant derivative
\begin{equation}
\nabla_\mu \equiv \partial_\mu + i E_\mu
\end{equation}
and a  field strength
\begin{equation}
\begin{gathered}
E_{\mu\nu} = \partial_\mu E_\nu - \partial_\nu E_\mu + i [E_\mu , E_\nu] \\[0.25cm]
E_{\mu\nu}(\Pi)  \to h(\Pi, g)\, E_{\mu\nu}(\Pi) \, h^\dagger(\Pi, g) \, .
\end{gathered}
\end{equation}
The covariant variables $d_\mu(\Pi)$, $E_{\mu\nu}(\Pi)$, and those formed by acting with the covariant derivative $\nabla_\mu$
are the building blocks of the chiral lagrangian.  

There are additional covariant structures, however, that can be constructed once
the external gauging of  a group $\calHp \subseteq \calG$ is turned on (eventually, we will be interested
in gauging a subgroup $SU(2)_L\times U(1)_Y$ identified with the SM electroweak group). In this
case all the above relations hold true provided one replaces
$\partial_\mu \to D_\mu = \partial_\mu + i A_\mu$
in eq.(\ref{eq:DefOfdE}), which thus becomes
\begin{equation} \label{eq:DefOfdEgauged}
-i\, U^\dagger  D_\mu U  = d_\mu^{\hat a} T^{\hat a} + E_\mu^a T^a \equiv d_\mu + E_\mu\, .
\end{equation}
The external gauge fields $A_\mu = A^{\hat a}_\mu T^{\hat a}+A^{a}_\mu T^{a}$ transform according to the usual
rule under (local) $\calHp$ transformations:~\footnote{To assign quantum numbers we formally take $A$ to gauge the whole $G$.} 
$A_\mu \to g A_\mu g^\dagger - i \, g\,  \partial_\mu g^\dagger$.
Two new covariant structures (i.e. variables that transform as representations of the local symmetry $\calH$)
can thus be constructed from the field strength of the external gauge fields as follows:
\begin{equation}
\begin{gathered}
f_{\mu\nu} = U^\dagger F_{\mu\nu} U = (f_{\mu\nu}^-)^{\hat a} T^{\hat a} + (f_{\mu\nu}^+)^{a} T^{a} 
 \equiv f_{\mu\nu}^- + f_{\mu\nu}^+ \\[0.3cm]
f_{\mu\nu}^\pm(\Pi)  \to  h(\Pi, g)\, f_{\mu\nu}^\pm(\Pi) \, h^\dagger(\Pi, g) \, .
\end{gathered}
\end{equation}
%

\subsection{The $SO(5)/SO(4)$  chiral lagrangian at $O(p^2)$ and its accidental symmetries}
\label{sec:Op2chiralLag}

\begin{figure}[t]
\begin{center}
\includegraphics[height=60mm]{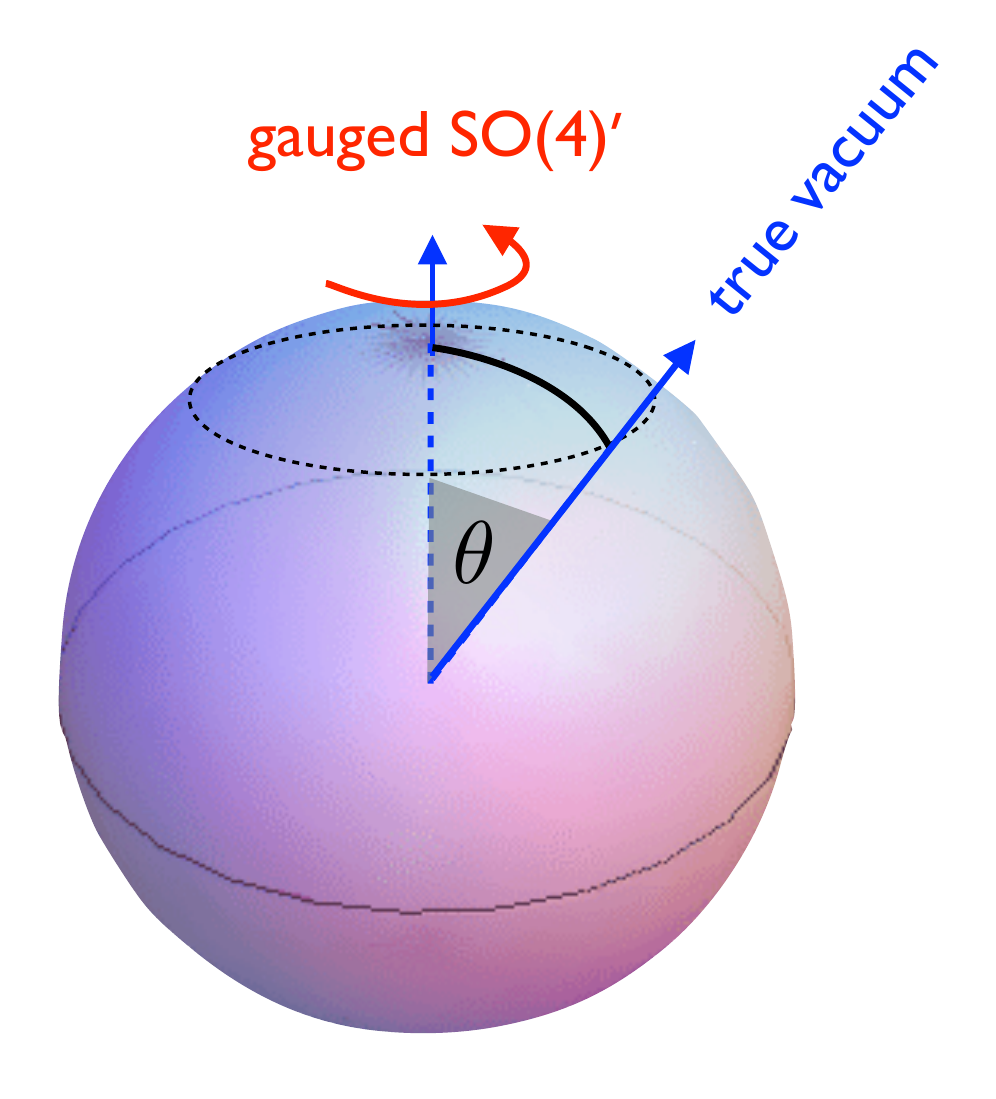} 
\caption{ \label{fig:S4}
\small
The NG bosons of $SO(5)/SO(4)$ live on the four-sphere $S^4$. A generic vacuum points in a direction forming
an angle $\theta$ with that fixed by the `gauged' $SO(4)^\prime$.  The electroweak symmetry breaking can be seen
as due to the misalignment $\theta$.
Even assuming no misalignment at the tree level, a non-vanishing $\theta = \langle \pi \rangle/f$ is
generated at the loop level after the NG 4-vector acquires a vev $\langle \pi \rangle \not =0$ (black curve).
}
\end{center}
\end{figure}

The lagrangian of composite Higgs models based on the $SO(5)/SO(4)$ coset can be easily
constructed by means of the CCWZ covariant variables defined above.
In this case there are four NG bosons associated to the breaking $SO(5)\to SO(4)$, $\pi^{\hat a}$ with
$\hat a = 1,2,3,4$, which live on the  four-sphere  ($SO(5)/SO(4) = S^4$). They transform as a $\mathbf{4}$
of $SO(4)$, or equivalently as a $(\mathbf{2,2})$ of $SU(2) \times SU(2) \sim SO(4)$.
The SM  electroweak vector bosons gauge a subgroup $SU(2)_L\times U(1)_Y \subset SU(2)_L \times SU(2)_R \sim SO(4)^\prime$
contained in $SO(5)$, such that $Y = T_{3R}$.~\footnote{In realistic models there is a larger pattern of global symmetries, 
 $SO(5)\times U(1)_X \to SO(4) \times U(1)_X$, and hypercharge is defined as $Y= T_{3R} + X$.  A non-zero $X$ charge
is required for the SM fermions to correctly reproduce their hypercharge. 
Since the NG bosons are neutral under the additional $U(1)_X$, this latter plays no role in the following discussion
and will be omitted for simplicity.
}
It is possible to parametrize the orientation of the `gauged' $SO(4)^\prime$ (i.e. that which contains the SM group
$SU(2)_L\times U(1)_Y$)
with respect to the linearly-realized global $SO(4)$
by an angle $\theta$.
For example, by representing the vacuum as a 5-dimensional unit vector $\Phi_0$, 
and letting the gauged $SO(4)'$ act on the first four entries, one has
$\Phi_0 = (0, 0, 0,  \sin\theta , \cos\theta)$.  The gauged $SO(4)'$ thus identifies a preferred direction inside $SO(5)$,
and the angle $\theta$ precisely measures the misalignment of the vacuum with respect to it, see Fig.~\ref{fig:S4}.
%
The field
\begin{equation} \label{eq:Phi}
\Phi = U(x) \Phi_0 = e^{i \sqrt{2} \, T^{\hat a}(\theta) \pi^{\hat a}(x)/f} \Phi_0 = 
\begin{pmatrix}
\hat \pi^1 \sin(\pi/f) \\[0.1cm]
\hat \pi^2 \sin(\pi/f) \\[0.1cm]
\hat \pi^3 \sin(\pi/f) \\[0.1cm]
\hat \pi^4 \sin(\pi/f) \cos\theta + \cos(\pi/f) \sin\theta \\[0.1cm]
-\hat \pi^4 \sin(\pi/f) \sin\theta + \cos(\pi/f) \cos\theta
\end{pmatrix}
\end{equation}
parametrizes the massless excitations around the vacuum, where we have defined $\pi= \sqrt{(\pi^{\hat a})^2}$ and
$\hat \pi^{\hat a} = \pi^{\hat a}/\pi$.~\footnote{The factor $\sqrt{2}$ in the exponent of eq.(\ref{eq:Phi}) has been introduced
to  match  the standard normalization adopted in the literature. It can be absorbed by a redefinition of $f$.
}
Here and in the following we denote the generators of $SO(5)\to SO(4)$ as $T^{a,\hat a} = T^{a,\hat a} (\theta)$,  
which are related to those of $SO(5)\to SO(4)^\prime$, where $ SO(4)^\prime$ is the gauged subgroup, by a 
rotation of an angle $\theta$, see Appendix~\ref{app:SO5generators}.

For $\theta=0$ the SM electroweak group is unbroken, being contained in the preserved global $SO(4)$, and the four NG bosons
form a complex doublet of $SU(2)_L$.
For $\theta \not =0$, on the other hand,  the SM vector bosons gauge (a combination of) the $SO(5)/SO(4)$ broken generators,
so that three NG bosons are  eaten to give mass to the $W$ and the $Z$, while a fourth one is identified with the Higgs boson.
This can be easily seen as follows. Since the gauged $SO(4)^\prime$ acts on the first four entries of the field $\Phi$ in eq.(\ref{eq:Phi}),
these can be conveniently rewritten as a modulus, $\phi_4$, times a unit 4-vector. The unit vector can in turn be expressed 
as a constant vector invariant under electromagnetic ($U(1)_{em}$) transformations times a phase $\exp(i \chi^i(x) A^i /v)$, 
where $A^i$ are $SO(4)^\prime/SO(3)$ generators. 
Considering that  $||\Phi|| =1$ implies $\phi_4 \leq 1$, and that in the vacuum $\langle \phi_4 \rangle= \sin\theta$,  
it is convenient to define $\phi_4(x) \equiv \sin(\theta + h(x)/f)$. Hence,
\begin{equation} \label{eq:Phi2}
\Phi = 
\begin{pmatrix}
\sin(\theta+h(x)/f)\;   e^{i \chi^i(x) A^i/v}\!
\begin{pmatrix}
0 \\ 0 \\ 0 \\ 1
\end{pmatrix} \\
\cos(\theta+h(x)/f)  
\end{pmatrix} \, .
\end{equation}
By construction, the three $\chi^i$ are the fields eaten after the $SU(2)_L \times U(1)_Y$ external gauging is turned on,
while $h$, which parametrizes $SO(4)^\prime$-invariant fluctuations around the vacuum~$\theta$, 
remains in the spectrum as a pseudo-NG boson. It is thus identified with the Higgs boson. 
By equating (\ref{eq:Phi}) and (\ref{eq:Phi2}) one obtains the (non-linear) field redefinition that relates the four NG bosons 
of $SO(5)/SO(4)$, $\pi^{\hat a}$, and the `physical' degrees of freedom, $\chi^i$, $h$: 
\begin{equation} \label{eq:fieldredef}
\begin{split}
& \sin(\theta+ h(x)/f) \, \hat \chi^i(x) \sin(\chi(x)/v) = \hat \pi^i(x) \sin(\pi(x)/f), \qquad\qquad  i=1,2,3 \\[0.25cm]
& \cos(\theta+ h(x)/f) = \cos(\pi(x)/f) \cos\theta - \hat \pi^4(x) \sin(\pi(x)/f) \sin\theta\, ,
\end{split}
\end{equation}
where $\chi \equiv \sqrt{(\chi^i)^2}$, $\hat \chi^i \equiv \chi^i/\chi$.

In realistic models, the  value of $\theta$  is  dynamically determined, and the breaking of the electroweak symmetry
can be seen as the result of a vacuum misalignment. Another point of view, however, is possible and sometimes
useful. If all the explicit breaking of the global $SO(5)$ comes from the $SU(2)_L \times U(1)_Y$ external  gauging and from the couplings
of other elementary fields (in particular the SM fermions),
then at  tree level the orientation of the vacuum is arbitrary and one can
suitably set $\theta = 0$ (so that $SO(4)^\prime = SO(4)$).  
With this choice, the four NG bosons of $SO(5)/SO(4)$ transform as a 
a complex doublet of the gauged $SU(2)_L$, and none of them is eaten.  Loop corrections 
will however generate a potential for the NG bosons and can lead to a non-vanishing vev for the modulus of the NG  4-vector: 
$\langle \pi \rangle \not = 0$ (see Fig.~\ref{fig:S4}).
As a result,  $SO(4)$ is spontaneously broken to (a custodial) $SO(3)$, and three of the original NG bosons are eaten.
The field $\Phi$ can  be recast in the form of eq.(\ref{eq:Phi2}) by  identifying $\theta = \langle \pi \rangle/f$ and the field $h(x)$ as the 
fluctuation of the modulus of the  NG 4-vector around its vev.
One can thus think of the electroweak symmetry breaking as a two-step process: a first spontaneous breaking, $SO(5)\to SO(4)$,
occurs at the scale $f$, giving rise to an $SU(2)_L$ doublet of NG bosons;  at a lower scale $v =   f \sin(\langle \pi \rangle/f) \equiv f \sin\theta$
the electroweak symmetry is spontaneously broken, $SO(4)\to SO(3)$, leaving an approximate custodial symmetry.

A simple way to derive the $SO(5)/SO(4)$ chiral lagrangian at $O(p^2)$ is by adopting the basis of fields $\{ \chi^i, h\}$ and
making use eq.(\ref{eq:Phi2}). One has (see Appendix~\ref{app:proofs}): 
\begin{equation} \label{eq:standardOp2}
\begin{split} 
{\cal L}^{(2)} & = \frac{f^2}{2} (D_\mu \Phi)^T  (D^\mu \Phi) \\[0.2cm]
& =  \frac{1}{2} (\partial_\mu h)^2 + \frac{f^2}{4} \Tr\!\left[  (D_\mu \Sigma)^\dagger (D^\mu \Sigma) \right] \sin^2\!\left( \theta + \frac{h(x)}{f} \right)\, ,
\end{split}
\end{equation}
where $\Sigma \equiv \exp(i \sigma^i \chi^i/v)$, and $\sigma^i$ are the Pauli matrices.
No covariant derivative acts on $h$, as one could have anticipated by noticing that
the fluctuations parametrized by this field are $SO(4)'$-invariant.
Choosing the unitary gauge, $\Sigma =1$, and expanding around $\theta$, one immediately finds the relation
$m_W^2 = (g^2 f^2 \sin^2\theta)/4$, which determines the value of the electroweak scale $v = f \sin\theta$, 
and the value of the Higgs couplings to the vector bosons.

The same expression for ${\cal L}^{(2)}$ can be obtained by using the CCWZ formalism. 
At the level of two derivatives, there is only one operator which can be formed:
\begin{equation} \label{eq:CCWZOp2}
{\cal L}^{(2)} = \frac{f^2}{4} \Tr\!\left[ d_\mu d^\mu \right]\, .
\end{equation}
The equivalence with eq.(\ref{eq:standardOp2}) is proved in Appendix~\ref{app:proofs}, but it can be quickly checked, for example, 
by monitoring the mass terms for the vector bosons.
In the case of eq.(\ref{eq:CCWZOp2}) these arise from the component of the gauge fields along the broken generators contained in $d_\mu$.
From eq.(\ref{eq:Udef}) and (\ref{eq:DefOfdEgauged}), after setting $\Pi(x) = \sqrt{2}\, T^{\hat a}(\theta) \pi^{\hat a}(x)/f$, one finds:
\begin{align}
\label{eqL:dmuforSO5}
d_\mu^{\hat a} &= A_\mu^{\hat a} + \frac{\sqrt{2}}{f} \left( D_\mu \pi\right)^{\hat a} + O(\pi^3) \\[0.15cm]
E_\mu^{a} & = A_\mu^a - \frac{i}{f^2} \big( \pi \overleftrightarrow{D}_{\!\!\mu\,} \pi \big)^{a} + O(\pi^4)\, .
\end{align}
For a generic $\theta$, the components of the external gauge fields are given by
\begin{align}
\label{eq:brokenA}
& A^{\hat a}_\mu :\quad \begin{cases}
A^{\hat i}_\mu = \displaystyle\frac{\sin\theta}{\sqrt{2}} \left( W_\mu^i - \delta^{i3} B_\mu \right), 
\qquad i=1,2,3 \\[0.3cm]
A^{\hat 4}_\mu = 0
\end{cases} 
\\[0.3cm]
\label{eq:unbrokenA}
& A^{a}_\mu :\quad \begin{cases}
A^{a_L}_\mu =  \left( \displaystyle\frac{1+\cos\theta}{2}\right) W_\mu^{a}  + 
   \delta^{a 3} \left( \displaystyle\frac{1-\cos\theta}{2}\right)  B_\mu \\[0.6cm]
A^{a_R}_\mu =  \left( \displaystyle\frac{1-\cos\theta}{2}\right) W_\mu^{a}  + 
 \delta^{a 3}  \left( \displaystyle\frac{1+\cos\theta}{2}\right) B_\mu \, ,
\end{cases}
\end{align}
where $W_\mu^a$, $B_\mu$ are the $SU(2)_L\times U(1)_Y$ vector bosons.
Using eq.(\ref{eqL:dmuforSO5}) and (\ref{eq:brokenA}) one can easily derive the $W$ and $Z$ mass terms from eq.(\ref{eq:CCWZOp2}).

Notice that in the case of $SO(5)/SO(4)$, the unbroken generators transform as a \textit{reducible} representation of 
$SO(4) \sim SU(2)_L \times SU(2)_R$, namely they form a $(\mathbf{3}, \mathbf{1}) + (\mathbf{1}, \mathbf{3})$.
Each of the variables $E_\mu$, $f^+_{\mu\nu}$ can thus be divided into two individually covariant structures, aligned
respectively  along the $ SU(2)_L$ and $ SU(2)_R$ generators:
\begin{equation}
\begin{aligned}
E^{L,R}_\mu & \equiv E^{a_{L,R}}_\mu T^{a_{L,R}}(\theta)  
 &  \hspace{0.5cm} &
E^{L,R}_\mu(\Pi)  \to h(\Pi,g) \, E^{L,R}_\mu(\Pi) \, h^\dagger(\Pi,g)  - i \, h(\Pi, g)  \partial_\mu h^\dagger(\Pi, g) \\[0.4cm]
f^{L,R}_{\mu\nu} & \equiv (f^+_{\mu\nu})^{a_{L,R}} T^{a_{L,R}}(\theta)
 &  &
f^{L,R}_{\mu\nu}(\Pi)  \to h(\Pi,g)\,  f^{L,R}_{\mu\nu}(\Pi) \, h^\dagger(\Pi,g)\, .
\end{aligned}
\end{equation}

In absence of the external gauging, the chiral lagrangian (\ref{eq:CCWZOp2}) has various discrete symmetries 
that lead to some important selection rules.  Although a partial gauging of $SO(5)$ eventually breaks explicitly 
some of these parities,  it is useful to identify them 
by thinking of $SO(5)$ as being fully gauged by external vectors, 
which is equivalent to formally assigning a transformation rule for $f^{\pm}_{\mu\nu}$. 
The following discrete symmetries, approximate or exact, are useful to describe the phenomenology of our model:

\vspace{0.5cm}
\textsf{Grading of the algebra, $R$:}
\vspace{0.3cm}

The quotient space $SO(5)/SO(4)$ is symmetric, that is, there exists an automorphism of the algebra  (grading), $R$, 
under which  the broken generators change sign:
\begin{align}
\label{eq:gradingR}
& \begin{aligned}
    T^{\hat a}(\theta) &\to - T^{\hat a}(\theta) \\[0.2cm]
    T^a(\theta)  &\to + T^a(\theta) 
    \end{aligned}
  &  
& R = \begin{pmatrix}
          -1 & & & & \\
          & -1 & & & \\
          & & -1 & & \\
          & & & -\cos(2\theta) & \sin(2\theta) \\
          & & &  \sin(2\theta) & \cos(2\theta)
          \end{pmatrix}\, .  \\
\intertext{The transformation rules of the fields are:}
& \pi^{\hat a} \to - \pi^{\hat a}\, ,   
 &  
& \begin{aligned}
    d_\mu^{\hat a} & \to - d^{\hat a}_\mu \\[0.1cm]
    E_\mu^{a} & \to + E^{a}_\mu \\[0.13cm]
    f_{\mu\nu}^\pm & \to \pm f_{\mu\nu}^\pm \, .
    \end{aligned}
\end{align}
Notice that for a generic $\theta$, the action of $R$  is linear on the $SO(5)/SO(4)$ NG bosons (all the $\pi$'s are odd), but
non-linear on the fields $\chi^i$, $h$. 
Since $R$ is an element of the unbroken global $SO(4)$ (for any value of $\theta$), i.e. it is an internal automorphism 
of the algebra, it will be an exact symmetry of the lagrangian at \textit{any}  chiral order in  absence of the weak gauging. 
In particular, it will be unbroken to all orders in the chiral lagrangian. This implies that any process involving an odd number of NG bosons will
vanish in the limit of zero electroweak couplings $g = g^\prime=0$. When the $SU(2)_L \times U(1)_Y$ gauging is turned on,
the grading $R$ is explicitly broken, except for $\theta =0, \pi$.

\vspace{0.5cm}
\textsf{LR parity, $P_{LR}$:}
\vspace{0.3cm}

The parity $P_{LR}$ exchanges the generators of the $SU(2)_L$ and $SU(2)_R$ subgroups of $SO(4)$, and changes
sign to the first three broken generators ($\eta^{\hat a} \equiv (+1, +1, +1, -1)$): 
\begin{align}
& \begin{aligned}
    T^{\hat a}(\theta)   &\to - \eta^{\hat a}\, T^{\hat a}(\theta)  \\[0.2cm]
    T^{a_L}(\theta)      &\leftrightarrow T^{a_R}(\theta) 
    \end{aligned}
  &  
& P_{LR} = \begin{pmatrix}
          -1 & & & & \\
          & -1 & & & \\
          & & -1 & & \\
          & & & +1 & \\
          & & & & +1 
          \end{pmatrix}\, .  \\
\intertext{This implies the following transformation rules for the NG bosons and the CCWZ covariant variables:}
&      \pi^{\hat a}  \to - \eta^{\hat a}\, \pi^{\hat a}\, , 
 &  
& \begin{aligned}
    d_\mu^{\hat a} & \to - \eta^{\hat a}\, d^{\hat a}_\mu \\[0.1cm]
    E_\mu^{L} & \leftrightarrow  E^{R}_\mu \\[0.13cm]
    f_{\mu\nu}^L & \leftrightarrow f_{\mu\nu}^R \\[0.13cm]
    (f_{\mu\nu}^-)^{\hat a} & \to - \eta^{\hat a}\, (f_{\mu\nu}^-)^{\hat a}\, .
    \end{aligned}
    \label{LRtransf}
\end{align}
The action of $P_{LR}$ is linear on the $SO(5)/SO(4)$ NG bosons, as well as on the $\chi^i$, $h$ fields, see eq.(\ref{eq:fieldredef}). 
In particular, one has: $\chi^i \to -\chi^i$ ($i=1,2,3$), and $h\to + h$. The $P_{LR}$ symmetry was invoked in the literature on 
composite Higgs models as a way to suppress large corrections to the $Zb\bar b$ vertex~\cite{Agashe:2006at}. 
By eq.(\ref{LRtransf})  we see that $P_{LR}$ is an invariance of the 
Goldstone boson lagrangian at $O(p^2)$ (see also \cite{Mrazek:2011iu}), but, since it is not an element of $SO(4)$, one expects that it will generally be broken at $O(p^4)$, even in absence of the
external gauging. In other words, $P_{LR}$ is an accidental symmetry of the chiral lagrangian at $O(p^2)$.
On the other hand, since $P_{LR} \in O(4)$, one could simply enforce it by requiring that the symmetry breaking pattern be
$O(5) \to O(4)$.  
In either case,  the accidental $P_{LR}$ invariance at $O(p^2)$ implies that 
processes involving an odd number of $\chi$'s, like the scattering of longitudinal vector bosons $W_L W_L \to Z_L h$,  vanish 
at leading chiral order for $g, g^\prime =0$. In particular, this shows that the scattering amplitude for $W_L W_L \to Z_L h$
will behave like a constant at large energies $E \gg m_W, m_h$, rather than growing like $E^2$. When the $SU(2)_L \times U(1)_Y$ gauging is turned on,
$P_{LR}$  is explicitly broken for any value of $\theta$.

\newpage
\textsf{Higgs parity, $P_{h}$:}
\vspace{0.3cm}

A further discrete symmetry can be constructed as the product of the grading $R$ and $P_{LR}$: $P_h = R\cdot P_{LR}$. Like $R$,  for generic $\theta$,  $P_h$ is a symmetry at $O(p^2)$  in the gaugeless limit $g,g'=0$.
Its action on the NG bosons is such that the $\pi^{i}$ are even ($i = 1,2,3$), while $\pi^{4}$ is odd:
$\pi^{i}\to + \pi^{i}$, $\pi^{4}\to - \pi^{4}$.
In general, the action of $P_h$ on the fields $\chi^i$, $h$ is non-linear, as can be deduced from their
definition in terms of the $\pi^{\hat a}$ in eq.(\ref{eq:fieldredef}). However  at linear order
$h = \pi^{4} + \dots$, so  that  $\pi^{4} $  interpolates an asymptotic state with one Higgs boson.
In this sense, $P_h$ acts like a `Higgs parity' for any value of $\theta$ when neglecting $g$ and $g'$.  For the special value $\theta = \pi/2$,  
we have  $P_h=P_5\equiv {\rm diag}(+1,+1,+1,+1,-1)$ which commutes with the gauged $SO(4)'$. In that case $P_h$ is exact at $O(p^2)$ even in the 
presence of gauging, and it may  as well be enforced as an exact symmetry to all chiral orders.~\footnote{If the 
couplings of the SM fermions to the strong sector preserve it, $P_h$ can be enforced as an exact invariance of the full theory.
This is possible for example in the MCHM4~\cite{Agashe:2004rs} where fermions are embedded into spinorial representations
of $SO(5)$,  while in the MCHM5~\cite{Contino:2006qr}, with fermions in fundamental representations of $SO(5)$, the fermion couplings 
break $P_h$. On the other hand, $v=f$ ($\theta = \pi/2$) is phenomenologically excluded in the MHCM5, as it implies vanishing fermion 
masses. An exact  $P_h$ invariance was invoked in Ref.~\cite{hosotani} as a way to have a stable Higgs boson playing the role of a dark 
matter candidate.}
Notice also that for $\theta =\pi/2$  the action becomes linear,
$\chi^i \to + \chi^i$, $h \to -h$.

\vspace{0.5cm}
In addition to those described above, there are additional discrete symmetries which leave the lagrangian (\ref{eq:CCWZOp2}) invariant.
Similarly to $P_h$, other three parities can be defined, $P_i$,
under which the $i$-th NG boson changes sign while the others are even. Together with $P_h$,
these parities correspond to the four $Z_2$'s contained in $O(4)$.
Eq.(\ref{eq:CCWZOp2}) is also invariant under the
spatial parity $P_0: (t, \vec x) \to (t, -\vec x)$, and the ordinary parity $P = P_0 \cdot P_{LR}$ under which the $\chi$'s 
transform as pseudo-scalars and the Higgs as a scalar.
From the previous discussion it is clear that the ordinary parity $P$  itself is accidental at order $O(p^2)$, and that it will be broken
at higher orders.  In the following we will not make any restrictive assumption of the symmetries possessed by the EWSB
sector, and in particular we will allow for parity violation.

\subsection{The $SO(5)/SO(4)$ chiral lagrangian at $O(p^4)$ and its contribution to the scattering amplitudes}

At the level of four derivatives, we find that a complete basis for the $SO(5)/SO(4)$ chiral lagrangian
is made of 11 operators:
\begin{equation} \label{eq:L4}
\begin{split}
{\cal L}^{(4)} = & \sum_i c_i O_i  \\[0.5cm]
O_1 =& \Tr[ d_\mu d^\mu]^2  \\[0.1cm]
O_2 =& \Tr[ d_\mu d_\nu] \Tr[ d^\mu d^\nu] \\[0.1cm]
O_3 =& \left( \Tr [ E_{\mu\nu}^L E^{L\, \mu\nu}] -\Tr [ E_{\mu\nu}^R E^{R\, \mu\nu}] \right) \\[0.5cm]
O_4^+ =& \Tr \left[ (f_{\mu\nu}^{L} +f_{\mu\nu}^{R})  \, i [d^\mu , d^\nu ] \right] \\[0.1cm]
O_5^+ =& \Tr\left[ (f_{\mu\nu}^-)^2 \right] \\[0.1cm]
O_4^- =&\Tr \left[ (f_{\mu\nu}^{L} - f_{\mu\nu}^{R}) \, i [d^\mu , d^\nu ] \right] \\[0.1cm]
O_5^- =& \Tr\left[  (f_{\mu\nu}^{L})^2 - (f_{\mu\nu}^{R})^2 \right]  \\[0.5cm]
O_6^+ =& \epsilon^{\mu\nu\rho\sigma}\Tr \left[ (f_{\mu\nu}^{L} +f_{\mu\nu}^{R})  \, i [d_\rho , d_\sigma ] \right] \\[0.1cm]
O_7^+ =& \epsilon^{\mu\nu\rho\sigma}\Tr\left[ f_{\mu\nu}^- f^-_{\rho\sigma} \right] \\[0.1cm]
O_6^- =&\epsilon^{\mu\nu\rho\sigma} \Tr \left[ (f_{\mu\nu}^{L} - f_{\mu\nu}^{R}) \, i [d_\rho , d_\sigma ] \right] \\[0.1cm]
O_7^- =& \epsilon^{\mu\nu\rho\sigma}\Tr\left[  f_{\mu\nu}^{L} f^L_{\rho\sigma} -  f_{\mu\nu}^{R} f^R_{\rho\sigma}  \right] 
\end{split}
\end{equation}
Any other operator can be rewritten in terms of those shown above by means of  identities
valid in the case of a generic $\calG/\calH$ symmetric space or specifically for the $SO(5)/SO(4)$ coset, 
see Appendices~\ref{app:SO5generators} and~\ref{app:GoHsymm}.

It is useful to classify the $O_i$ according to their behavior under $P_{LR}$ and the ordinary parity $P$
(the quantum numbers under $P_h$ and $P_0$ can be easily derived in turn).
One has that:  $O_1$, $O_2$, $O_4^+$, $O_5^+$ are even under both $P_{LR}$ and $P$;
$O_3$, $O_4^-$, $O_5^-$ are odd under both $P_{LR}$ and $P$; $O_6^+$, $O_7^+$ are $P_{LR}$ even
and $P$ odd;  $O_6^-$, $O_7^-$ are $P_{LR}$ odd but $P$ even.

We will be concerned in what follows with the size of the coefficients of the above operators. For that purpose it is worth reminding the reader that 
since those operators represent a subleading effect in the chiral lagrangian their coefficients evolve  at 1-loop. In 1-loop accuracy we can thus 
conveniently write the value of the $c_i$ at the weak scale as
\begin{equation}
c_i(m_Z)=c_i(\mu_{UV}) +\frac{b_i}{16\pi^2}\ln\frac{m_Z}{\mu_{UV}}
\end{equation}
where $\mu_{UV}$ is some UV scale. The $b_i$ are calculable $O(1)$ coefficient, fully determined by the $O(p^2)$ chiral lagrangian. In our discussion we shall mostly be concerned with physical situations where $c_i(\mu_{UV})$ dominate the above equation.

Only the first three operators in eq.(\ref{eq:L4}) do not vanish in the limit in which the weak $SU(2)_L \times U(1)_Y$ gauging
is turned off, and  thus contribute at leading order to the scattering of NG bosons. The other operators are thus not relevant
for the following analysis and will not enter our discussion. The only exception is $O_5^+$, which is the operator corresponding
to the Peskin-Takeuchi $S$ parameter~\cite{ST}. It can be recast in the form (see Appendix~\ref{app:proofs}):
\begin{equation} \label{eq:O5}
O_5^+ = \frac{1}{2} \sin^2\!\left(\theta + \frac{h(x)}{f} \right) \Big(  (W^a_{\mu\nu})^2 + (B_{\mu\nu})^2 
-   \Tr\left[ \Sigma^\dagger \, W_{\mu\nu}^a \sigma^a \, \Sigma \, B_{\mu\nu} \sigma^3 \right] \Big)\, ,
\end{equation}
where, we recall, $\Sigma = \exp(i \chi^i(x) \sigma^i/v)$. From the last term in the parenthesis it follows
\begin{equation} \label{eq:Spar}
\Delta \hat S = 2 g^2\sin^2\theta\,c_5^+ \, .
\end{equation}
The logarithmically divergent contribution to this operator was discussed in Ref.~\cite{Barbieri:2007bh}, while Ref.~\cite{Giudice:2007fh} 
focussed on the UV-saturated contribution.

As a last remark we notice that the operator $O_3$, being $P_{LR}$ and $P_h$ odd, would have the right
quantum numbers to contribute to the scattering $W_L W_L \to Z_L h$, which does not arise at leading chiral order
because it violates both $P_{LR}$ and $P_h$. However, by switching off the external gauging and expanding $O_3$ 
one finds that the leading  term with four  NG bosons identically vanishes, that is: $O_3 \sim O(\pi^6)$. This is simply because, 
up to integration by parts, the only $P_{LR}$-odd structure one can form with four NG bosons and four derivatives is 
$\epsilon^{IJKL}\partial_\mu \pi^I\partial^\mu\pi^J\partial_\nu \pi^K\partial^\nu\pi^L$, which trivially vanishes by Bose symmetry.
This implies that $O_3$ does  not in fact contribute to $2\to 2$ scattering amplitudes of NG bosons
(although it will contribute to $2\to 4$ ones). One can check that for $g=g^\prime =0$ the first non-vanish contribution to 
$W_L W_L \to Z_L h$ arises at $O(p^6)$, for instance from the operator  $\epsilon^{IJKL}d_\mu^I(\nabla_\rho d^{\mu})^Jd_\nu^K(\nabla^\rho d^\nu)^L$.

Notice also in passing that for $SO(5)/SO(4)$ no Wess-Zumino-Witten term~\cite{WZW} can be constructed, since the fifth de Rham cohomology 
group of $SO(5)/SO(4)$ vanishes. Anyway such term would only affect amplitudes with at least five legs.

At this point we are ready to derive the contribution of the $O(p^4)$ lagrangian
to the scattering amplitudes for $\pi \pi \to \pi\pi$. 
Since at linear order $\pi^i = \chi^i + \dots$ and $\pi^4 = h + \dots$  (see eq.(\ref{eq:fieldredef})),
by virtue of the Equivalence Theorem~\cite{equivtheorem}, 
the  $\pi \pi \to \pi\pi$ amplitude reproduces the large-energy behavior of the amplitudes for the scattering among  the Higgs 
and longitudinal vector bosons (see Appendix~\ref{app:scattampl}).
The general amplitude can be decomposed in terms of \textit{two}  functions:
\begin{equation} \label{eq:NGBscattampldef}
{\cal A}(\pi^a \pi^b \to \pi^c \pi^d) = A(s,t,u)\, \delta^{ab}\delta^{cd} +
A(t,s,u)\, \delta^{ac}\delta^{bd} + A(u,t,s)\, \delta^{ad}\delta^{bc} 
+ B(s,t,u)\, \epsilon^{abcd} \, ,
\end{equation}
where $s$, $t$, $u$ are the Mandelstam variables. The functions $A(s,t,u)$ and $B(s,t,u)$ are respectively associated with  $P_{LR}$-even and 
$P_{LR}$-odd transitions. In particular, $B$ parametrizes the amplitude for $W_L W_L \to Z_L h$ and crossed
processes. Crossing symmetry requires that
$A(s,t,u)=A(s,u,t)$, and that $B(s,t,u)$ be antisymmetric under the exchange of any two Mandelstam variables.
The latter requirement implies that the lowest-order contribution to $B$ arises at $O(p^6)$, that is
$B\propto (s-u)(u-t)(t-s)$, in accordance with our previous discussion.
From the $O(p^4)$ lagrangian given in
eqs.(\ref{eq:CCWZOp2}) and (\ref{eq:L4}), one can easily derive:
\begin{equation} \label{eq:scattamplfromci}
\begin{split}
A(s,t,u) & = \frac{s}{f^2} + \frac{4}{f^4} \left[ 2 c_1 \, s^2 + c_2 \left(t^2+u^2 \right) \right] \\[0.1cm]
B(s,t,u) & = 0\, .
\end{split}
\end{equation}
Some comments concerning the sign and size of the coefficients $c_{1,2}$ are in order.~\footnote{We assume $c_{1,2}(\mu_{UV})$ dominates 
over the 1-loop running contribution to $c_{1,2}$.} 
Unitarity and causality are known to imply the constraint $c_1+c_2 >0$, $c_2>0$~\cite{causality}. 
However, that does not translate into a definite sign for the 
interference with the leading term when considering the squared amplitude. This can be seen by considering the plots of the following section 
where we consider the contributions to $c_{1,2}$ from heavy resonances, which satisfy the unitarity and causality bound. For instance, while the 
exchange of a vector enhances the $W^+W^-$ final state and depletes  $hh$, the converse is true for the exchange of a heavy singlet scalar. 
Let us consider instead the size of $c_{1,2}$.
Notice that, by factoring out the leading term $s/f^2$, the subleading corrections are characterized by the scales
$f/ \sqrt{c_{1,2}}$, that can be interpreted as the next physics threshold. 
If the only contribution to the chiral coefficients $c_{1,2}$ came from physics at the maximally allowed cutoff scale, $\Lambda_{\rm max} \sim 4\pi f$, the one at which $\pi\pi$ scattering becomes maximally strong,
then a naive estimate gives $c_{1,2} \sim 1/16\pi^2$. In that case the contribution of $O_{1,2}$ to the scattering amplitudes is always
subleading for energies $E\lesssim 4\pi f$.  If however some of the resonances of the strong sector where  the global
symmetry breaking takes place is somewhat lighter than the cutoff, then its contribution can become relevant. 
This possibility will be studied in the following sections.

\section{Lagrangian for spin-1 and spin-0 resonances}
\label{philosophy}

We shall now consider the case in which one single resonance is lighter than the cutoff scale. 

Before going into the details we should  briefly illustrate our  assumptions.
By necessity, the cut-off $\Lambda$ of our effective $\sigma$-model description should satisfy $\Lambda \leq 4 \pi f$, since the latter is the scale 
where it becomes strongly coupled. We could also  trade $\Lambda$ for the $\sigma$-model coupling at that scale $g_*\equiv \Lambda/f$. At around 
the scale $\Lambda$ we  expect new states, and if the coupling strength is universal, they will couple to each other with strength $g_*$. 
Our assumption now is that there is one further resonance $\Phi$ with mass $m_\Phi< \Lambda$,
whose interactions to the NG bosons and to itself are described by a set of couplings $\{g_\Phi^{i}\}$. 
We can think of two plausible criteria to constrain the couplings $\{g_\Phi^{i}\}$ and to assess their most likely range.
The first and less conservative criterion is given by the request that the new couplings stay weak up to the cut-off scale $\Lambda$.  
The second, similar but stronger, is that the couplings should  not exceed, and preferably saturate, the $\sigma$-model coupling 
$g_*\equiv \Lambda/f$ at the scale $\Lambda$. 
This second criterion is  perhaps more plausible than the first one, as it is basically realized at large $N$ or in weakly-coupled 5D field theories, 
with $g_*$ equalling respectively $4\pi/\sqrt N$ and the Kaluza-Klein coupling $g_{KK}$. Moreover, as will become clear below,
one interesting  consequence of couplings that saturate the second criterion is that at $s> m_\Phi^2$ the 
leading linear growth of ${\cal A}(\pi\pi\to \pi\pi)$ with $s$ changes by $O(1)$ with respect to the pure $\sigma$-model result at $s\ll m_\Phi^2$.
This means that $\Phi$ is as  important as the heavier resonances  in the UV completion  of $\pi\pi$ scattering, or, in other words, that its  
exchange {\it partially UV completes} $\pi\pi$ scattering. In view of that implication we shall from now on refer to the second (stronger) criterion as 
Partial UV Completion (PUVC).  Notice that the request that the new resonance $\Phi$  ``unitarizes'' or minimizes the amplitude, as for instance 
requested in Ref.~\cite{Barbieri:2008cc}, is a very special case of partial UV completion, in particular it leads to couplings of similar size. PUVC, 
as we defined it, 
is however distinguished from ``unitarization'', as it does not necessarily imply a reduced scattering amplitude, or if so only a partial reduction.
Keeping in mind  our goal of parametrizing unknown physics in as much generality as possible, PUVC seems a reasonable criterion to 
adopt.~\footnote{It must however be noted that Ref.~\cite{Barbieri:2008cc}  emphasizes how in low-energy QCD the $\rho$ vector meson seems 
to remarkably closely adhere to the more restricted request of ``unitarization".}
In view of its more interesting properties, and in view of the parameter reduction  
it entails, we will adopt PUVC in    our phenomenological analysis in section 4.

We will study only the case of resonances which have the right quantum numbers under $SO(4)$ to be exchanged in  $\pi\pi$ scattering. 
Since the NG bosons transform as a $\mathbf{4}$ of $SO(4)$, the resonance must fall in one of the following representations:
$\mathbf{4}\times \mathbf{4} = \mathbf{1}+ \mathbf{6} +\mathbf{9}$ of $SO(4)$, that is $(\mathbf{1},\mathbf{1}) + 
(\mathbf{3},\mathbf{1}) + (\mathbf{1},\mathbf{3}) + (\mathbf{3},\mathbf{3})$ of $SU(2)\times SU(2)$.
Bose symmetry then implies that the possible spin assignments are (we consider only spin 0 and spin 1 resonances):
\begin{itemize}
\item[--] $\eta = (\mathbf{1},\mathbf{1})$, spin$\,=0$ [isospin$\, = 0$]
\item[--] $\rho^L=(\mathbf{3},\mathbf{1})$, $\rho^R =(\mathbf{1},\mathbf{3})$, spin$\,=1$  [isospin$\, = 1$]
\item[--] $\Delta = (\mathbf{3},\mathbf{3})$, spin$\,=0$  [isospin$\, = 0+1+2$]
\end{itemize}
In parenthesis we have indicated the decomposition under $SO(3)$ (isospin) quantum numbers; for example, the scalar $\Delta$ contains
three components respectively with isospin $0$, $1$, $2$.

In the following we will consider each of above resonances, 
writing down its chiral lagrangian and computing its contribution to the $\pi\pi\to \pi\pi$ scattering amplitude.
Let us start with the case of a spin-1 $\rho^L$.

\subsection{$\rho^L = (3,1)$}
\label{subsec:rhoL}

We follow the vector formalism and describe the spin-1 resonance with a field $\rho_\mu$ transforming 
non-homogeneusly,~\footnote{See for example Ref.~\cite{Ecker:1989yg} for a comparison among different formalisms.}
\begin{equation}
\rho_\mu \to h(\Pi, g)\, \rho_\mu \, h^\dagger(\Pi, g) - i \, h(\Pi, g)  \partial_\mu h^\dagger(\Pi, g) \, ,
\end{equation}
so that a field strength can be defined as $\rho_{\mu\nu} = \partial_\mu \rho_\nu - \partial_\nu \rho_\mu + i[\rho_\mu , \rho_\nu]$. We are interested in a 
physical situation where the interactions among the NG bosons and the $\rho$ at a scale of the order of its mass $m_\rho$ are weak and described by 
the lowest terms in a derivative expansion. For instance, a plausible assumption is that the derivative expansion is controlled by $\partial/ \Lambda$,  
the cut-off scale satisfying $m_\rho \ll \Lambda \leq 4\pi f$. It is thus convenient to write the effective lagrangian for the $\rho$ in the form
\begin{equation}
\label{eq:rhoLag}
{\cal L}^{(\rho_L)} = 
-\frac{1}{4g_{\rho_L}^2} \rho_{\mu\nu}^{a_L} \rho^{a_L \, \mu\nu} + \frac{m_{\rho_L}^2}{2g_{\rho_L}^2} \left(
\rho_\mu^{a_L} - E_\mu^{a_L} \right)^2 + \sum_i \alpha_i\, Q_i \, ,
\end{equation}
where $Q_i$ denote the (infinite) series of higher-order (in field and derivative) operators. 
Without these latter, ${\cal L}^{(\rho_L)}$ coincides with the lagrangian originally 
considered in $SO(4)/SO(3)$ models of Hidden Local Symmetry (HLS)~\cite{Bando:1984ej,BESS,Georgi:1989xy}.
For  phenomenological applications we are interested in operators that affect $\pi\pi$ scattering as well as operators that  play a role in 
precision electroweak tests by modifying the electroweak vector boson propagators. Within this limited but reasonable perspective, a straightforward 
operator analysis singles out four leading independent operators
\begin{equation} \label{eq:Qi}
\begin{split}
Q_1 &= \Tr \left(  \rho^{\mu\nu} \, i [d_\mu, d_\nu] \right) \\[0.1cm]
Q_2 &=  \Tr \left(  \rho^{\mu\nu}  f_{\mu\nu}^+  \right) \\[0.3cm]
Q_{3} &= \epsilon^{\mu\nu\rho\sigma} \Tr \left( \rho_{\mu\nu} \, i[d_\rho, d_\sigma]  \right) \\
Q_{4} &= \epsilon^{\mu\nu\rho\sigma} \Tr \left(  \rho_{\mu\nu}   f^+_{\rho\sigma}  \right). \\ 
\end{split}
\end{equation}
In Appendix~\ref{sec:operators} we give more details on the analysis that leads to the above result and also list 
additional operators that can affect other processes, like for example $\rho\pi\to\rho\pi$. Operators that  do not involve the $\rho_{\mu\nu}$ field 
strength are written in terms of $\bar \rho_\mu \equiv \rho_\mu -E_\mu$. For instance, in order to illustrate our working assumption, it is useful to consider 
the operator $Q_{1(2)}=(\nabla^\mu \bar \rho_\mu)^2$, which we did not list among the leading ones. This operator has the same structure as the $\rho$ 
mass term, but with two additional derivatives. According to our assumption that the derivative expansion is controlled by the scale 
$\Lambda = g_* f \gg m_\rho$,
we must have  $\alpha_{1(2)}\lesssim 1/g_*^2$.
This is sensible since, at it is well known, $Q_{1(2)}$ implies the presence of a scalar ghost with mass $m_\rho /(g_\rho \sqrt {\alpha_{1(2)}})$.

Among the leading operators, $Q_1$ and $Q_3$ in principle  affect the $\pi\pi\rho$ vertex, and thus $\pi\pi$ scattering, but in fact
only $Q_1$ does,  since the contribution from $Q_3$ is easily shown to vanish because of Bose symmetry. 
The mass term also includes a contribution to the $\pi\pi\rho$ vertex, as well as a 
contact interaction among four NG bosons. 
Upon expanding in the number of NG fields and using the commutation relations of eq.(\ref{eq:SO5algebra}), 
one has ($i,j,k=1,2,3$, $a,b = 1,2,3,4$):
\begin{equation} \label{eq:rhopipivertex}
\begin{split}
{\cal L}^{(\rho_L)}  \supset 
  & \, - \frac{a^2_{\rho_L} g_{\rho_L}}{2} \left[  \epsilon^{ijk} \pi^i \partial^\mu \pi^j \rho_\mu^{k} 
        + \left( \pi^k \partial^\mu \pi^4 - \pi^4 \partial^\mu \pi^k \right) \rho_\mu^k \right] \\[0.15cm]
  & -  \frac{a^2_{\rho_L}}{8 f^2} \left[ \left(\pi^a \partial_\mu \pi^a \right)^2 - \left(\pi^a \partial_\mu \pi^b \right)^2 \right] \\[0.15cm]
  & - \frac{2g_\rho \alpha_1}{f^2} \left[  \epsilon^{ijk}  \partial^\mu \pi^i \partial^\nu \pi^j  \partial_\mu\rho_\nu^{k} -
     \left( \partial^\mu \pi^4 \partial^\nu \pi^k - \partial^\mu \pi^k \partial^\nu \pi^4 \right)  \partial_\mu \rho_\nu^{k} \right]
  + \dots \, ,
\end{split}
\end{equation}
where we have rescaled the $\rho$ field in order to canonically normalize its kinetic term and we have defined
\begin{equation} \label{eq:arhodef}
a_{\rho_L}  \equiv \frac{m_{\rho_L} }{g_{\rho_L} f}\, .
\end{equation}

At energies $E\ll m_{\rho_L} \ll \Lambda$ one can integrate out the $\rho$ by solving the equations of motion
at lowest order in the derivative expansion. One has~\footnote{Notice that in the power counting of the low-energy effective 
lagrangian $\rho_\mu$ counts as one derivative.}
\begin{equation} \label{eq:rhoEOM}
\rho^L_\mu = E^L_\mu + O\!\left( \frac{E^2}{m_{\rho_L}^2} \right)\, .
\end{equation}
This implies, first of all, that the $O(p^2)$ lagrangian in eq.(\ref{eq:CCWZOp2}) is not renormalized by the exchange of the $\rho$~\cite{Weinberg:1968de}.
As concerns the $O(p^4)$ lagrangian in eq.(\ref{eq:L4}), we notice that $Q_5$ and $Q_6$ discussed in the Appendix, as well as all the operators with 
two or more $\bar \rho's$,  vanish at leading chiral order on the equations of motions, since they contain at least one power of $\bar \rho$. 
Those operators therefore do not affect the $O(p^4)$ low-energy lagrangian. The latter is only affected by $Q_1$-$Q_4$.
Plugging (\ref{eq:rhoEOM}) into (\ref{eq:rhoLag}) gives rise to the effective lagrangian  of eq.(\ref{eq:L4}) with coefficients
\begin{equation} \label{eq:cifromrhoL}
\begin{split}
c_1 &= - c_2 = \frac{1}{2} c_3 = -\frac{1}{4} \left( \alpha_1 + \frac{1}{4g_{\rho_L}^2} \right) \\[0.1cm]
c_4^{+} &= \frac{1}{2} \left(\alpha_1-\alpha_2\right) + \frac{1}{4g_{\rho_L}^2} \\[0.1cm]
c_5^+ &= \frac{1}{2} \left( \frac{1}{4g_{\rho_L}^2} - \alpha_2 \right) \\[0.1cm]
c_4^- &= -c_5^- = -\frac{1}{2} \left( \alpha_2+\alpha_1\right) \\[0.1cm]
c_6^+ &=   \frac{1}{2} \left(\alpha_3-\alpha_4\right)\\[0.1cm]
c_7^+ &=   -\frac{\alpha_4}{2}\\[0.1cm]
c_6^- &= -c_7^- = -\frac{1}{2} \left( \alpha_3+\alpha_4\right)\, .
\end{split}
\end{equation}
Similarly, it can be shown that by integrating out a $\rho^R = (\mathbf{1},\mathbf{3})$, one obtains
the same value for all the coefficients of the $P_{LR}$-even operators ($c_1$, $c_2$, $c_4^+$, $c_5^+$, $c_6^+$, $c_7^+$),
and the same value but opposite sign for the the coefficients of the $P_{LR}$-odd operators
($c_3$, $c_4^-$, $c_5^-$, $c_6^-$, $c_7^-$).

By the above equations and from the results of the previous section, we can read the effects of the various parameters
on low-energy quantities: $\alpha_2$ affects the $S$ parameter while $\alpha_1$ and $g_\rho$ determine the $O(s^2)$ 
corrections to the $\pi\pi$ scattering amplitude. More explicitly we have
\begin{equation} \label{eq:ABfromrhoatOp4}
\begin{split}
A(s,t,u) & = \frac{s}{f^2} + \frac{1}{f^4} \left( \alpha_1 - \frac{1}{4g_{\rho_L}^2} \right)  \left( 2 s^2 -t^2-u^2 \right) \\[0.1cm]
B(s,t,u) & = 0
\end{split}
\end{equation}
and
\begin{equation} \label{eq:SparfromrhoL}
\Delta \hat S = a_{\rho_L}^2 \frac{m_W^2}{m_{\rho_L}^2} - 4 \alpha_2\frac{m_W^2}{v^2}\xi\, .
\end{equation}
The amplitude at energies of order $m_\rho$ depends on just one additional parameter, $m_\rho$ itself, and we shall present it later. 

In order to assess which are the most plausible regions of parameter space, we should study the theoretical constraints on the above parameters. 
Those are mostly associated to the strength of the interactions at energies around or above $m_\rho$. In the latter regime the longitudinal and 
transverse polarizations of the $\rho$ behave differently. The power counting is more readily done by parametrizing the longitudinal polarizations, 
$\rho_L$, in terms of an additional set of eaten NG bosons. The convenient way to proceed is to first formally take the limit $g_\rho\to 0$ with 
$m_\rho/g_\rho$ fixed, characterize the $\sigma$-model so obtained, and then re-introduce
the transverse $\rho_T$ field with coupling $g_\rho$ by gauging a subgroup of the global symmetry. It is rather clear that the $\sigma$-model 
one obtains in the first step can be written as $SO(5)\times SU(2)_H/SU(2)'_L\times SU(2)_R$, where, in an obvious notation, the embedding is
\begin{equation}
SU(2)_L\times SU(2)_R\sim SO(4)\subset SO(5) \qquad\qquad SU(2)_L'=\left [SU(2)_L\times SU(2)_H\right ]_{\rm diag}.
\end{equation}
This coset is parametrized by a set of 7 Goldstone bosons transforming as $(2,2)\oplus (3,1)$ under $SU(2)'_L\times SU(2)_R$, which we indicate as 
$\pi^{\hat i} =(2,2)$ (${\hat i}=1,\dots,4$) and $\eta^a=(3,1)$ ($a=1,2,3$). As explained in Appendix~\ref{sec:cosetrho}, a suitable parametrization of the 
coset when performing the CCWZ construction is $U=e^{i\pi}e^{i\eta}$.
This way  the $d$ symbols depend on $\pi$ and $\eta$ according to
\begin{align}
(2,2)\quad\rightarrow\quad d_\mu^{\hat i}&\equiv d_\mu^i(\pi)\sim \partial_\mu \pi +\pi^2\partial_\mu \pi+\dots \\[0.2cm]
(3,1) \quad\rightarrow\quad    \tilde d_\mu^{a}&\equiv \tilde d_\mu^a(\pi,\eta)\sim \partial_\mu \eta +\pi\partial_\mu \pi+\eta^2\partial_\mu \eta +\dots
\end{align} 
Notice that in $\tilde d$ there is an $O(\pi^2)$ term corresponding to the fact that the coset is not a symmetric space.
Finally, since the coset is reducible, the $O(p^2)$ lagrangian  is determined by two independent parameters (decay constants):
\begin{equation}
\label{Lnewcoset}
{\cal L}= f^2\, d_\mu d^\mu +f_\rho^2\, \tilde d_\mu\tilde d^\mu\, .
\end{equation}
By gauging $SU(2)_H$ by the introduction of $\rho_\mu$, the second term in the above equation is mapped into the $\rho$ mass term, 
hence the identification
\begin{equation}
m_\rho = g_\rho f_\rho \qquad\rightarrow\qquad a_\rho=\frac{f_\rho}{f}\, .
\end{equation}
We have now all the ingredients to estimate the strength of the interactions among $(W_L,\rho_L,\rho_T)\sim (\pi,\eta,\rho_T)$. 
Focussing on the $\rho$ kinetic lagrangian and on $Q_1$, we find the following result:

\vspace{0.2cm}
\begin{center}
 \includegraphics[width=28mm]{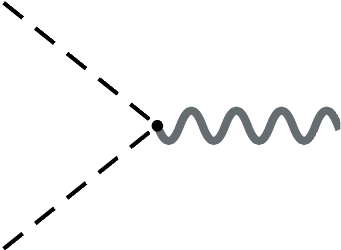}  \hspace{1cm}
\begin{minipage}{0.4\linewidth}
\vspace*{-1.7cm}
$\displaystyle  {\cal A}(\pi\pi\rho_T) \sim  g_\rho\left [\frac{f_\rho^2}{f^2}+\alpha_1\frac{E^2}{f^2}\right]$
\end{minipage}  \\[0.5cm]
\includegraphics[width=28mm]{rhopipi.pdf}  \hspace{1cm}
\begin{minipage}{0.4\linewidth}
\vspace*{-1.7cm}
$\displaystyle  {\cal A}(\pi\pi\rho_L) \sim  \frac{f_\rho E}{f^2}$
\end{minipage}  \\[0.5cm]

\includegraphics[width=28mm]{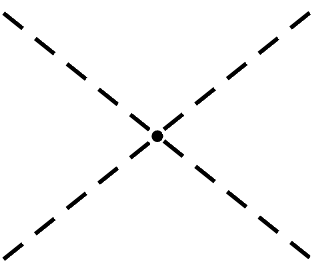}  \hspace{1cm}
\begin{minipage}{0.4\linewidth}
\vspace*{-2.15cm}
$\displaystyle  {\cal A}(\pi^4) \sim \frac{f_\rho^2E^2}{f^4}+\frac{E^2}{f^2}$
\end{minipage}  \\[0.5cm]
\includegraphics[width=28mm]{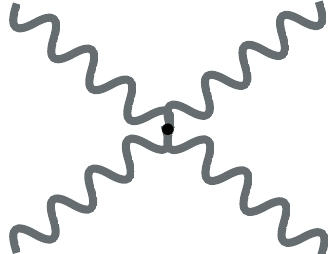}  \hspace{1cm}
\begin{minipage}{0.4\linewidth}
\vspace*{-1.8cm}
$\displaystyle  {\cal A}(\rho_L^4) \sim \frac{E^2}{f_\rho^2 }$
\end{minipage} 
\end{center}
\vspace{0.05cm}
Assuming now the  cut-off scale is $\Lambda\equiv g_* f\ll 4\pi f$, we can apply the two criteria discussed at the beginning of this section. 
The less conservative criterion correponds to the request  that the above interactions stay weak below $\Lambda$. This leads 
to the constraints
\begin{equation} 
\frac{\Lambda}{4\pi f} \lesssim a_\rho \lesssim \frac{4\pi f}{\Lambda}\, ,\qquad\qquad \alpha_1< \frac{1}{g_\rho g_*}\frac{4\pi}{g_*}\, .
\end{equation}
We thus conclude that $a_\rho$ is allowed to vary in a range around $1$. The coefficient $\alpha_1$ is bounded from above, but as long as
$g_*^2<4\pi g_\rho$, it can be as large as $1/g_\rho^2$, in which case it contributes significantly to $\pi\pi$ scattering also at and below the 
$\rho$ threshold (see equations in footnote~\ref{ftn:fullamplitude}). 

If we make the stronger, and maybe more realistic request that all couplings do not exceed $g_*$ at $g_*f$,~\footnote{More precisely, we require
the trilinear couplings to be $<g_*$ and the quadrilinears to be $< g_*^2$.}  we simply find
\begin{equation}
a_\rho \sim 1\, ,\qquad\qquad \alpha_1< \frac{1}{g_\rho g_*}<\frac{1}{g_\rho^2}\,.
\end{equation}
where in the last inequality we used that $g_\rho<g_*$, which follows from the hypothesis $m_\rho <g_* f$ for $a_\rho \sim 1$. 
Using the above  relations and  eqs.(\ref{rhoamplitude2}), (\ref{eq:ABfromrhofull}) one can check that
the contribution of the $\rho$  partially UV completes  ${\cal A}(\pi\pi\to \pi\pi)$, i.e. it changes
by $O(1)$ the rate of growth of the leading term (see also the estimate of the $\pi^4$ contact interaction in the figure above). 
That motivates us to simply call the second (stronger) criterion Partial UV Completion.
Notice indeed that in this case the $\rho$ behaves approximately like an elementary gauge boson, with the form factor parameter $\alpha_1$ negligible 
around the $\rho$ peak (see the $\pi\pi\rho$ amplitude estimates in the figure).

One final constraint concerns the parameter $\alpha_2$. 
In principle the operator $Q_2$ belongs to a different class
compared to $Q_1$, as it involves a coupling between the $\rho$ and the 
elementary gauge fields. In this sense it might not 
be subject to the request of having a strength $< g_*$ at the scale $\Lambda$, as imposed by PUVC.
There is however a constraint that $\alpha_2$ must satisfy for consistency: this parameter contributes an off-diagonal kinetic term for the vectors, 
so that when it is sufficiently large one vector will turn into a ghost. In order to avoid that, a bound must be satisfied:
\begin{equation}
\alpha_{2}< \frac{1}{g_\rho g_{SM}}\, .
\end{equation}
This bound leaves plenty of space for $\alpha_2$ to give a sizeable contribution to $S$. In particular, in the  range $\alpha_2\sim 1/g_\rho^2$ and for
$\alpha_2 >0$,  the contribution to $S$ can compensate the positive one from $\rho$ exchange (see eq.(\ref{eq:SparfromrhoL})). In our study,
we shall also consider the possibility to have rather low values of the $\rho$ mass, $m_\rho\sim 1\,$TeV, in a region where it is disfavored by the bounds 
on $S$. 
It is thus reassuring that we can compensate for $S$ with  a slight tuning of  $\alpha_2$. 
It is also interesting to notice that a value $\alpha_2\sim 1/g_\rho^2$ can lead to Vector Meson Dominance (VMD) in the form factor of the
pion to the external $SU(2)_L \times U(1)_Y$ gauge fields (the analog of the electromagnetic form factor of the pion in QCD) for a generic
$a_\rho \sim O(1)$.~\footnote{For example, 
by making use of the lagrangian ${\cal L}^{(\rho_L)}$ and keeping only the $\rho$ kinetic and mass terms and $Q_2$, 
we find that for $\theta =0$ the form factor that parametrizes the interaction of the pion with the  $SU(2)_L$ gauge field is given by
\begin{equation}
F(q^2) = 1- a_\rho^2 - \frac{a_\rho^2 m_\rho^2}{q^2-m_\rho^2}\left( 1- 2 g_\rho^2 \alpha_2  \frac{q^2}{m_\rho^2} \right)\, .
\end{equation}
The third term is due to the exchange of the $\rho$, while the second follows from the direct $W\pi\pi$ interaction that comes
from the $\rho$ mass term. The request of Vector Meson Dominance, namely that $F(-\infty) =0$, is satisfied for 
\begin{equation} \label{eq:alpha2VMD}
\alpha_2 = \frac{a_\rho^2-1}{2a_\rho^2} \, \frac{1}{g_\rho^2}\, .
\end{equation}
It thus follows that for $a_\rho > 1$ (so that $\alpha_2 > 0$) a reduction in $S$ is possible.
More precisely, by substituting eq.(\ref{eq:alpha2VMD}) in eq.(\ref{eq:SparfromrhoL}) one obtains $\Delta \hat S =  (2 - a_\rho^2) m_W^2/m_\rho^2$. 
An exact cancellation of $S$ then occurs for $a_\rho^2 = 2$.
Notice that for $a_\rho^2 =2$ one also obtains the two relations
\begin{align}
\label{eq:universality}
g_{\rho \pi\pi} &= g_\rho \\
\label{eq:KSFR}
m_\rho^2 &= 2 g_{\rho\pi\pi}^2 f^2\, ,
\end{align}
where $g_{\rho\pi\pi}$ denotes the coupling of the $\rho$ to the NG bosons. 
The analog relations in QCD go respectively under the name of 
`coupling universality' and (a possible formulation of) the  KSFR relation (see for example Ref.~\cite{Bando:1984ej}).
In general, from eq.(\ref{eq:rhopipivertex}) and (\ref{eq:arhodef}) it follows (after neglecting the subleading 
contribution  to $g_{\rho\pi\pi}$ from $\alpha_1$): $g_{\rho\pi\pi} = a_\rho^2 g_\rho/2$ and $m_\rho^2 =  4 g_{\rho\pi\pi}^2 f^2/a_\rho^2$.
Notice also that for a vanishing $\alpha_2$,   the request of VMD can be satisfied for $a_\rho^2 =1$, for which the relations (\ref{eq:universality}), 
(\ref{eq:KSFR}) do not hold.
This is in contrast with the case of QCD, where VMD, universality of couplings and the KSFR relation can all be obtained from a lagrangian
containing just the kinetic and mass term of the $\rho$.  
}

While $\alpha_2 \sim 1/g_\rho^2$ is possible under the assumption of PUVC and 
also compatible with the request of Vector Meson Dominance in the $SU(2)_L$ form factor of the NG bosons,
perhaps a more reasonable estimate of $\alpha_2$ from physics at the scale $g_*f$ is precisely $\alpha_2\sim 1/g_*^2$, 
in which case it would be subleading.
One should also recall that in explicit models based on 5D theories or deconstruction, where $\rho$ is followed by a tower of vectors, 
the total contribution to $\alpha_2$ and $c_5^+$ from the heavy resonances always corresponds to a positive $S$. 
But within our limited perspective we can only parameterize $S$, rather than calculate it.

It is useful to compare with the approach adopted by the authors of Ref.~\cite{Barbieri:2009tx}, who studied
the effect of the exchange of a vector resonance $V$ in the context of an $SO(4)/SO(3)$  theory.
They require, as a rationale for estimating the coefficients of the operators in the  chiral lagrangian
of the resonance, that the scattering amplitudes for $\pi\pi \to VV$ do not grow faster than $E^2$,
and that those for $\psi\bar \psi \to VV$ (where $\psi$ is a SM fermion) be at most constant at large
energies. In our notation that corresponds to requiring  negligible  coefficients for all the $Q_i$, which is also true in PUVC. 
However PUVC is more restrictive as it further constrains the coefficients of the amplitude at $O(E^2)$.

Having discussed in detail our assumptions and their implications on the structure of the lagrangian,
we now turn to the study of the contribution of the $\rho$ to $\pi\pi$ scattering. We do so by assuming 
PUVC  and thus neglecting
the contribution of the operators $Q_i$.
Using eq.(\ref{eq:rhopipivertex}), it is straightforward to derive the expression of the scattering amplitudes valid up to
energies  $E \ll \Lambda$. We find:~\footnote{\label{ftn:fullamplitude}
For completeness we report the contribution to the $\pi\pi\to\pi\pi$ amplitudes generated by an
insertion of the operator $Q_1$. This is the only additional operator  contributing to the scattering amplitude that is not negligible under the less 
stringent assumption of perturbativity up to a scale $g_* f> m_\rho$. We find ($\widetilde \alpha_1\equiv \alpha_1 g_{\rho_L}^2$)
\begin{equation}
\begin{split}
A(s,t,u)&=\frac{a_{\rho_L}^2\widetilde \alpha_1}{f^2}\left[\frac{u^2-s^2}{t-m_{\rho_L}^2}+\frac{t^2-s^2}{u-m_{\rho_L}^2}\right]+
 \frac{a_{\rho_L}^2\widetilde\alpha_1^2}{f^2}\left[\frac{t^2(u-s)}{m_{\rho_L}^2(t-m_{\rho_L}^2)}+\frac{u^2(t-s)}{m_{\rho_L}^2(t-m_{\rho_L}^2)}\right]\, ,\label{rhoamplitude2}
\\[0.2cm]
B(s,t,u)&=-a_{\rho_L}^2\widetilde \alpha_1(1-\widetilde \alpha_1)\frac{m_{\rho_L}^2}{f^2}
 \left[\frac{(t-u)(u-s)(s-t)}{(s-m_{\rho_L}^2)(t-m_{\rho_L}^2)(u-m_{\rho_L}^2)} \right].
\end{split}
\end{equation}
In order to get a more compact expression, we have dropped the decay widths in the propagators, as they can be reintroduced straightforwardly.
Notice that $B$ is completely antisymmetric in the Mandelstam variables as required by Bose symmetry.
}
\begin{equation} \label{eq:ABfromrhofull}
\begin{split}
A(s,t,u) 
 =& \frac{s}{f^2} \left( 1 - \frac{3}{4} a_{\rho_L}^2 \right) 
        - \frac{1}{4} a_{\rho_L}^2 \frac{m^2_{\rho_L}}{f^2}\, \times \\[0.1cm]
   & \times \left[ \frac{s-u}{t-m_{\rho_L}^2 \! + i \Gamma_{\rho_L} m_{\rho_L} \theta(t) } 
       + \frac{s-t}{u-m_{\rho_L}^2 \! +  i \Gamma_{\rho_L} m_{\rho_L} \theta(u)}  \right] \\[0.6cm]
B(s,t,u) 
 =&   \frac{1}{4} a_{\rho_L}^2 \frac{m^2_{\rho_L}}{f^2}  \bigg[ \frac{u-t}{s-m_{\rho_L}^2 \! + i \Gamma_{\rho_L} m_{\rho_L} \theta(s)}  
         + \frac{s-u}{t-m_{\rho_L}^2 \! +  i \Gamma_{\rho_L} m_{\rho_L} \theta(t)} \\[0.1cm]
    &  + \frac{t-s}{u-m_{\rho_L}^2 \! + i \Gamma_{\rho_L} m_{\rho_L} \theta(u)} \bigg]\, ,
\end{split}
\end{equation}
where we have included the imaginary part in the denominators to take into account the finite width, $\Gamma_{\rho_L}$, of the $\rho^L$. 
As expected $B(s,t,u)\sim (E/m_{\rho_L})^6$ at energies $E\ll m_{\rho_L}$.
The expression of $A(s,t,u)$ coincides with that 
previously obtained in the literature for a vector of $SO(3)$ in $SO(4)/SO(3)$ models,
see for example Ref.\cite{Bagger:1993zf}.~\footnote{A 
meaningful comparison requires to  define the parameters $g_\rho$ and $f$ in both theories by means of physical observables.
In particular, $g_\rho$ can be defined as the strength of the interaction among three $\rho$'s.}
As  for the $SO(4)/SO(3)$ case~\cite{Barbieri:2008cc}, we also find that the $\pi\pi$ elastic scattering can be ``perturbatively unitarized'' for 
$a_{\rho_L}= 2/\sqrt{3}$ (for the same value, however, the inelastic channels remain strongly coupled).
In the case of a $\rho^R = (\mathbf{1},\mathbf{3})$, the $\pi^4\pi^i\rho^i$ interaction (the second term in  square brackets in the second 
line of eq.(\ref{eq:rhopipivertex})) changes sign, which implies that the amplitude $B(s,t,u)$ also changes sign, while $A(s,t,u)$ 
is the same as for a $\rho^L$.

\begin{figure}[!t]
\begin{center}
\hspace*{-0.65cm} 
\begin{minipage}{0.5\linewidth}
\begin{center}
\hspace*{-3.0cm} 
\fbox{\footnotesize $W^+W^- \to W^+ W^-$} \\[0.05cm]
\includegraphics[width=80mm]{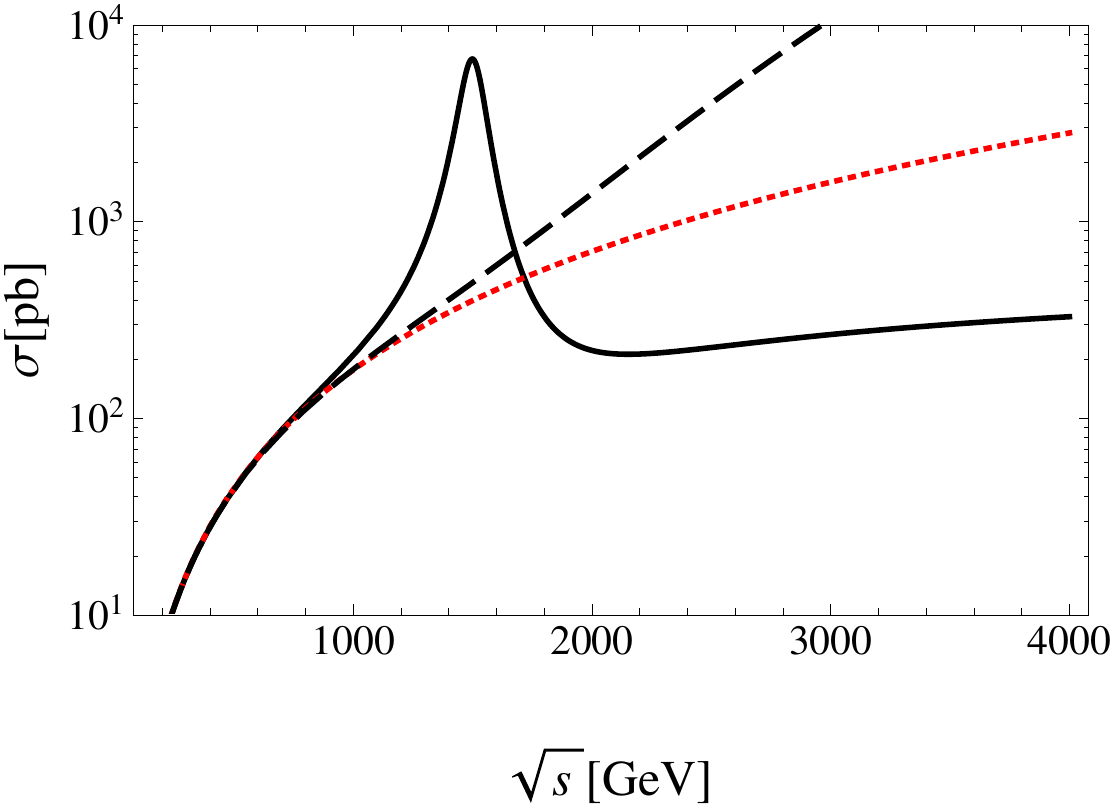}
\end{center}
\end{minipage}
\hspace{0.25cm}
\begin{minipage}{0.5\linewidth}
\begin{center}
\hspace*{-3.8cm} 
\fbox{\footnotesize $W^+W^- \to hh$} \\[0.05cm]
\includegraphics[width=80mm]{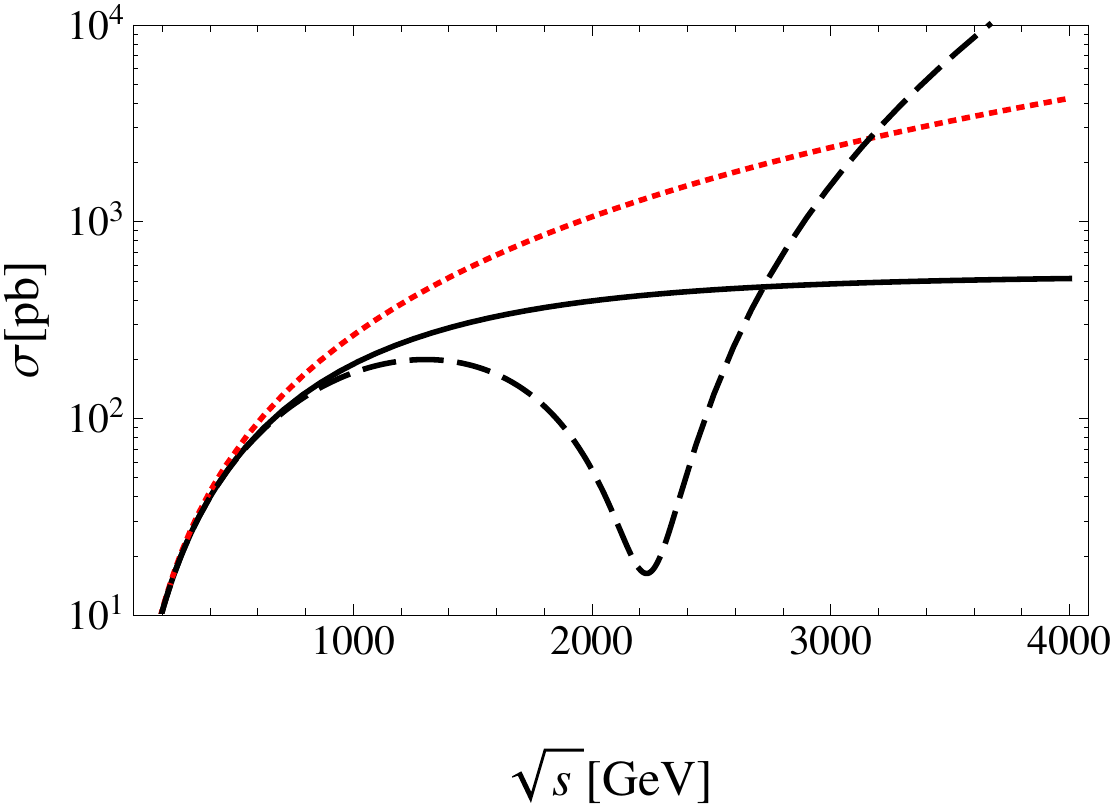}
\end{center}
\end{minipage}
\\[0.65cm]
\hspace*{-0.35cm} 
\begin{minipage}{0.5\linewidth}
\begin{center}
\hspace*{-3.75cm} 
\fbox{\footnotesize $W^+W^- \to Zh$} \\[0.05cm]
\includegraphics[width=80mm]{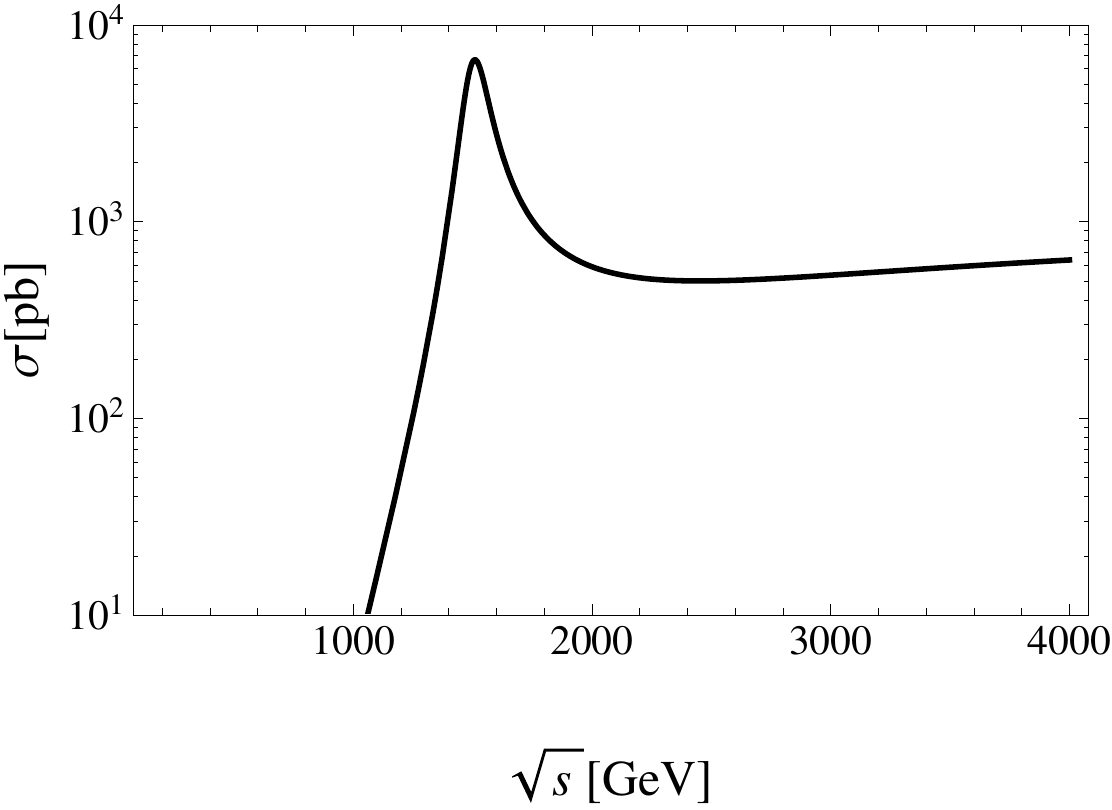}
\end{center}
\end{minipage}
\end{center}
\vspace*{-0.2cm}
\caption{\label{fig:rhopart} 
\small
Contribution of $\rho^L$ to the $W_L^+W_L^- \to W_L^+ W_L^-$ (upper left), $W_L^+W_L^- \to hh$ (upper right)
and $W_L^+ W_L^- \to Z_L h$ (lower panel) cross sections for $\xi = 0.5$, $m_{\rho_L} = 1.5\,$TeV and $a_{\rho_L} = 2/\sqrt{3}$,
which implies $\Gamma_{\rho_L} = 123\,$GeV.
The dotted red and dashed black curves respectively show the $O(p^2)$ and $O(p^4)$ predictions, 
as obtained by using eq.(\ref{eq:ABfromrhoatOp4}) with $\alpha_1=0$.
The solid black curve shows  the full effect of the $\rho^L$ exchange, as computed by means of eq.(\ref{eq:ABfromrhofull}).
}
\end{figure}

Figure~\ref{fig:rhopart} reports the cross section of $W_L^+W_L^- \to W_L^+ W_L^-$, $W_L^+W_L^- \to hh$
and $W_L^+ W_L^- \to Z_L h$, where all the gauge bosons are assumed to be longitudinally polarized (see Appendix~\ref{app:scattampl}
for the expressions of the scattering amplitudes among longitudinal vector bosons).
The following values of  input parameters have been chosen: 
$\xi \equiv (v/f)^2 =\sin^2\theta = 0.5$, $m_{\rho_L} = 1.5\,$TeV and $a_{\rho_L} = 2/\sqrt{3}$. The $\rho^L$ decay constant
has been computed including only the decays to the NG bosons. We find
\begin{equation}
\Gamma_{\rho_L} = \frac{a_{\rho_L}^2 m^3_{\rho_L}}{96\pi f^2}\, ,
\end{equation}
which  gives $\Gamma_{\rho_L} = 123\,$GeV for the above set of numerical inputs.
In Fig.~\ref{fig:rhopart} we have not shown the cross sections of $W_L^\pm Z_L\to W_L^\pm Z_L$ and $W_L^+ W_L^+\to W_L^+ W_L^+$ because
they are qualitatively equal  to those of  $W_L^+W_L^- \to W_L^+ W_L^-$ and $W_L^+W_L^- \to hh$ respectively.

As one could have anticipated (see for example Ref.~\cite{chanowitz}), the contribution of the $\rho$ enhances those processes
where it can be exchanged in $s$ channel, while it suppresses those where it only enters via the $t$ and $u$ channels.
More specifically, $W_L^+ W_L^- \to W_L^+ W_L^-$ and $W_L^\pm Z_L\to W_L^\pm Z_L$ are enhanced, while $W_L^+W_L^- \to hh$
and $W_L^+ W_L^+\to W_L^+ W_L^+$ get suppressed compared to the $O(p^2)$ result.
Notice that the $t$- and $u$-channel contributions of $\rho^L$ to $W_LW_L\to hh$  follow from the existence of
the coupling $\rho^L \chi h$. In the case of  $SO(4)/SO(3)$, with $h$ an ordinary neutral scalar, that coupling can be forbidden by imposing parity 
invariance. That is why the authors of Ref.~\cite{Hernandez:2010iu} find no contribution to $W_LW_L\to hh$ from the exchange of a vector resonance.

\subsection{$\eta = (\mathbf{1}, \mathbf{1})$}
\label{subsec:eta}

We now consider the case of a 
a scalar resonance $\eta$ transforming like a $(\mathbf{1}, \mathbf{1})$ of $SU(2) \times SU(2)$ (a singlet of $SO(4)$).
As done for the $\rho$, we write the effective lagrangian by focussing on the leading operators in a derivative
expansion $\partial/\Lambda$ that are relevant for $\pi\pi$ scattering. 
We find the lagrangian
\begin{equation} \label{eq:etaLag}
{\cal L}^{(\eta)} = \frac{1}{2} \left( \partial_\mu \eta \right)^2 - \frac{1}{2} m_\eta^2 \eta^2 
+\frac{f^2}{4} \left( 2 a_\eta \frac{\eta}{f} + b_\eta \frac{\eta^2}{f^2} \right)  \Tr\left[ d_\mu d^\mu\right] \, ,
\end{equation}
where cubic and quartic self-interactions for $\eta$ have been omitted because not relevant for the following.
It contains the following $\eta\pi\pi$ interaction term:
\begin{equation}
{\cal L}^{(\eta)}  \supset a_\eta \frac{\eta}{f} \partial_\mu\pi^a \partial^\mu \pi^a + \dots
\end{equation}
We can now bound the size of the couplings under the same assumptions of the previous section.
The PUVC request that no coupling exceeds $g_*$ at the putative cut-off scale $g_* f$  implies
\begin{equation}
a_\eta\, ,b_\eta \lesssim O(1)\, .
\end{equation}
More specifically, for  the  range $a_\eta =O(1)$ the exchange of $\eta$ does partially UV complete $\pi\pi$ scattering as shown in eq.(\ref{eq:ABfrometafull}) below.
Indeed, for $a_\eta = b_\eta =1$ the lagrangian (\ref{eq:etaLag}) describes a linear sigma model where the scalar $\eta$ and the $SO(5)/SO(4)$ 
NG bosons fit together in a fundamental (linearly-transforming) representation of $SO(5)$. 
For that particular choice all the scattering amplitudes are perturbatively unitarized provided $\eta$ is lighter than the cutoff~\cite{Barbieri:2007bh}.
On the other hand, by making the more conservative request that couplings be just perturbative at $g_* f$ we would obtain
\begin{equation}
a_\eta^2\, ,b_\eta \lesssim O(16\pi^2/g_*^2)\, ,
\end{equation}
by which the contribution from $a_\eta$ could well dominate $\pi\pi$ scattering around the $\eta$ peak.

It may be instructive to contemplate the possible effects of higher derivative terms, not displayed in eq.(\ref{eq:etaLag}).
For instance the term $\eta \partial^2 (d_\mu d_\mu)$, involves two extra powers of $p$ with respect to the term proportional to $a_\eta$. 
Its role is thus analogous to that  of $Q_1$ in the $\rho \pi\pi$ vertex. Of course under the  request that all couplings do not exceed $g_*$ at $g_*f$ 
this term, as all higher derivative terms, is subleading. By making the weaker
request that the couplings be just perturbative at $g_*f$ 
one finds that the new operator can affect at $O(1)$ the amplitude around ${\sqrt s}\sim m_\eta$ only if $g_*^3< 4\pi (m_\eta/f)^2$.
The latter constraint  implies very little separation between  $m_\eta$
an the cutoff $g_* f$.  In view of the above, in the following we will neglect  higher-derivative terms.

Notice that the lagrangian (\ref{eq:etaLag}) is obviously invariant under $P_{LR}$,
by assigning $\eta$ positive  $P_{LR}$ parity.
This means that the exchange of $\eta$ will not mediate any $P_{LR}$-violating process, like $W_L W_L \to Z_L h$, nor
will it generate $P_{LR}$-odd operators once integrated out at low energy.

At energies $E\ll m_\eta$ the field  $\eta$ can be integrated out by solving the equations of motion at leading order. This gives
\begin{equation} 
\eta = \frac{a_\eta f}{2 m_\eta^2}  \Tr\left[ d_\mu d^\mu\right] \left( 1 + O\left(\frac{E^2}{m_\eta^2} \right) \right)\, .
\end{equation}
By plugging the above solution into eq.(\ref{eq:etaLag}), one obtains a 
low-energy chiral lagrangian of the form (\ref{eq:L4}) with
\begin{equation}
c_1 = \frac{a_\eta^2 f^2}{8 m_\eta^2}
\end{equation}
and all the remaining chiral coefficients equal to zero. From eq.(\ref{eq:scattamplfromci}) then it follows that, for $E \ll m_\eta$,
\begin{equation} \label{eq:ABfrometaatOp4}
\begin{split}
A(s,t,u) & = \frac{s}{f^2} + a_\eta^2 \,\frac{s^2}{m_\eta^2 f^2}   \\[0.1cm]
B(s,t,u) & = 0\, .
\end{split}
\end{equation}
Full inclusion of the contribution  from the $\eta$ exchange up to energies $E \ll \Lambda$ gives
\begin{equation} \label{eq:ABfrometafull}
\begin{split}
A(s,t,u) =& \frac{s}{f^2} - a_\eta^2\, \frac{s}{f^2} \,\frac{s}{s-m_\eta^2 + i \Gamma_\eta m_\eta \theta(s)} \\
B(s,t,u) =& 0\, .
\end{split}
\end{equation}
As for the $\rho$, we have added the imaginary part in the denominator of the above formula to take into
account the finite width, $\Gamma_\eta$, of the $\eta$. By including only the decay channels to NG bosons,
we find:
\begin{equation}
\Gamma_\eta = \frac{a_\eta^2 m_\eta^3}{8\pi f^2}\, .
\end{equation}
The expression for $A(s,t,u)$ in eq.(\ref{eq:ABfrometafull}) coincides with that previously derived in the literature for a scalar singlet of $SO(3)$
in $SO(4)/SO(3)$ theories, see for example Ref.~\cite{Bagger:1993zf}

\begin{figure}[!t]
\begin{center}
\hspace*{-0.4cm} 
\begin{minipage}{0.48\linewidth}
\begin{center}
\hspace*{-2.85cm} 
\fbox{\footnotesize $W^+W^+ \to W^+ W^+$} \\[0.05cm]
\includegraphics[width=80mm]{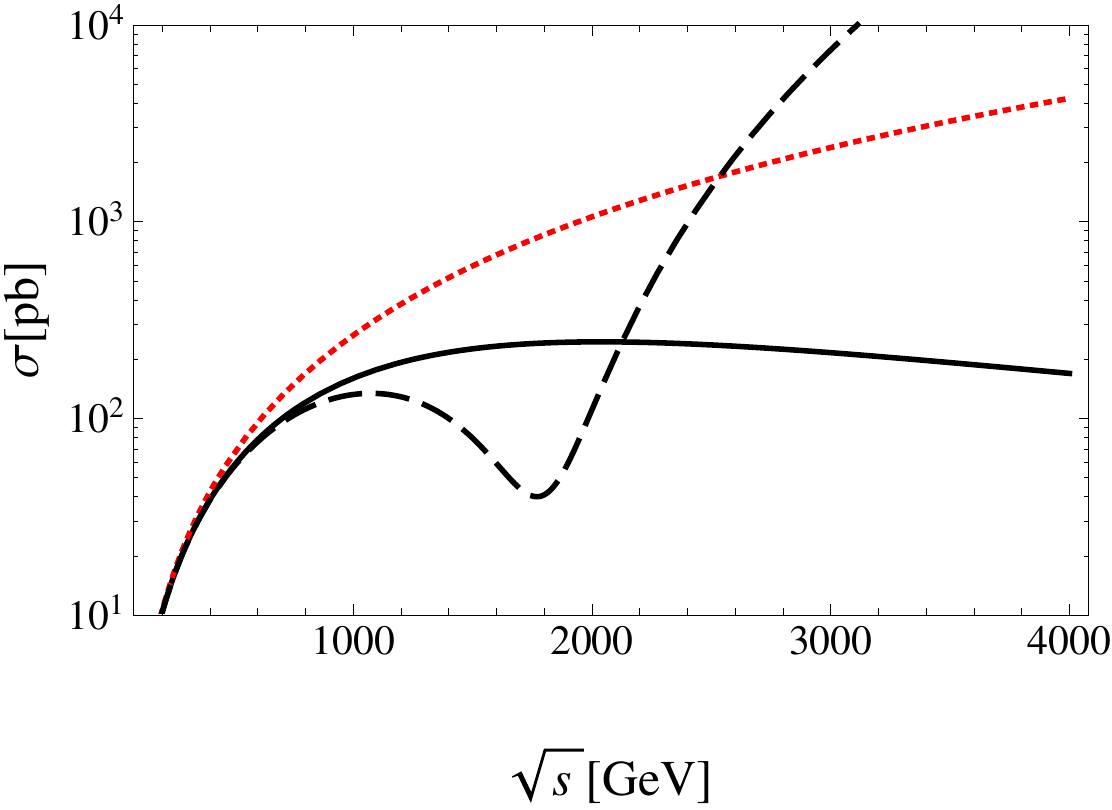}
\end{center}
\end{minipage}
\hspace{0.3cm}
\begin{minipage}{0.48\linewidth}
\begin{center}
\hspace*{-3.65cm} 
\fbox{\footnotesize $W^+W^- \to hh$} \\[0.05cm]
\includegraphics[width=80mm]{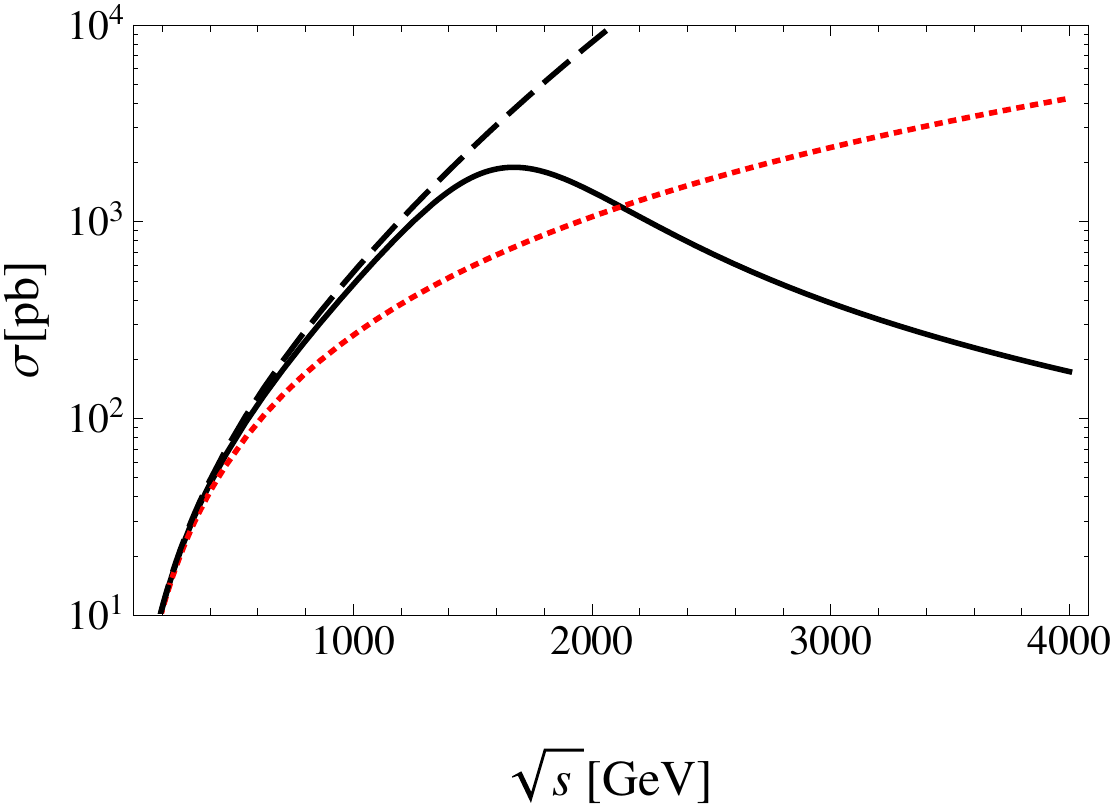}
\end{center}
\end{minipage}
\end{center}
\vspace*{-0.2cm}
\caption{\label{fig:etapart} 
\small
Contribution of $\eta$ to the $W_L^+W_L^+ \to W_L^+ W_L^+$ (left) and $W_L^+W_L^- \to hh$ (right) cross sections
 for $\xi = 0.5$, $m_{\eta} = 1.5\,$TeV and $a_{\eta} = 1$, which implies $\Gamma_\eta = 1.1\,$TeV.
The dotted red and dashed black curves respectively show the $O(p^2)$ and $O(p^4)$ predictions, 
as obtained from eq.(\ref{eq:ABfrometaatOp4}).
The solid black curve shows  the full effect of the $\eta$ exchange, as computed by means of eq.(\ref{eq:ABfrometafull}).
}
\end{figure}

Figure~\ref{fig:etapart} reports the cross sections of  $W_L^+W_L^- \to hh$ and $W_L^+W_L^+ \to W_L^+ W_L^+$
for $\xi = 0.5$, $m_{\eta} = 1.5\,$TeV and $a_{\eta} = 1$, which implies $\Gamma_\eta = 1.1\,$TeV.
We do not show the cross section of $W_L^\pm Z_L\to W_L^\pm Z_L$ since it is qualitatively equal to that of $W_L^+W_L^+ \to W_L^+ W_L^+$,
as well as $W_L^+W_L^- \to W_L^+ W_L^-$ and $Z_LZ_L\to Z_LZ_L$ have the same qualitative behavior of $W_L^+W_L^- \to hh$.
Notice that $Z_LZ_L\to Z_LZ_L$ identically vanishes at $O(p^2)$, but can be generated by the $\eta$ exchange.
As expected, the effect of  $\eta$ is that of enhancing those processes, like $W_L^+W_L^- \to W_L^+ W_L^-$,
$W_L^+W_L^- \to hh$, $Z_LZ_L\to Z_LZ_L$, where it is exchanged in $s$ channel, while the others are suppressed compared to the $O(p^2)$ result.

\begin{figure}[!t]
\begin{center}
\vspace*{0.4cm}
\hspace*{0.1cm} 
\begin{minipage}{0.5\linewidth}
\begin{center}
\hspace*{-3.75cm} 
\fbox{\footnotesize $W^+W^- \to hh$} \\[0.05cm]
\includegraphics[width=80mm]{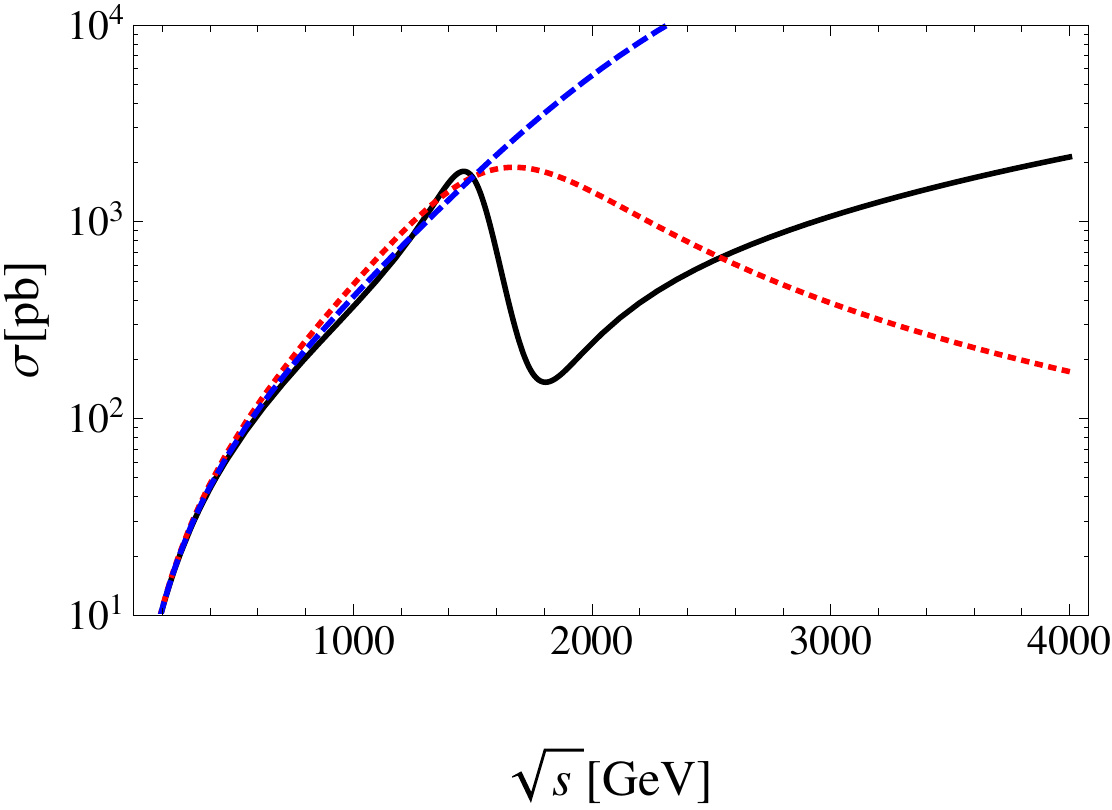}
\end{center}
\end{minipage}
\end{center}
\vspace*{-0.2cm}
\caption{\label{fig:PMtohhvsaeta} 
\small
Contribution of $\eta$ to the $W_L^+W_L^- \to hh$  cross section
for $a_\eta = 0.5$ (black solid curve), $a_\eta=1$ (red dotted curve) and $a_\eta =2$ (blue dashed curve).
The other input parameters are fixed as in Fig.~\ref{fig:etapart}.
}
\end{figure}

Figure~\ref{fig:PMtohhvsaeta} shows how the $W_L^+W_L^- \to hh$ cross section is affected by
varying $a_\eta$. 
In particular, we find that for $a_\eta < 1$ there is a destructive interference between the $\eta$ contribution
and the $O(p^2)$ term which gives a suppression of the cross section right after the resonance peak.
That is similar to the $Z$-photon interference observed at LEP.

\subsection{$\Delta = (\mathbf{3}, \mathbf{3})$}

In the case of $SO(5)/SO(4)$, since the broken generators $T^{\hat a}$ transform as a $\mathbf{4}$ of $SO(4)$,
the product $T^{\hat a} T^{\hat b}$ can be decomposed into the direct sum $\mathbf{1} + \mathbf{6} + \mathbf{9}$.
In particular, the traceless symmetric tensor
\begin{equation}
\left( K^{\hat a \hat b} \right)_{ij}= \{T^{\hat a}, T^{\hat b}\}_{ij} - \frac{1}{2} \delta^{\hat a \hat b} 
 \sum_{\hat c =1}^4 \left( T^{\hat c}  T^{\hat c} \right)_{ij}
\end{equation}
corresponds to the $\mathbf{9}$ and can  be used to describe our scalar $\Delta$.
We thus define a $5\times 5$ matrix
\begin{equation}
\left(\Delta_5(x) \right)_{ij} = \Delta^{\hat a\hat b}(x) \left( K^{\hat a \hat b} \right)_{ij} \, ,
\end{equation}
which contains 9 real scalar fields, denoted as $\Delta^{\hat a\hat b}(x)$.
Under a global  transformation $g\in SO(5)$, $\Delta_5$ transforms as
\begin{equation}
\Delta_5 \to h(\Pi, g)\, \Delta_5 \, h^\dagger(\Pi, g)\, .
\end{equation}
The lagrangian for $\Delta_5$  reads
\begin{equation} \label{eq:DeltaLag}
\begin{split}
{\cal L}^{(\Delta)} =& \frac{1}{4} \Tr\!\left[ \left(\nabla_\mu \Delta_5 \right)^2 \right]  -\frac{1}{4} m_\Delta^2 \Tr\!\left[ \left(\Delta_5\right)^2 \right]
+ \frac{f^2}{4} \Tr\!\left[ \left( 2\sqrt{2}\, a_\Delta \frac{\Delta_5}{f} + b_\Delta \frac{\Delta_5^2}{f^2} \right) d_\mu d^\mu \right] \\[0.1cm]
 &+ \frac{c_\Delta}{4} \, \Tr\!\left[ \Delta_5 d_\mu \Delta_5 d^\mu \right]\, ,
\end{split}
\end{equation}
where $a_\Delta$, $b_\Delta$, $c_\Delta$ are free parameters. 
We have included only the leading operators in  $\partial/\Lambda$ relevant for $\pi\pi$ scattering, and omitted
cubic and quartic self-interactions for $\Delta_5$ since they are not relevant for the following.
Upon expanding in the NG fields, one obtains the following $\Delta\pi\pi$ interaction term:
\begin{equation}
{\cal L}^{(\Delta)}  \supset \frac{a_\Delta}{\sqrt{2} f} \left( \sum_{a} \Delta^{aa} \partial_\mu\pi^a\partial^\mu\pi^a + 
 2 \sum_{b < a} \Delta^{ab} \partial_\mu\pi^a\partial^\mu\pi^b  \right) + \dots
\end{equation}
Again, requiring that the couplings do not exceed $g_*$ at the cut off scale $g_* f$ implies
\begin{equation}
a_\Delta, b_\Delta, c_\Delta \lesssim O(1)\, .
\end{equation}
The weaker request of perturbativity up to the scale $g_* f$ corresponds instead  to 
\begin{equation}
a_\Delta^2, b_\Delta, c_\Delta \lesssim O\left(\frac{16\pi^2}{g_*^2}\right)\, .
\end{equation}
Similarly to  the case of the $\eta$, also
the lagrangian (\ref{eq:DeltaLag}) is  $P_{LR}$ invariant,   with $\Delta$ even.
This  means that the exchange of the $\Delta$ does not mediate any $P_{LR}$-violating process, nor does it
generates $P_{LR}$-odd operators once integrated out at low energy.

As before, at energies $E\ll m_\Delta$ the field $\Delta_5$ can be integrated out by solving the equations of
motion at leading chiral order. We find~\footnote{The following identity is useful to solve the equations of motion:
\begin{equation}
\Tr\!\left[  K^{\hat a \hat b} T^{\hat c} T^{\hat d}  \right] = \frac{1}{2} \,\Tr\!\left[ K^{\hat a \hat b} K^{\hat c\hat d} \right] =
\frac{1}{4} \left( \delta^{\hat a \hat d} \delta^{\hat b \hat c} 
+  \delta^{\hat a \hat c} \delta^{\hat b \hat d} 
- \frac{1}{2} \delta^{\hat a \hat b} \delta^{\hat d \hat d}  \right) \, .
\end{equation}
}
\begin{equation}
\left( \Delta_5 \right)_{ij} = \frac{\sqrt{2} a_\Delta f}{m_\Delta^2} \left( (d_\mu d^\mu)_{ij} - \frac{1}{4} \sum_{\hat a=1}^{4} (T^{\hat a} T^{\hat a})_{ij} 
 \,\Tr\left[d_\mu d^{\mu}\right] \right) \left( 1 + O\left( \frac{E^2}{m_\Delta^2} \right) \right)\, .
\end{equation}
Plugging the above solution into eq.(\ref{eq:DeltaLag}) and keeping only $O(p^4)$ terms gives a low-energy chiral
lagrangian of the form (\ref{eq:L4}) with
\begin{equation} \label{eq:cifromDelta}
c_1 = -\frac{1}{4} c_2 = -\frac{a_\Delta^2 f^2}{32 m_\Delta^2}
\end{equation}
and all the remaining chiral coefficients equal to zero.~\footnote{
This  result can be compared with the  formulas known for the $SO(4)/SO(3)$ case.
Indeed, the scalar $\Delta$ contains  three components,  $\sigma$, $\zeta^i$, $\varphi^{ij}$, 
respectively with $SO(3)$ isospin $0$, $1$ and $2$. 
When integrating out a heavy resonance in an $SO(4)/SO(3)$ theory,
two chiral operators are generated at the $O(p^4)$ level:
\begin{equation}
\bar c_1\, \Tr\!\left[  ( D_\mu \Sigma^\dagger  D^\mu\Sigma ) \right]^2 +
\bar c_2\,  \Tr\!\left[  ( D_\mu \Sigma^\dagger  D_\nu\Sigma ) \right] \Tr\!\left[  ( D^\mu \Sigma^\dagger  D^\nu\Sigma ) \right]\, ,
\end{equation}
where $\Sigma = \exp(i \sigma^i \chi^i/v)$.
These operators are respectively contained in $O_1$ and $O_2$ of eq.(\ref{eq:L4}) provided one identifies $\bar c_{1,2} = c_{1,2}$
and $f=v$, as it can be  checked for example by expanding in the number of NG boson fields.
By integrating out the $\sigma$ and the $\varphi^{ij}$ in an $SO(4)/SO(3)$ theory we find
\begin{alignat}{4} 
\label{eq:csigma}
c_1|_\sigma =& +\frac{4}{3} \pi \frac{v^4}{m_\sigma^5} \Gamma_\sigma 
& \qquad
c_2|_\sigma =& 0 \\[0.1cm]
\label{eq:cphi}
c_1|_\varphi =& -\frac{4}{3} \pi \frac{v^4}{m_\varphi^5} \Gamma_\varphi   
&
 c_2|_\varphi =& -3 c_1|_\varphi\, ,
\end{alignat}
while the $\zeta^i$ does not contribute.  In the $SO(5)/SO(4)$ case, the $\sigma$ component inside $\Delta$ also couples to the Higgs 
(according to eq.(\ref{eq:DeltaLag})), and its decay width gets enhanced by a factor $4$ compared to the $SO(4)/SO(3)$ case.
The coefficients in (\ref{eq:cifromDelta}) are thus obtained as the sum of the $\sigma$ and $\varphi$
contributions in eqs.(\ref{eq:csigma}),(\ref{eq:cphi}) after replacing $\Gamma_\sigma \to (1/4) \Gamma_\sigma$ in 
$c_1|_\sigma$ and setting $m_\sigma = m_\varphi = m_\Delta$,  $\Gamma_\sigma = \Gamma_\varphi =  \Gamma_\Delta$.
Equation (\ref{eq:csigma}) agrees with the $SO(4)/SO(3)$ result for $\sigma$ reported in  Ref.~\cite{Donoghue:1988ed},
but eq.(\ref{eq:cphi}) disagrees with the formulas given in the same paper for the isospin tensor $\varphi$.
}

From eq.(\ref{eq:cifromDelta}) and (\ref{eq:scattamplfromci}) then it follows that, for $E \ll m_\Delta$,
\begin{equation} \label{eq:ABfromDeltaatOp4}
\begin{split}
A(s,t,u) & = \frac{s}{f^2} + \frac{a_\Delta^2}{4m_\Delta^2}  \left( 2 (t^2+u^2) - 2 s^2 \right) \\[0.1cm]
B(s,t,u) & = 0\, .
\end{split}
\end{equation}
Full inclusion of  the $\Delta$ exchange up to energies $E \ll \Lambda$ gives
\begin{equation} \label{eq:ABfromDeltafull}
\begin{split}
A(s,t,u) =& \frac{s}{f^2} + \frac{a_\Delta^2}{4 f^2} \bigg[  
 \frac{s^2}{s-m_\Delta^2 \! + i \Gamma_\Delta m_\Delta \theta(s)}
 -  \frac{2 t^2}{t-m_\Delta^2 \! + i \Gamma_\Delta m_\Delta \theta(t)} \\[0.15cm]
 & -  \frac{2 u^2}{u-m_\Delta^2\! + i \Gamma_\Delta m_\Delta \theta(u)} \bigg]
\\[0.2cm]
B(s,t,u) =& 0\, .
\end{split}
\end{equation}
Notice that, contrary to what we have found for the other resonances,  $\Delta$ cannot unitarize perturbatively
the $\pi\pi$ scattering amplitudes for any choice of the input parameters.
As before, the imaginary part in the denominators has been added to take into
account the finite width, $\Gamma_\Delta$, of $\Delta$. By including only the decay channels to NG bosons,
we find:
\begin{equation}
\Gamma_\Delta = \frac{a_\Delta^2 m_\Delta^3}{32\pi f^2}\, .
\end{equation}

\begin{figure}[!t]
\begin{center}
\hspace*{-0.6cm} 
\begin{minipage}{0.48\linewidth}
\begin{center}
\hspace*{-2.8cm} 
\fbox{\footnotesize $W^+W^- \to W^+ W^-$} \\[0.05cm]
\includegraphics[width=80mm]{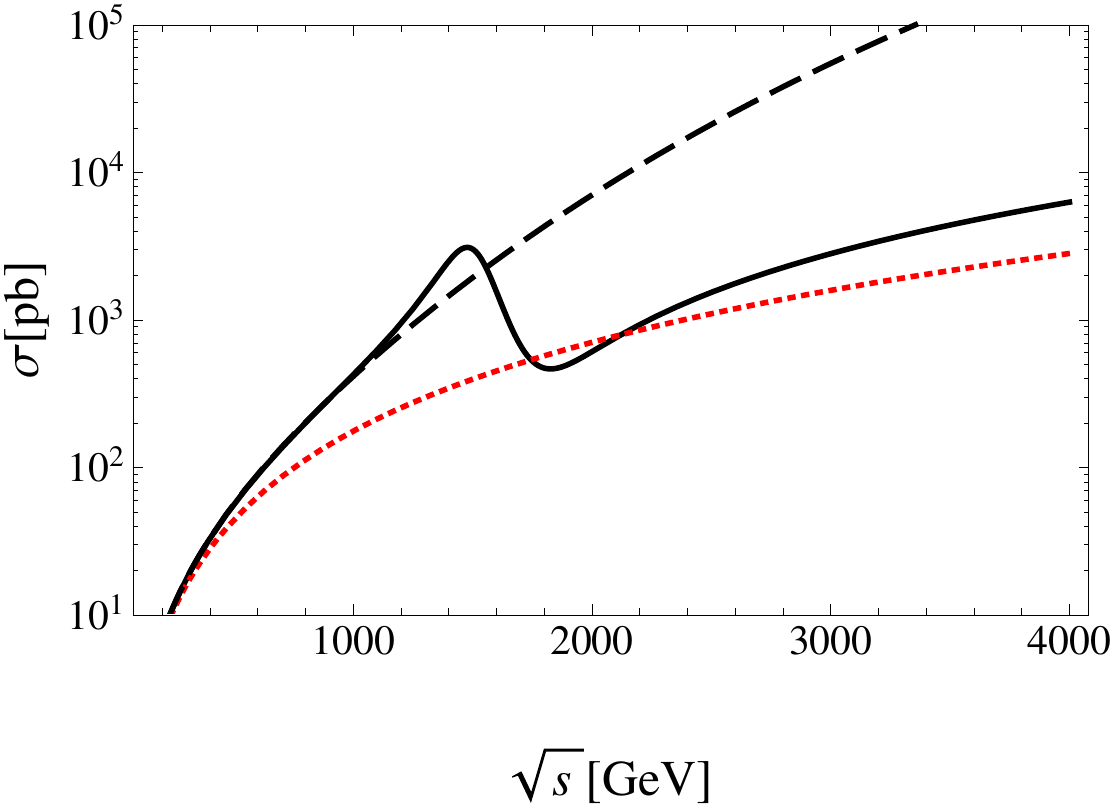}
\end{center}
\end{minipage}
\hspace{0.2cm}
\begin{minipage}{0.48\linewidth}
\begin{center}
\hspace*{-3.65cm} 
\fbox{\footnotesize $W^+W^- \to hh$} \\[0.05cm]
\includegraphics[width=80mm]{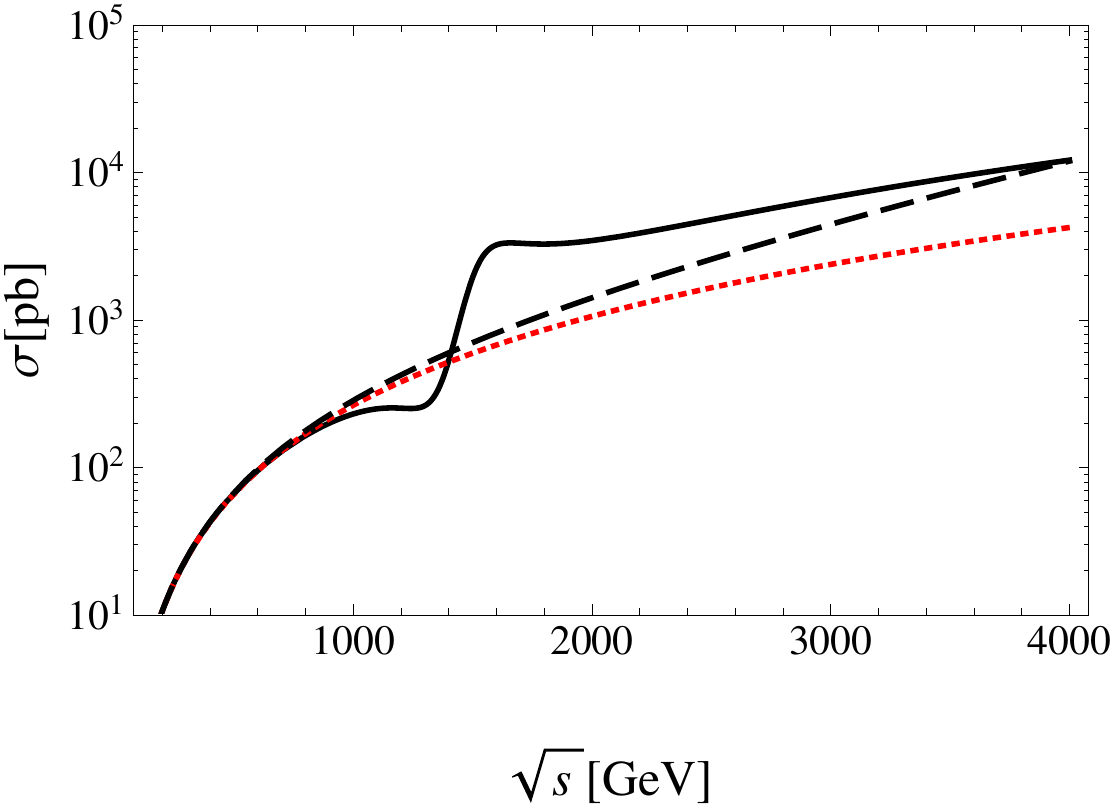}
\end{center}
\end{minipage}
\\[0.6cm]
\hspace*{-0.6cm} 
\begin{minipage}{0.48\linewidth}
\begin{center}
\hspace*{-2.8cm} 
\fbox{\footnotesize $W^+W^+ \to W^+ W^+$} \\[0.05cm]
\includegraphics[width=80mm]{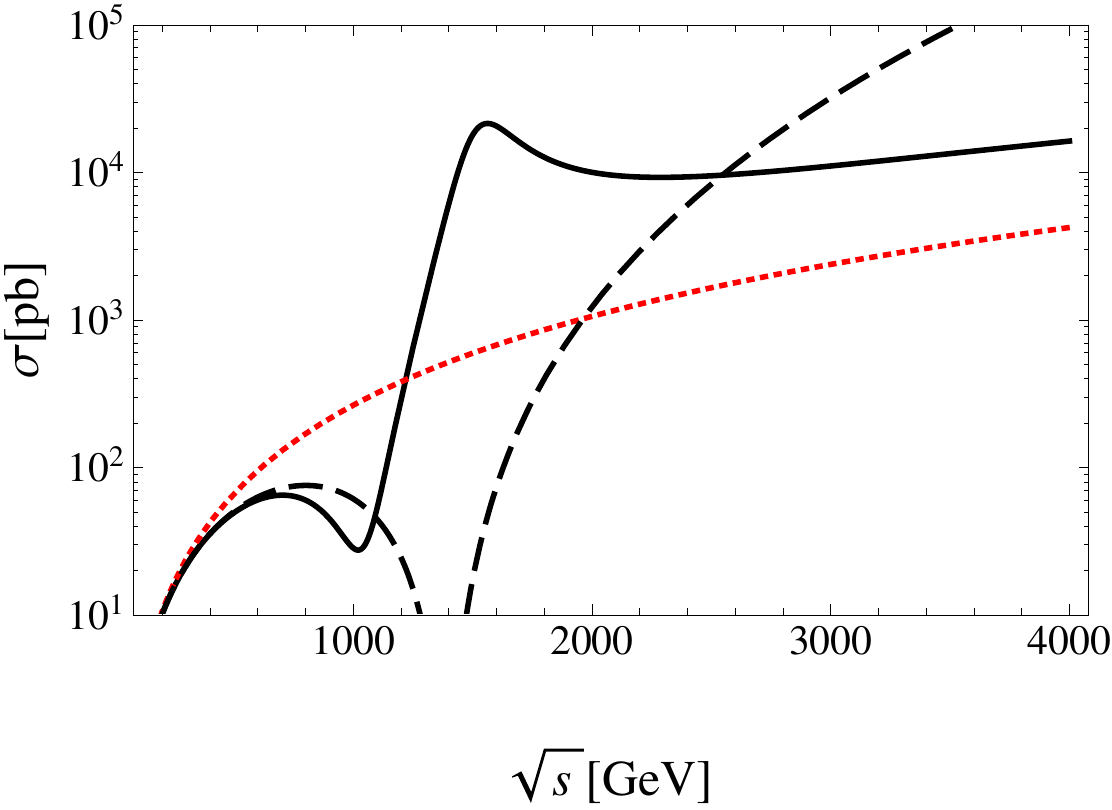}
\end{center}
\end{minipage}
\hspace{0.2cm}
\begin{minipage}{0.48\linewidth}
\begin{center}
\hspace*{-3.55cm} 
\fbox{\footnotesize $W^+Z \to W^+ Z$} \\[0.05cm]
\includegraphics[width=80mm]{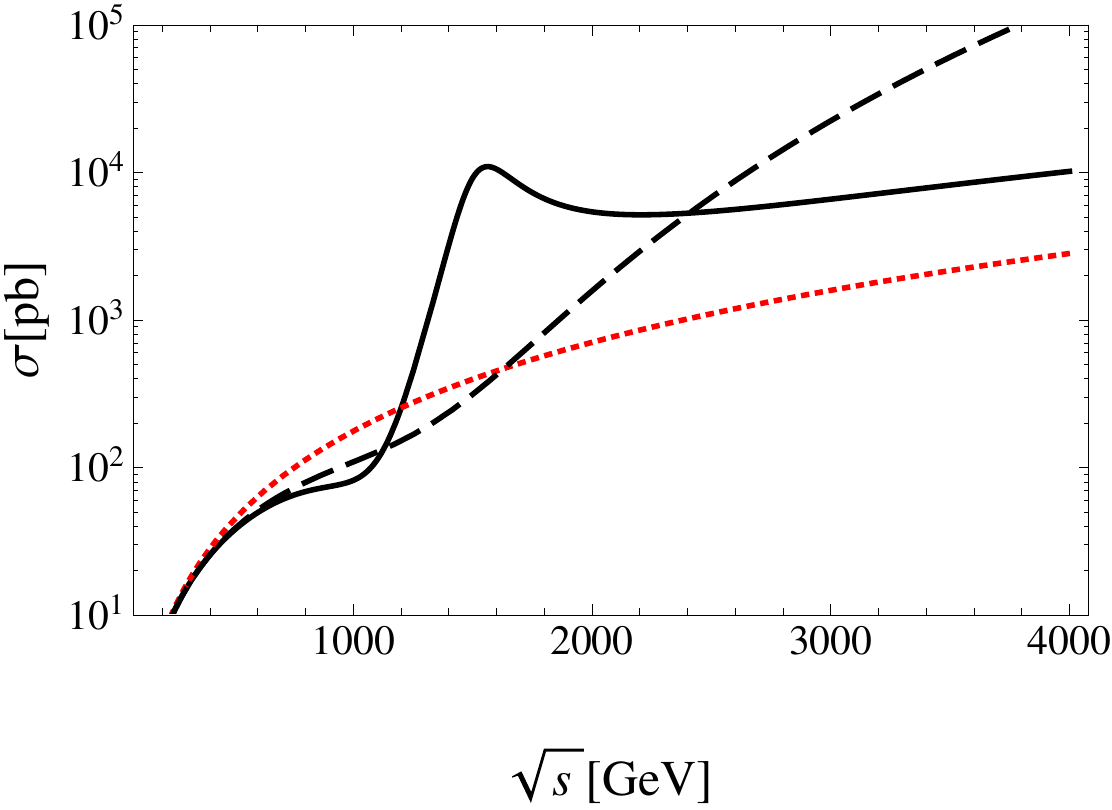}
\end{center}
\end{minipage}
\end{center}
\vspace*{-0.2cm}
\caption{\label{fig:Deltapart} 
\small
Contribution of $\Delta$ to the $W_L^+W_L^- \to W_L^+ W_L^-$ (upper left), 
$W_L^+W_L^- \to hh$ (upper right),  $W_L^+W_L^+ \to W_L^+ W_L^+$ (lower left)
and $W_L^+Z_L \to W_L^+ Z_L$ (lower right) cross sections
for $\xi = 0.5$, $m_{\Delta} = 1.5\,$TeV and $a_{\Delta} = 1$, which implies $\Gamma_\Delta = 277\,$GeV.
The dotted red and dashed black curves respectively show the $O(p^2)$ and $O(p^4)$ predictions, 
as obtained from eq.(\ref{eq:ABfromDeltaatOp4}).
The solid black curve shows  the full effect of the $\Delta$ exchange, as computed by means of eq.(\ref{eq:ABfromDeltafull}).
}
\end{figure}

Figure~\ref{fig:Deltapart} reports the cross sections of  $W_L^+W_L^- \to W_L^+ W_L^-$, 
$W_L^+W_L^- \to hh$,  $W_L^+W_L^+ \to W_L^+ W_L^+$ and $W_L^+Z_L \to W_L^+ Z_L$
for $\xi = 0.5$, $m_{\Delta} = 1.5\,$TeV and $a_{\Delta} = 1$ (which gives $\Gamma_\Delta = 277\,$GeV).
In this case, due to its isospin $0$, $1$ and $2$ components,  $\Delta$ can be exchanged in $s$ channel in all processes, 
which are thus all enhanced compared to their $O(p^2)$ value.

\section{WW scattering  at the LHC}

In this section we study the effect of the resonances on $WW$ scattering processes at the LHC.
For simplicity we will make use of the Effective $W$ Approximation (EWA)~\cite{EWA},  according to which a Vector Boson Fusion (VBF) process
can be effectively described as the slow emission of two vector bosons from the partons inside the protons, which subsequently
undergo a (fast) hard scattering.  This picture neglects the small degree of off-shellness of the emitted vector bosons (of the order
of their transverse momentum), and thus allows one to compute the differential cross section for the VBF process $pp\to X+jj$
as the convolution of a partonic cross section $\hat \sigma(V_i V_j \to X)$ times a luminosity factor 
$\rho_{V_i V_j}(m_X^2/s,Q^2)$:
\begin{equation} \label{eq:factorizedxsec}
\begin{split}
& \frac{d\sigma}{dm_X^2} = \frac{1}{m_X^2}\, \hat\sigma(V_i V_j \to X) \,\rho_{V_i V_j}(m_X^2/s,Q^2) \\[0.3cm]
& \rho_{V_i V_j}(\tau ,Q^2) = \tau \int_0^1 \!\! dx_1 \!\int_0^1 \!\! dx_2 \; f_{q_A}(x_1,Q^2) f_{q_B}(x_2,Q^2) \, \times \\[0.1cm]
    & \hspace{2.5cm} \times \int_0^1 \!\! dz_1 \!\int_0^1 \!\! dz_2\;  P_{q_A}^i(z_1) P^j_{q_B}(z_2) \, \delta(x_1 x_2 z_1 z_2 - \tau)\, .
\end{split}
\end{equation}
An implicit sum over all partons $q_A$, $q_B$ and over transverse and longitudinal vector boson polarizations
$i,j = T,L$ is understood.
In our case the final states of interest are $X = V_L V_L, hh$, with $V_L = W_L, Z_L$.
The luminosity factor $\rho_{V_i V_j}(\tau,Q^2)$ is defined as the probability for a pair of vector bosons $V_i V_j$ with invariant mass
$m_X^2 = \tau s$ to be emitted from the two colliding protons~\cite{Contino:2010mh}, where $\sqrt{s}=14\,$TeV is the LHC center of mass energy 
and $Q$ is the factorization scale.
As shown in eq.(\ref{eq:factorizedxsec}), it can be computed by integrating over the quark PDF's $f_{q}(x,Q^2)$,
and the vector boson splitting functions $P^i_q(z)$. These latter are given by~\cite{EWA}
\begin{equation}
\begin{split}
P^T(z) =& \, \frac{g_A^2+g_V^2}{4\pi^2} \, \frac{1+(1-z)^2}{2z} \log\left[ \frac{\bar p_T^2}{(1-z) m_W^2} \right]\\
P^L(z) =& \, \frac{g_A^2+g_V^2}{4\pi^2} \, \frac{1-z}{z}\, ,
\end{split}
\end{equation}
where, $z$ indicates the fraction of energy carried by the vector boson and $\bar p_T$ is the largest value allowed for its transverse momentum.
They depend upon the parton flavor and the vector boson species through the vectorial and axial couplings $g_V$, $g_A$:
\begin{equation}
\begin{aligned}
(g_V)_{qW} =&  - (g_A)_{qW} = \frac{g}{2\sqrt{2}}  & \quad
  (g_V)_{uZ} =& \frac{g}{\cos\theta_W} \left(\frac{1}{4} - \frac{2}{3} \sin^2\theta_W \right) \\[0.2cm]
(g_A)_{uZ} =&  - (g_A)_{dZ} = \frac{g}{4\cos\theta_W}  & 
  (g_V)_{dZ} =& \frac{g}{\cos\theta_W} \left(-\frac{1}{4} + \frac{1}{3} \sin^2\theta_W \right) \, .
\end{aligned}
\end{equation}

By making use of eq.(\ref{eq:factorizedxsec}) one can easily compute the total cross section for each final state
of interest by integrating over all values of $m_X$.
As a further simplification, we neglected the contribution from the transverse vector bosons and we used the expressions
of the longitudinal cross sections  derived in the previous section at leading
chiral order by means of the Equivalence Theorem.
We expect this latter approximation to work only at sufficiently large $m_X$, a limit in which  the contribution
of the longitudinal polarizations dominates and the vector bosons can be effectively treated as massless NG bosons.
Restricting to events with large $m_X$ is also required in order to reduce the model-dependent effects in $V_LV_L\to hh$ 
due to the Higgs cubic self-coupling, as shown by the analysis of Ref.~\cite{Contino:2010mh}.~\footnote{Non-derivative couplings involving 
$h$ and  $\eta$ (or $\Delta$),  which are induced by explicit $SO(5)$ breaking from SM couplings, are also a source of model dependency. 
At around the $\eta$ (or $\Delta$) mass their relative importance is only $O(g_{SM}^2/g_\rho^2)$, which can be safely neglected.}
We thus imposed the following cut:
\begin{equation} \label{eq:mXcut}
m_X \geq m_\text{cut} = 800\,\text{GeV}\, .
\end{equation}
A smaller value of the cut is still appropriate for the $V_LV_L$ final states, but it does not change substantially our conclusions.
We will discuss in detail the dependence of our results on $m_\text{cut} $ in the following.

In order to monitor the effect of a light resonance $\Phi = \eta,\rho,\Delta$, we construct the ratio
\begin{equation}
R(\Phi, \xi, m_\text{cut}) = \frac{\sigma(\Phi, \xi, m_\text{cut})}{\sigma(\text{LET}, \xi, m_\text{cut})}
\end{equation}
for each process of interest,
where $\sigma(\Phi, \xi, m_\text{cut})$ fully includes the contribution of the resonance 
while $\sigma(\text{LET}, \xi, m_\text{cut})$ is obtained by computing the partonic cross section using the $SO(5)/SO(4)$ 
chiral lagrangian at $O(p^2)$.
The ratio is largely insensitive to the effect of the acceptance cuts on the forward jets which will be applied in a realistic analysis
and that have not been included in the calculation of the individual cross sections by means of eq.(\ref{eq:factorizedxsec}).
This directly follows from the fact that the slow emission of the vector bosons factorizes from the subsequent
hard scattering, hence the kinematic distributions of the forward jets are universal.
As discussed in more detail in the following, we have estimated the efficiency for a standard set of acceptance cuts,
\begin{equation} \label{eq:AC}
p_T(j) > 30\,\text{GeV}\, , \qquad |\eta(j)| \leq 5\, , \qquad \Delta R(jj) \geq 0.7\, ,
\end{equation}
by utilizing the event generator MadGraph~\cite{MG-ME}. We found $\eps_{acc} \sim  0.6- 0.45$ in the range 
$m_\text{cut} \sim 0 - 3\,$TeV (see Fig.~\ref{fig:effvsmhhcut} below).

As already stated before, we shall focus on the more restrictive hypothesis of Partial UV Completion where the couplings do not exceed $g_*$ at the cut-off scale $g_*f$. According to  PUVC
the contribution of each resonance, $\Phi = \{ \rho , \eta , \Delta \}$, is controlled at leading order  by two parameters: 
its mass, $m_\Phi$, and its coupling to two NG bosons, $a_\Phi$.
For each resonance $\Phi$ we will thus show four contour plots in the plane $(a_\Phi, m_\Phi)$ reporting the ratio $R(\Phi, \xi, m_\text{cut})$ 
for the processes $pp\to X jj$ with $X = W^+_LW^-_L, hh, W^+_L W^+_L, W^+_L Z_L$.
We impose the cut of eq.(\ref{eq:mXcut}) and fix $\xi = (v/f)^2 = 0.5$.~\footnote{As for the quark PDF's, we used the 
CTEQ611 set and fixed the factorization scale to $Q = m_W$.}
For these values, in the absence of new resonances (or $m_\Phi \to \infty$), we obtain:
\vspace{0.1cm}
\begin{center}
\begin{tabular}{r|cccc}
& \multicolumn{4}{c}{process} \\[0.1cm]
 & $W^+_LW^-_L$ & $hh$ & $W^+_L W^+_L$ & $W^+_L Z_L$ \\[0.15cm]
\hline
&&&& \\[-0.3cm]
$\sigma(\text{LET}, \xi, m_\text{cut})$ [fb] & 2.46 & 2.37 & 1.87 & 1.47 
\end{tabular}
\end{center}
\vspace{0.25cm}
which can be used as reference values for a quick estimate of the absolute cross sections for given values of the ratio $R$.
In each contour plot we superimpose the isocurves of constant $\Gamma_\Phi/m_\Phi$, whose value can be used to 
monitor the strength of the coupling of the resonance to the NG bosons: the region $\Gamma_\Phi/m_\Phi \gtrsim \pi$ is where 
the coupling $\Phi \pi\pi$ grows strong 
and our perturbative calculation cannot be trusted.

\begin{figure}[!t]
\begin{center}
\begin{minipage}{0.4\linewidth}
\begin{center}
\hspace*{-1.95cm} 
\fbox{\footnotesize $pp\to W^+ W^- jj$} \\[0.05cm]
\includegraphics[width=68mm]{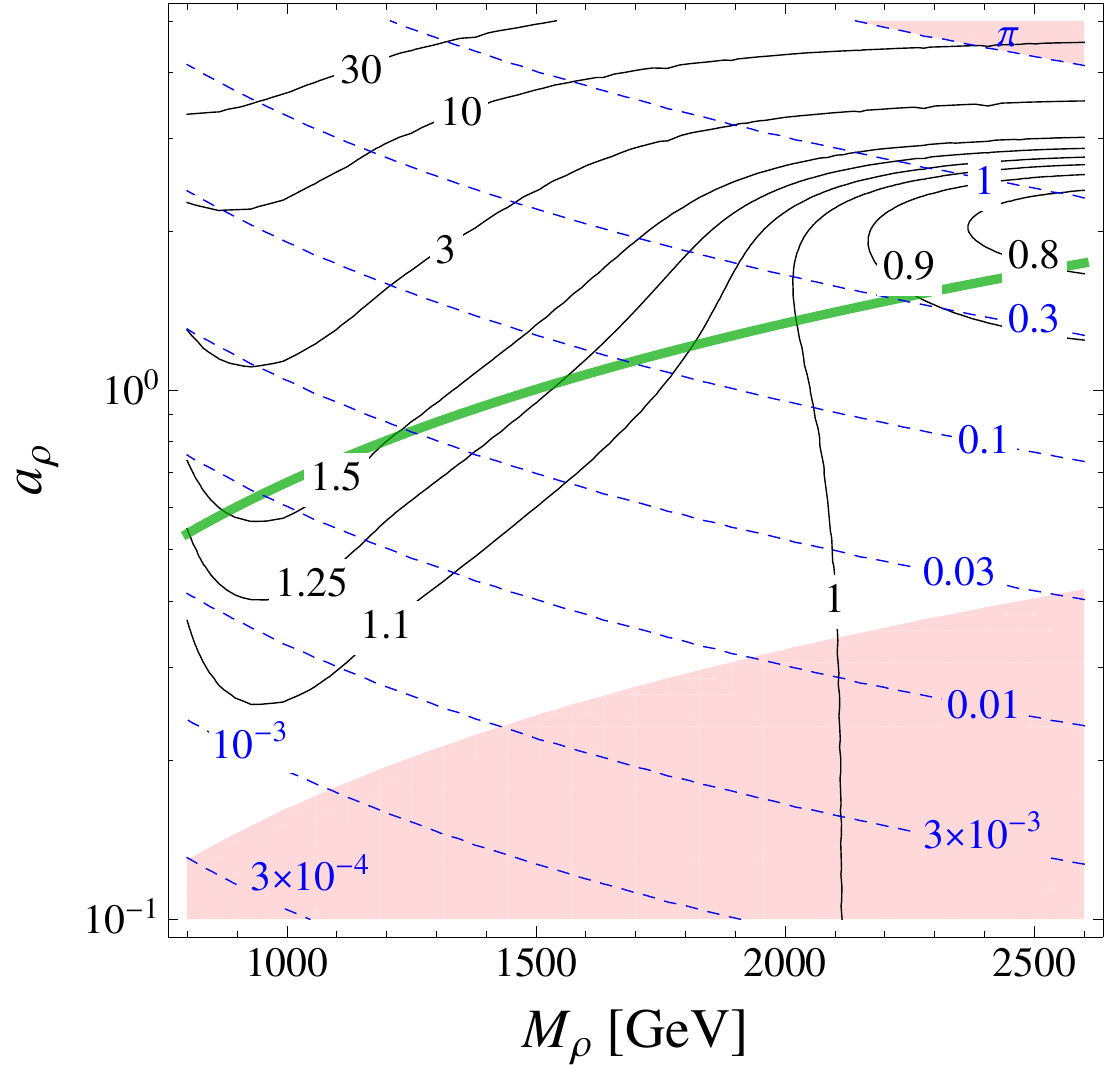} 
\end{center}
\end{minipage}
\hspace{0.6cm}
\begin{minipage}{0.4\linewidth}
\begin{center}
\hspace*{-2.8cm} 
\fbox{\footnotesize $pp\to hh jj$} \\[0.05cm]
\includegraphics[width=68mm]{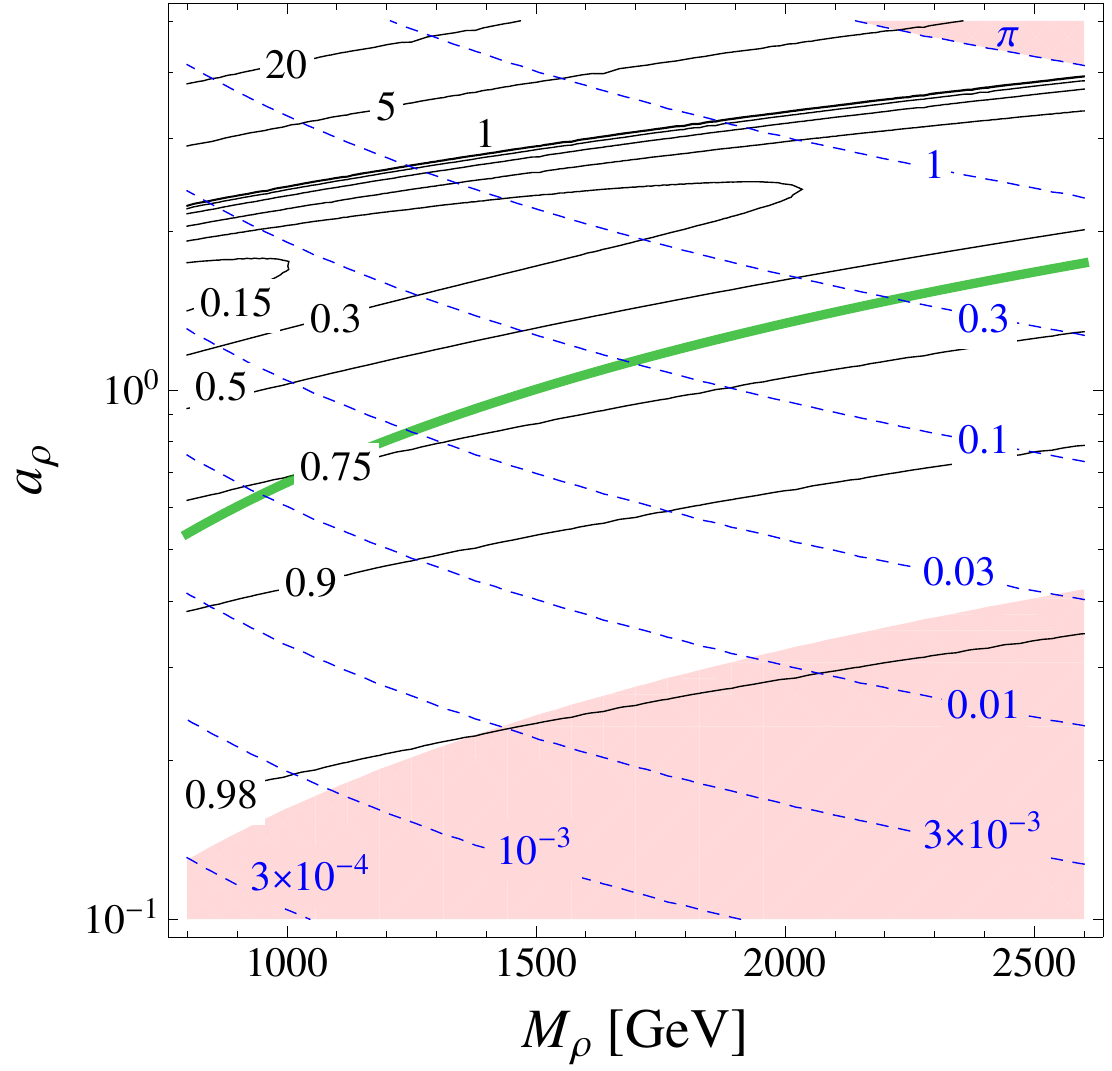} 
\end{center}
\end{minipage}
\\[0.6cm]
\begin{minipage}{0.4\linewidth}
\begin{center}
\hspace*{-1.92cm} 
\fbox{\footnotesize $pp\to W^+ W^+ jj$} \\[0.05cm]
\includegraphics[width=70mm]{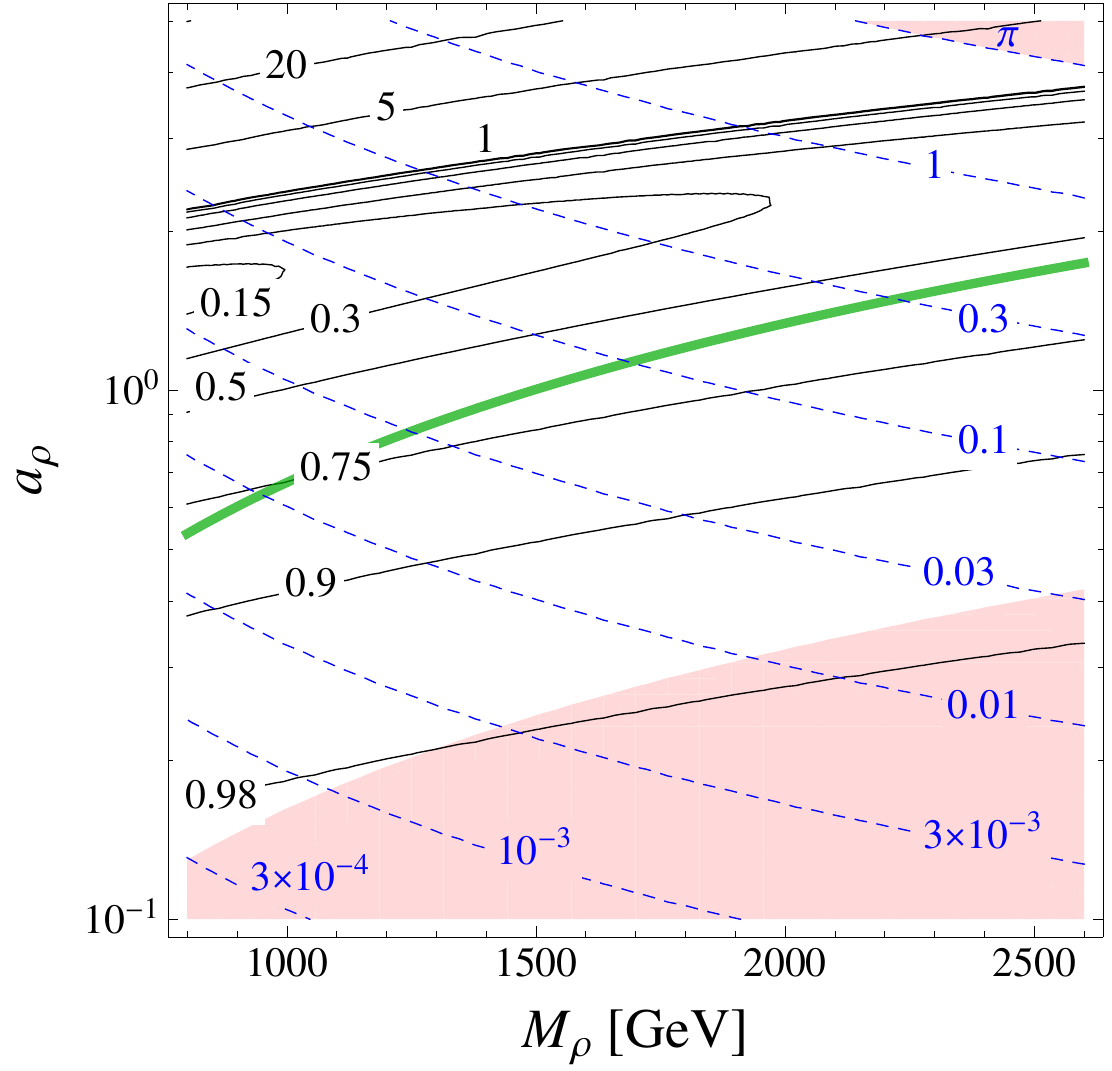}  
\end{center}
\end{minipage}
\hspace{0.6cm}
\begin{minipage}{0.4\linewidth}
\begin{center}
\hspace*{-2.3cm} 
\fbox{\footnotesize $pp\to W^\pm Z jj$} \\[0.05cm]
\includegraphics[width=70mm]{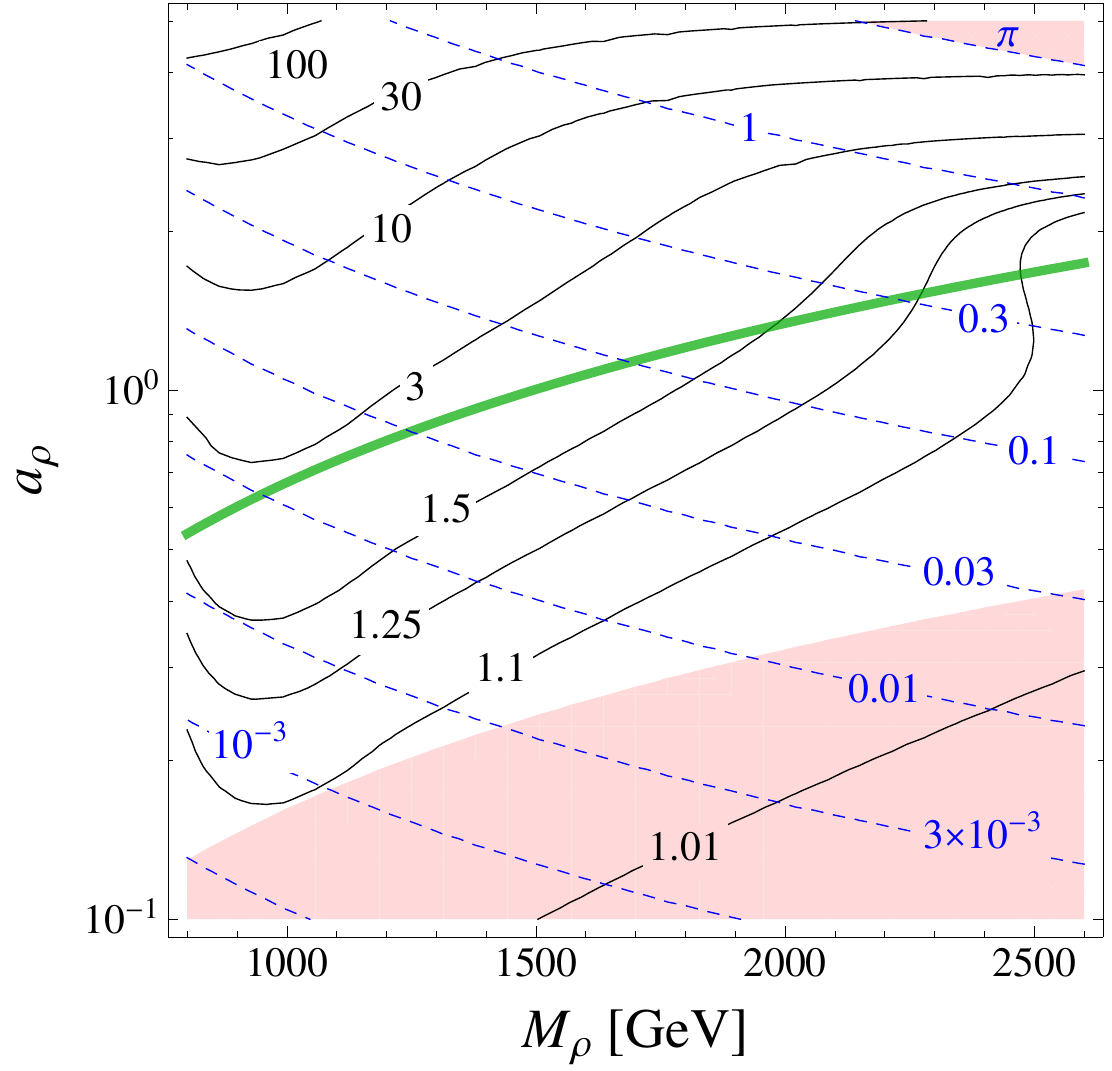}  
\end{center}
\end{minipage}
\end{center}
\caption{\label{fig:rhoLcontours} 
\small
Contours of constant $R(\rho^L,\xi,m_\text{cut})$ (continuous black lines) and constant $\Gamma_{\rho_L}/m_{\rho_L}$ (dashed blue lines)
in the plane ($a_{\rho_L}, m_{\rho_L}$) for $\xi =0.5$ and $m_\text{cut} = 800\,$GeV.  
The  pink (darker)  areas in the upper right corner and in the lower part of the plane respectively
correspond to the  regions where $\Gamma_{\rho_L}/m_{\rho_L} > \pi$ and $g_{\rho_L} > 4\pi$. 
In these regions our perturbative calculation cannot be trusted, see text.
The region below the thick green line is that where $\Delta \hat S <2.9 \times 10^{-3}$ for $\alpha_2 =0$ (see text).
}
\end{figure}

\vspace{0.6cm}
Figure~\ref{fig:rhoLcontours} shows the results for a spin-1 resonance $\rho^L$.
The solid black  and dashed blue lines  denote respectively the isocurves of constant $R(\rho^L,\xi,m_\text{cut})$  and
constant $\Gamma_{\rho_L}/m_{\rho_L}$ in the plane $(a_{\rho_L}, m_{\rho_L})$, where $a_{\rho_L}$ is defined in eq.(\ref{eq:arhodef}).
The region below the thick green line is that where $\Delta \hat S$, as computed by means of eq.(\ref{eq:SparfromrhoL}) with $\alpha_2=0$,
satisfies the bound $\Delta\hat S < 2.9 \times 10^{-3}$.~\footnote{This is the bound implied at $99\%$ CL by the LEP electroweak data
on $\Delta \hat S$ in presence of an arbitrary $\Delta\hat T$ correction, as obtained using the EW fit of Ref.~\cite{Agashe:2005dk} 
with $m_h=120\,$GeV.}
In most of the region shown  the $\rho^L$ has a width small compared to its mass. In the pink (darker) region
in the upper right corner of the plane, however, one has  $\Gamma_{\rho_L}/m_{\rho_L} > \pi$, which indicates
that the $\pi\pi\rho$ coupling is non-perturbative and consequently our calculation cannot be trusted.
The other pink (darker) area in the lower part of the plane  corresponds instead to the region where $g_{\rho_L} > 4\pi$ and
perturbation theory also breaks down. These two bounds represent the zeroth order request of the viability of our effective lagrangian, and are 
weaker than any of the two criteria discussed at the beginning of section~\ref{philosophy}.
On the other hand, PUVC  singles out the region  around $a_\rho \sim 1$.
Points further away from $a_\rho \sim 1$ can  still be allowed if one adopts the first criterion discussed at the beginning of section \ref{philosophy}. 
However,  as discussed in section~\ref{subsec:rhoL},  one should bear in mind that in that case neglecting the contribution 
of the operator $Q_1$ to the scattering amplitudes, as done in our calculation, is justified only for $g_*^2< 4\pi g_{\rho_L}$.

In the case of $pp\to W^+_L W^-_L jj$ and $pp\to W^+_L Z_L jj$, the ratio $R$ gets larger if one  fixes $m_{\rho_L}$ and increases  $a_{\rho_L}$.
This is expected, considering that in these processes $\rho^L$ can be exchanged in the $s$-channel and thus
tends to increase the cross section, as previously discussed. 
In the case of $pp\to W^+_L W^+_L jj$ and $pp\to hh jj$, on the other hand,  in the region of small $a_{\rho_L}$
the ratio $R$ decreases if $a_{\rho_L}$ is increased.
As also discussed in the previous section, this is the effect of the exchange of the $\rho^L$ in the $t$- and $u$-channels,
which tends to unitarize the scattering and thus decreases the cross section.
Such behavior however changes if one moves to values $a_{\rho_L} \gg 2/\sqrt{3}$:  in this case the contribution of $\rho^L$
dominates the scattering amplitude (it `overshoots' the $O(p^2)$ term and no longer unitarizes), thus
quickly driving the cross section to larger values.
Summarizing, for reference values $m_{\rho_L} = 1.5\,$TeV, $a_{\rho_L} = 1$, we predict a suppression $R= 0.7$ in the 
$W^+_L W_L^+$, $hh$ channels, and an enhancement $R =1.3$ and $R = 2.0$ respectively in the channels $W_L^+W_L^-$, $W^+_L Z_L$.

The ratio $R$ depends upon $\xi$ uniquely through the resonance's decay width, since both the scattering 
amplitude at $O(p^2)$ and the resonance contribution are proportional to $\xi$ for fixed $a_\rho$ and $m_{\rho_L}$,
see eq.(\ref{eq:ABfromrhofull}). This property also holds in the case of the other resonances.
As a consequence, the dependence of $R$ upon $\xi$ is absent  in those processes where the resonance is exchanged
only in $t$ and $u$ channel, whereas it can be significant when a resonance exchanged in $s$-channel is light. In the latter case, 
for sufficiently narrow width
 the dependence on the parton distributions  drops and one basically has $R\propto 1/\xi$. For sizeable $\xi$ this scaling is partly compensated by 
the ``spreading'' of the resonance to $s<m_\rho^2$ where the PDF's grow steeply.
This is clearly illustrated by the following tables, which report the value of $R$ in the channels 
$hh$ (left) and $W^+_L Z_L$ (right) for $a_{\rho_L} = 2/\sqrt{3}$:

\vspace{0.1cm}
\begin{center}
\begin{tabular}{rc|ccc}
$hh$ & & \multicolumn{3}{c}{$m_{\rho_L}\,$[GeV]} \\[0.15cm]
& $R$ & $1500$ & $2000$ & $2500$  \\[0.1cm]
\hline
&&&& \\[-0.3cm]
\multirow{3}{*}{$\xi$}\hspace{0.1cm}  & 0.1 & 0.59 & 0.71 & 0.78 \\[0.1cm]
                                                           & 0.5 & 0.59 & 0.71 & 0.78 \\[0.1cm]
                                                           & 0.8 & 0.59 & 0.71 & 0.78
\end{tabular}
\hspace{1.2cm}
\begin{tabular}{rc|ccc}
$W^+_L Z_L$ & & \multicolumn{3}{c}{$m_{\rho_L}\,$[GeV]} \\[0.15cm]
& $R$ & $1500$ & $2000$ & $2500$  \\[0.1cm]
\hline
&&&& \\[-0.3cm]
\multirow{3}{*}{$\xi$}\hspace{0.1cm}  & 0.1 & 8.8 & 3.7 & 2.0 \\[0.1cm]
                                                           & 0.5 & 2.3 & 1.4 & 1.1 \\[0.1cm]
                                                           & 0.8 & 1.6 & 1.1 & 1.0
\end{tabular}
\end{center}
\vspace{0.25cm}

\begin{figure}[!t]
\begin{center}
\begin{minipage}{0.4\linewidth}
\begin{center}
\hspace*{-1.95cm} 
\fbox{\footnotesize $pp\to W^+ W^- jj$} \\[-0.05cm]
\includegraphics[width=68mm]{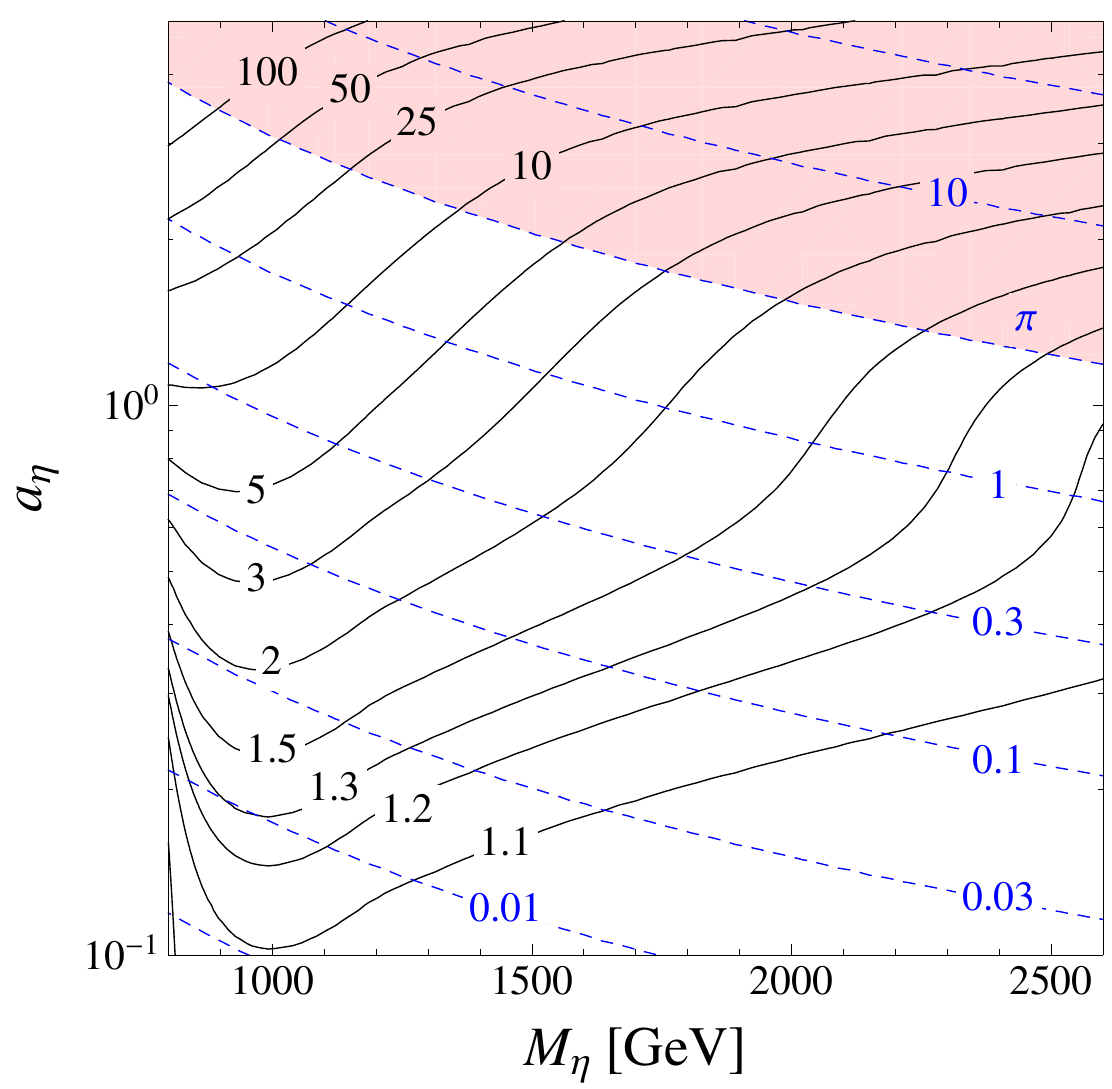} 
\end{center}
\end{minipage}
\hspace{0.6cm}
\begin{minipage}{0.4\linewidth}
\begin{center}
\hspace*{-2.8cm} 
\fbox{\footnotesize $pp\to hh jj$} \\[-0.05cm]
\includegraphics[width=68mm]{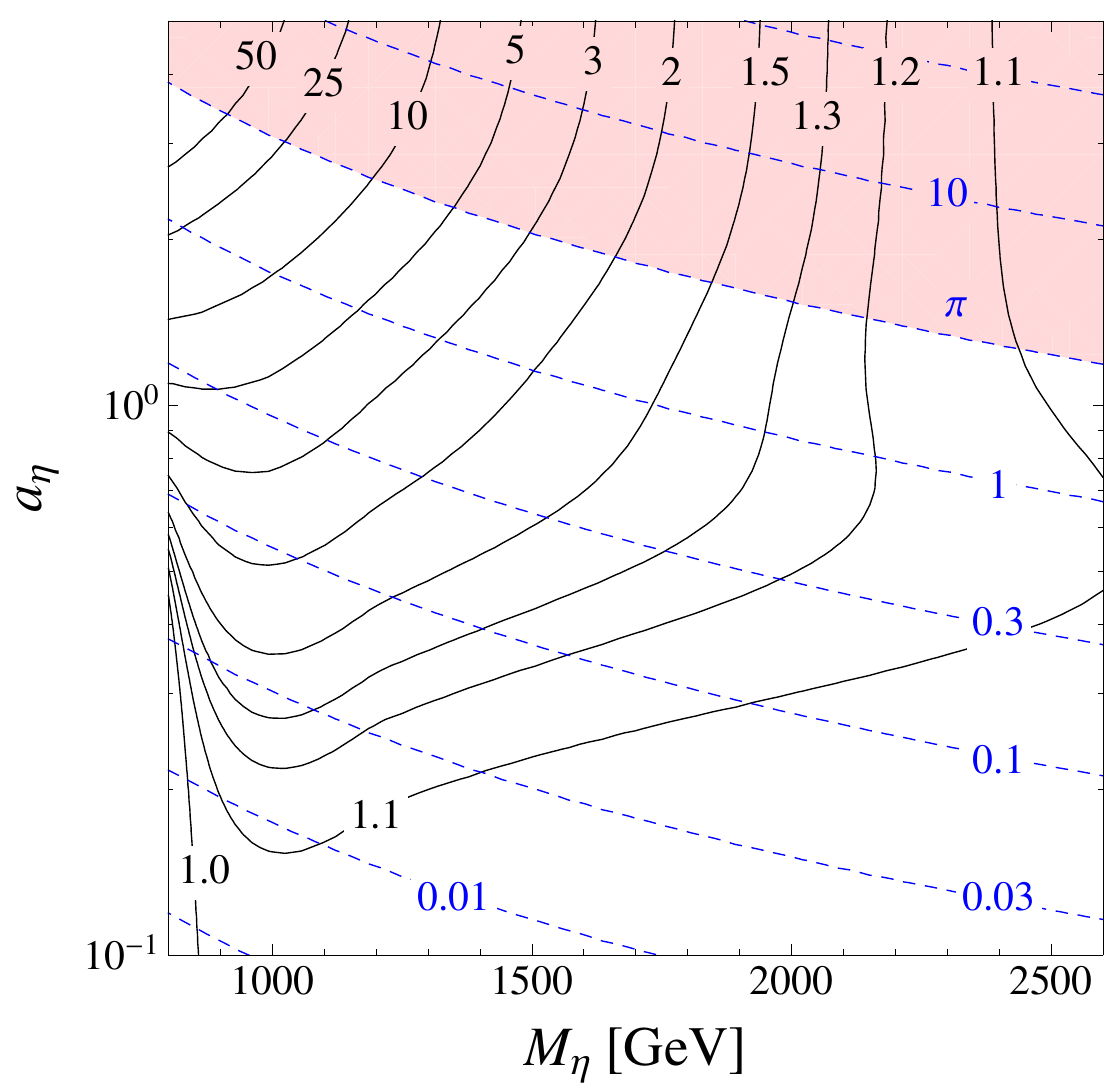} 
\end{center}
\end{minipage}
\\[0.6cm]
\begin{minipage}{0.4\linewidth}
\begin{center}
\hspace*{-1.92cm} 
\fbox{\footnotesize $pp\to W^+ W^+ jj$} \\[-0.05cm]
\includegraphics[width=70mm]{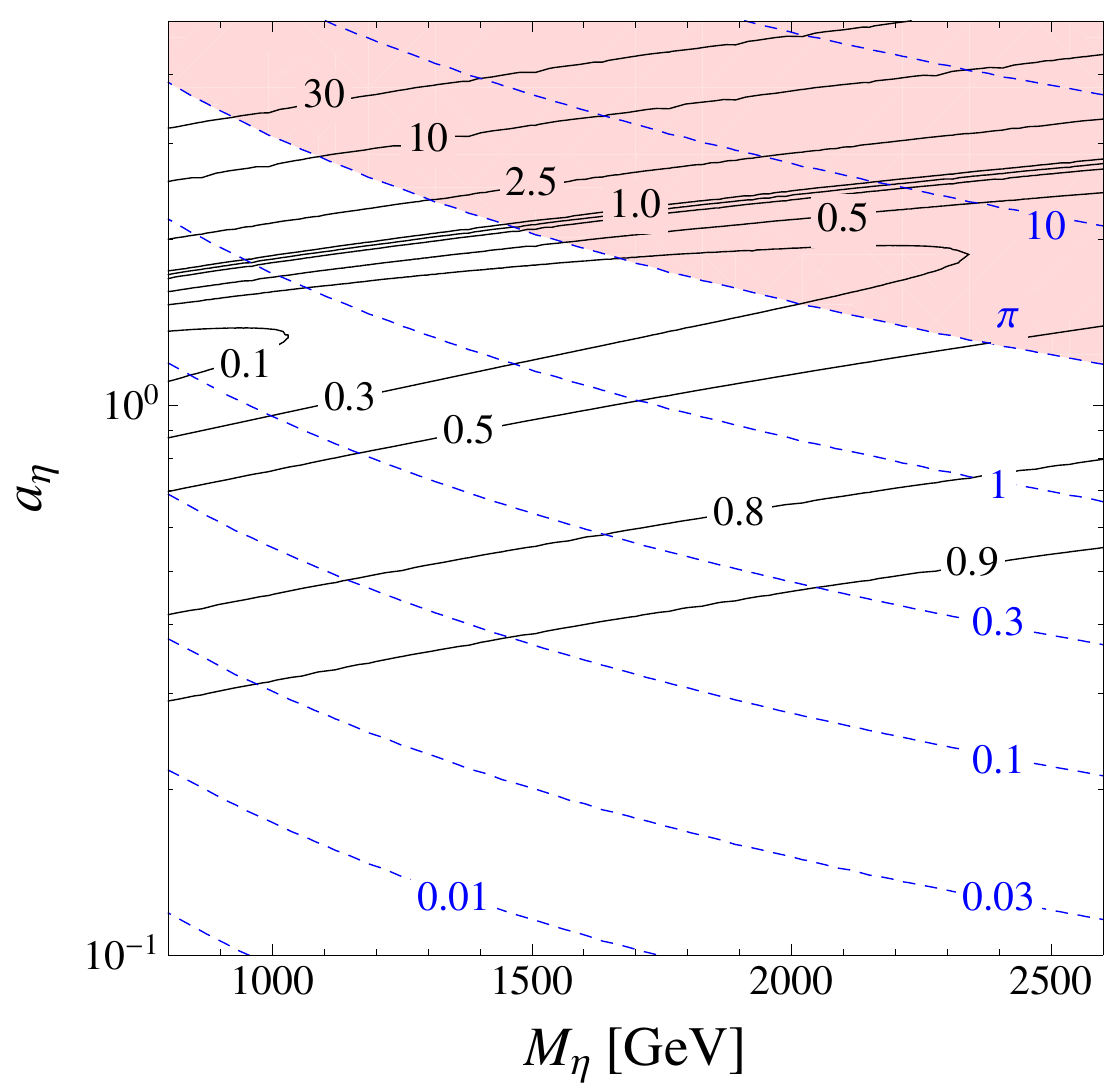}  
\end{center}
\end{minipage}
\hspace{0.6cm}
\begin{minipage}{0.4\linewidth}
\begin{center}
\hspace*{-2.3cm} 
\fbox{\footnotesize $pp\to W^\pm Z jj$} \\[-0.05cm]
\includegraphics[width=70mm]{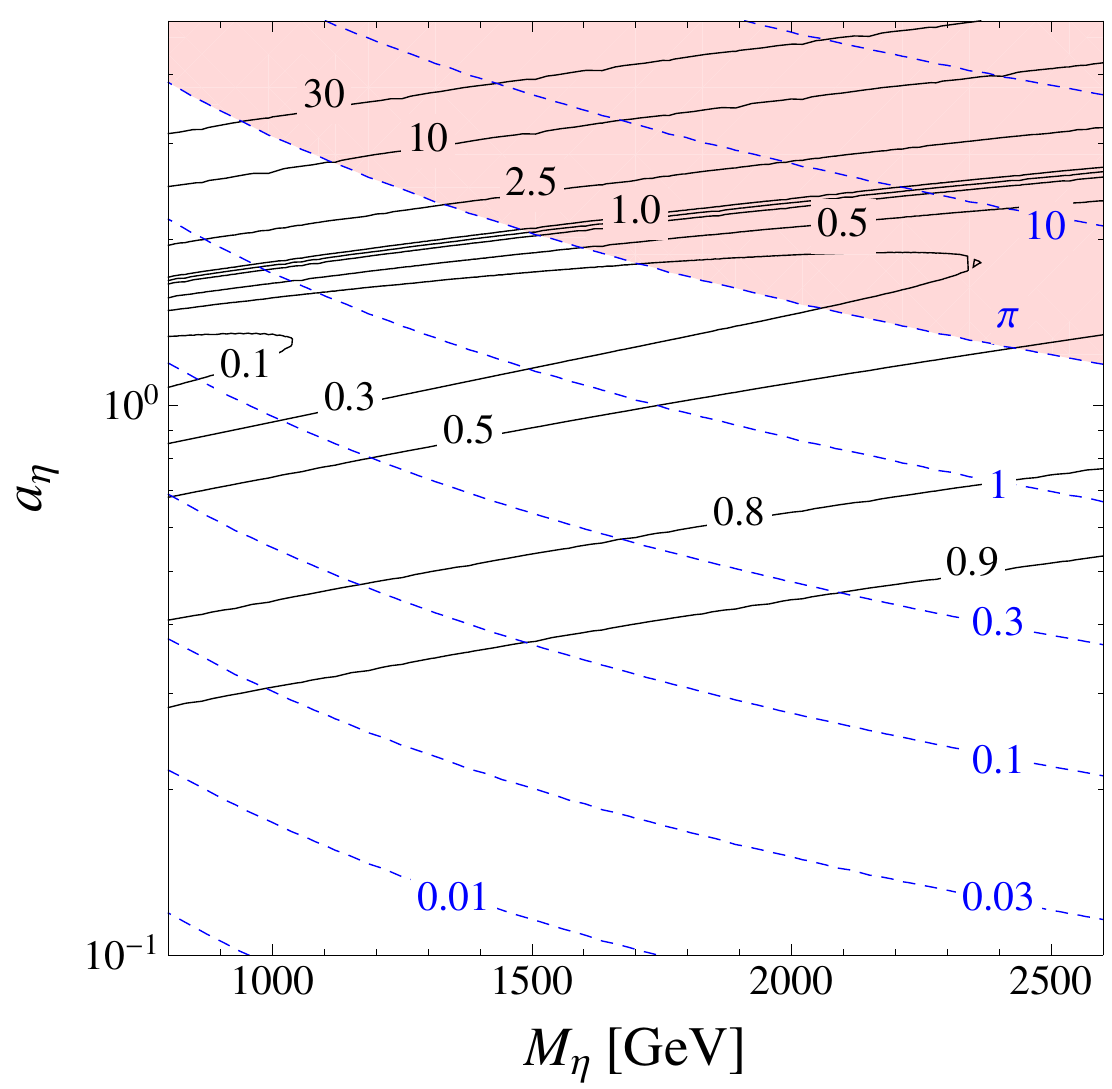}  
\end{center}
\end{minipage}
\end{center}
\caption{\label{fig:etacontours} 
\small
Contours of constant $R(\eta,\xi,m_\text{cut})$ (continuous black lines) and constant $\Gamma_\eta/m_\eta$ (dashed blue lines)
in the plane ($a_\eta, m_\eta$) for $\xi =0.5$, $m_\text{cut} = 800\,$GeV.  In the  pink (darker)  region  $\Gamma_\eta/m_\eta > \pi$,
and our perturbative calculation cannot be trusted, see text.
}
\end{figure}

\vspace{0.6cm}
Figure~\ref{fig:etacontours} shows the results for the  $\eta$.
In this case the contribution of the resonance tends to enhance the channels $hh$, $W^+_L W^-_L$
and to suppress $W^+_L W^+_L$, $W^+_L Z_L$.  Compared to the $\rho$, the $\eta$ has a larger decay width,
which implies a more extended region where $\Gamma_\eta/m_\eta > \pi$ and the coupling $\pi\pi\eta$ gets non-perturbative
(it corresponds to the pink (darker) area in Fig.~\ref{fig:etacontours}).
For reference values $m_{\eta} = 1.5\,$TeV, $a_{\eta} = 1$, we predict a suppression $R=0.5$ and $R=0.4$ respectively in the 
$W^+_L W_L^+$ and $W^+_L Z_L$ channels, and an enhancement $R =2.9$, $R = 1.9$ respectively in $W_L^+W_L^-$ and $hh$.
As expected from the above discussion, 
varying $\xi$ does not change the ratio $R$ in the case of the final states $W^+_L W^+_L$, $W^+_L Z_L$, while it has an important 
effect for $hh$, $W^+_L W^-_L$, where the resonance in exchanged in $s$-channel. 
This is clearly illustrated by the 
following tables, which report the value of $R$ in the channels $hh$ (left) and $W^+_L W^+_L$ (right) for $a_{\eta} = 1$:
\vspace{0.1cm}
\begin{center}
\begin{tabular}{rc|ccc}
$hh$ & & \multicolumn{3}{c}{$m_{\eta}\,$[GeV]} \\[0.15cm]
& $R$ & $1500$ & $2000$ & $2500$  \\[0.1cm]
\hline
&&&& \\[-0.3cm]
\multirow{3}{*}{$\xi$}\hspace{0.1cm}  & 0.1 & 8.4 & 3.6 & 2.1 \\[0.1cm]
                                                           & 0.5 & 1.9 & 1.3 & 1.1 \\[0.1cm]
                                                           & 0.8 & 1.4 & 1.1 & 1.0
\end{tabular}
\hspace{1.2cm}
\begin{tabular}{rc|ccc}
$W^+_L W^+_L$ & & \multicolumn{3}{c}{$m_{\eta}\,$[GeV]} \\[0.15cm]
& $R$ & $1500$ & $2000$ & $2500$  \\[0.1cm]
\hline
&&&& \\[-0.3cm]
\multirow{3}{*}{$\xi$}\hspace{0.1cm}  & 0.1 & 0.45 & 0.59 & 0.69 \\[0.1cm]
                                                           & 0.5 & 0.45 & 0.59 & 0.69 \\[0.1cm]
                                                           & 0.8 & 0.45 & 0.59 & 0.69
\end{tabular}
\end{center}
\vspace{0.25cm}

\begin{figure}[!t]
\begin{center}
\begin{minipage}{0.4\linewidth}
\begin{center}
\hspace*{-2.1cm} 
\fbox{\footnotesize $pp\to W^+ W^- jj$} \\[-0.04cm]
\includegraphics[width=68mm]{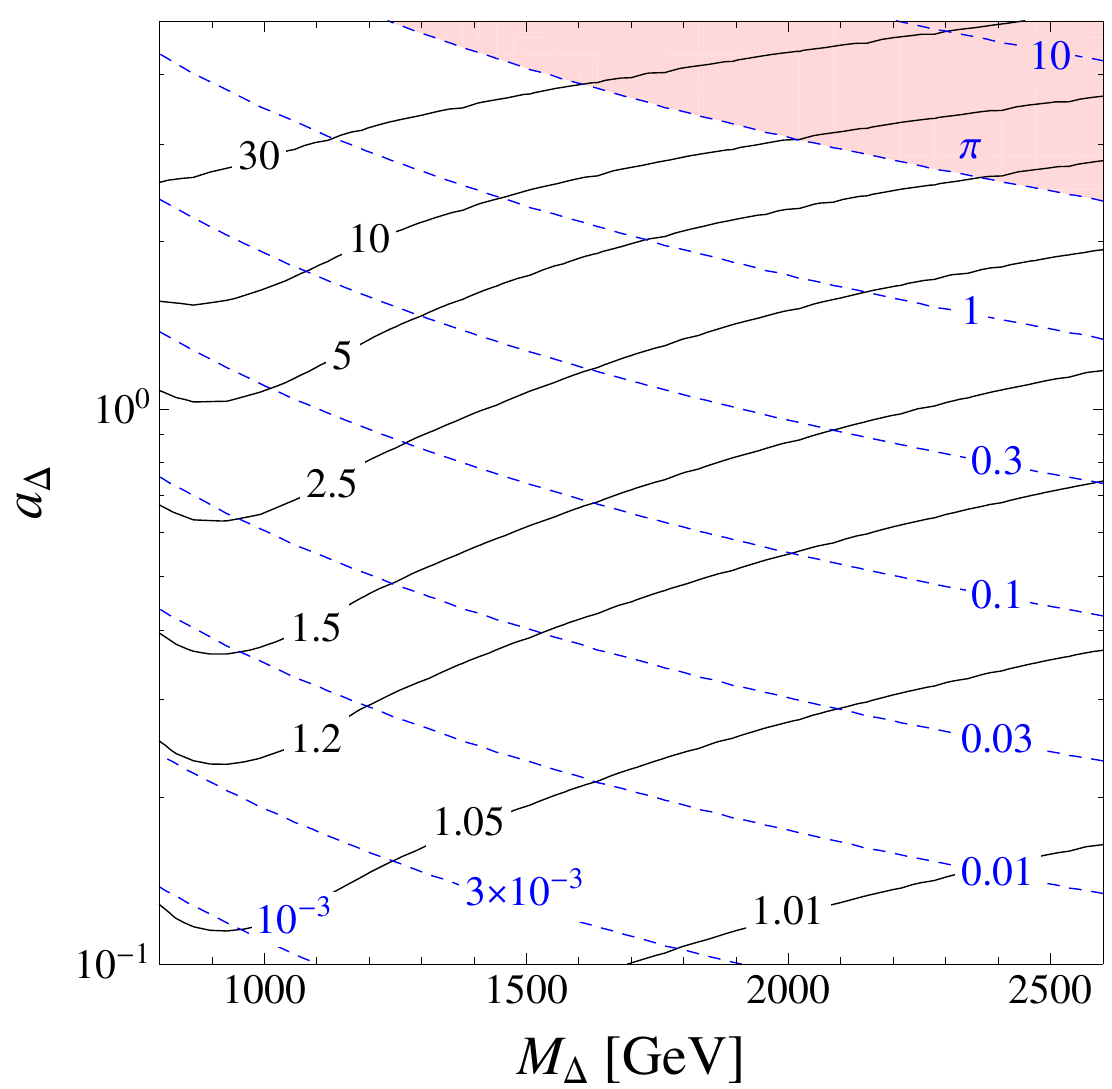} 
\end{center}
\end{minipage}
\hspace{0.6cm}
\begin{minipage}{0.4\linewidth}
\begin{center}
\hspace*{-2.9cm} 
\fbox{\footnotesize $pp\to hh jj$} \\[-0.04cm]
\includegraphics[width=68mm]{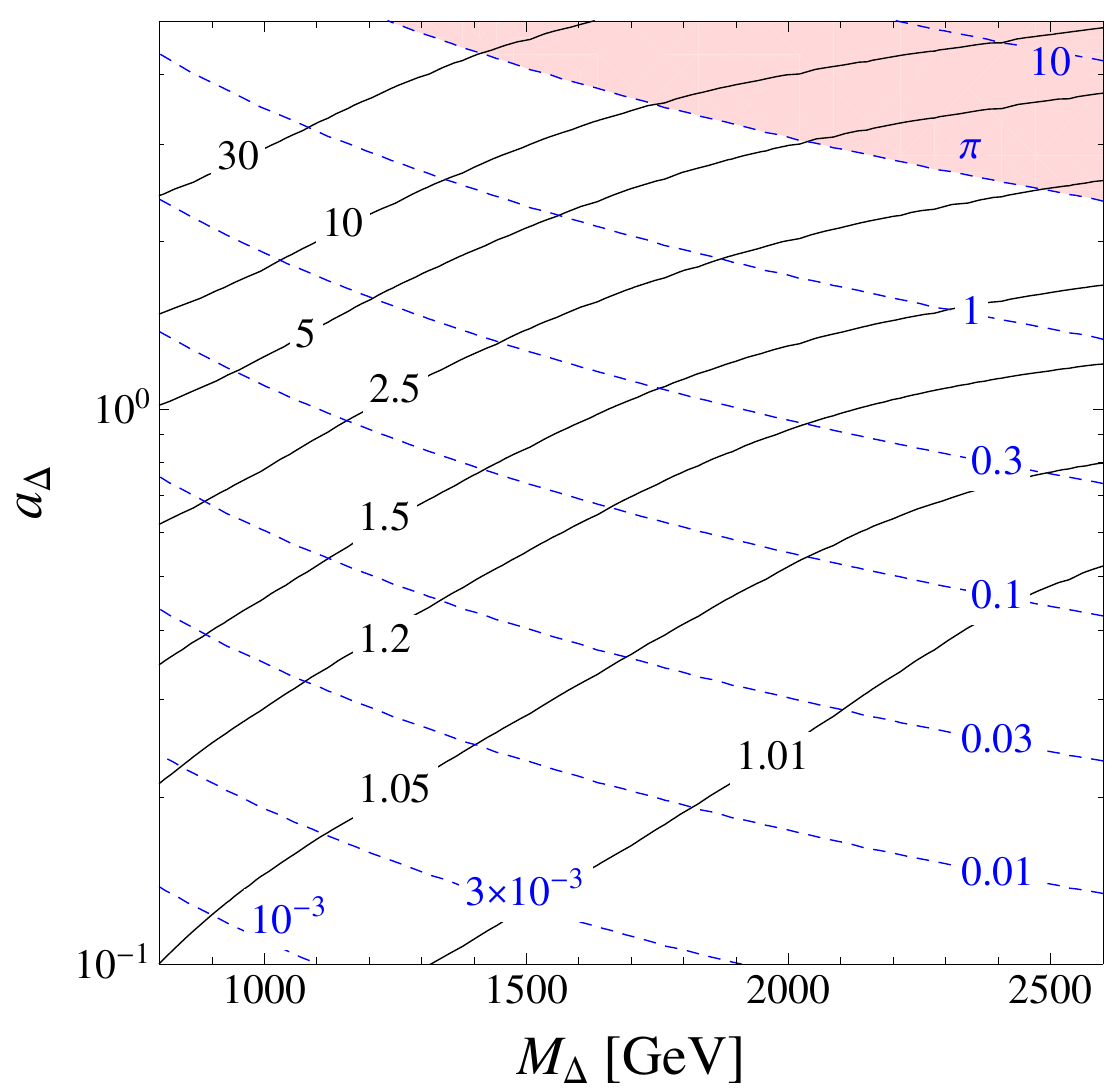} 
\end{center}
\end{minipage}
\\[0.6cm]
\begin{minipage}{0.4\linewidth}
\begin{center}
\hspace*{-2.02cm} 
\fbox{\footnotesize $pp\to W^+ W^+ jj$} \\[-0.04cm]
\includegraphics[width=70mm]{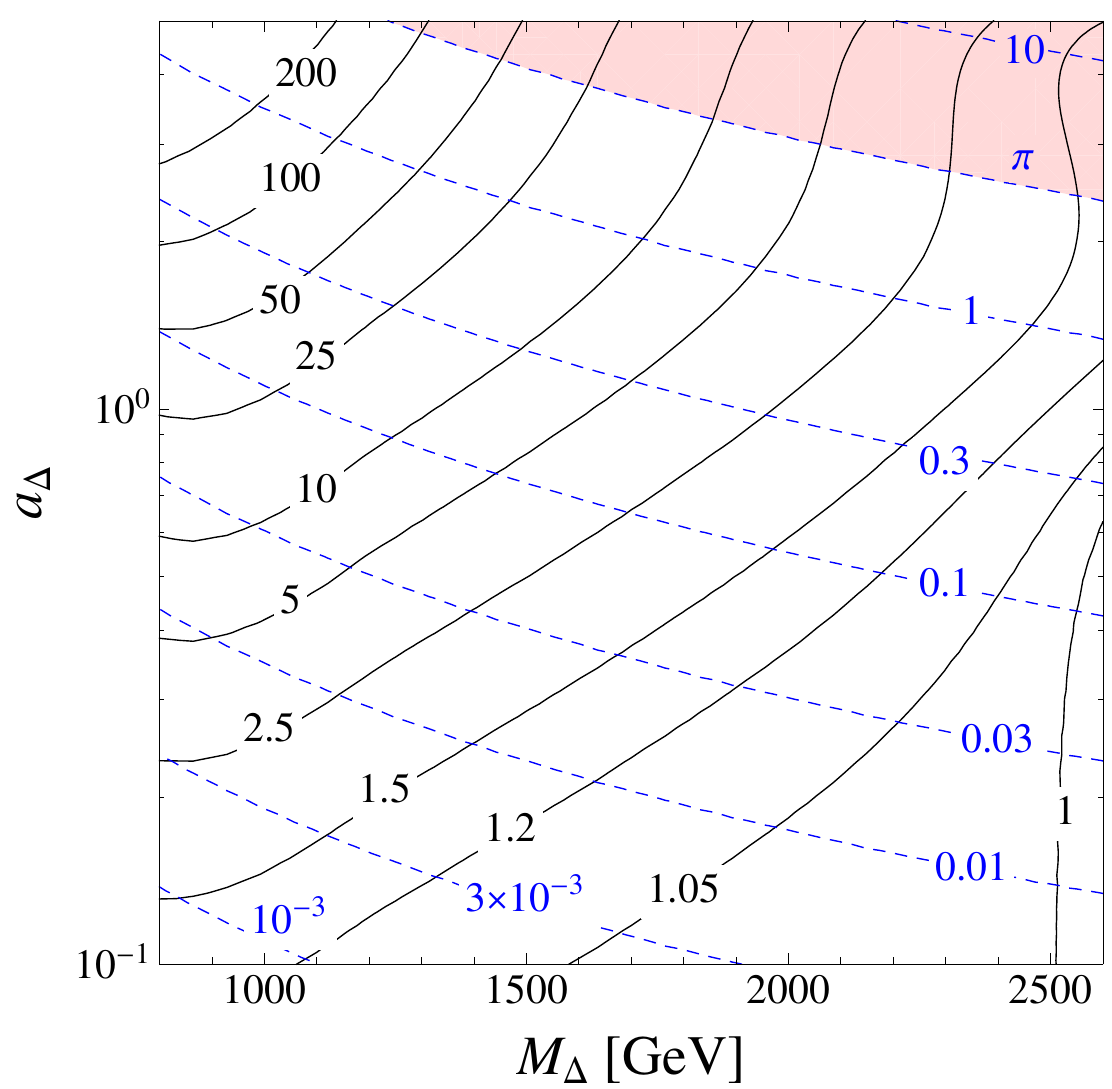}  
\end{center}
\end{minipage}
\hspace{0.6cm}
\begin{minipage}{0.4\linewidth}
\begin{center}
\hspace*{-2.4cm} 
\fbox{\footnotesize $pp\to W^\pm Z jj$} \\[-0.04cm]
\includegraphics[width=70mm]{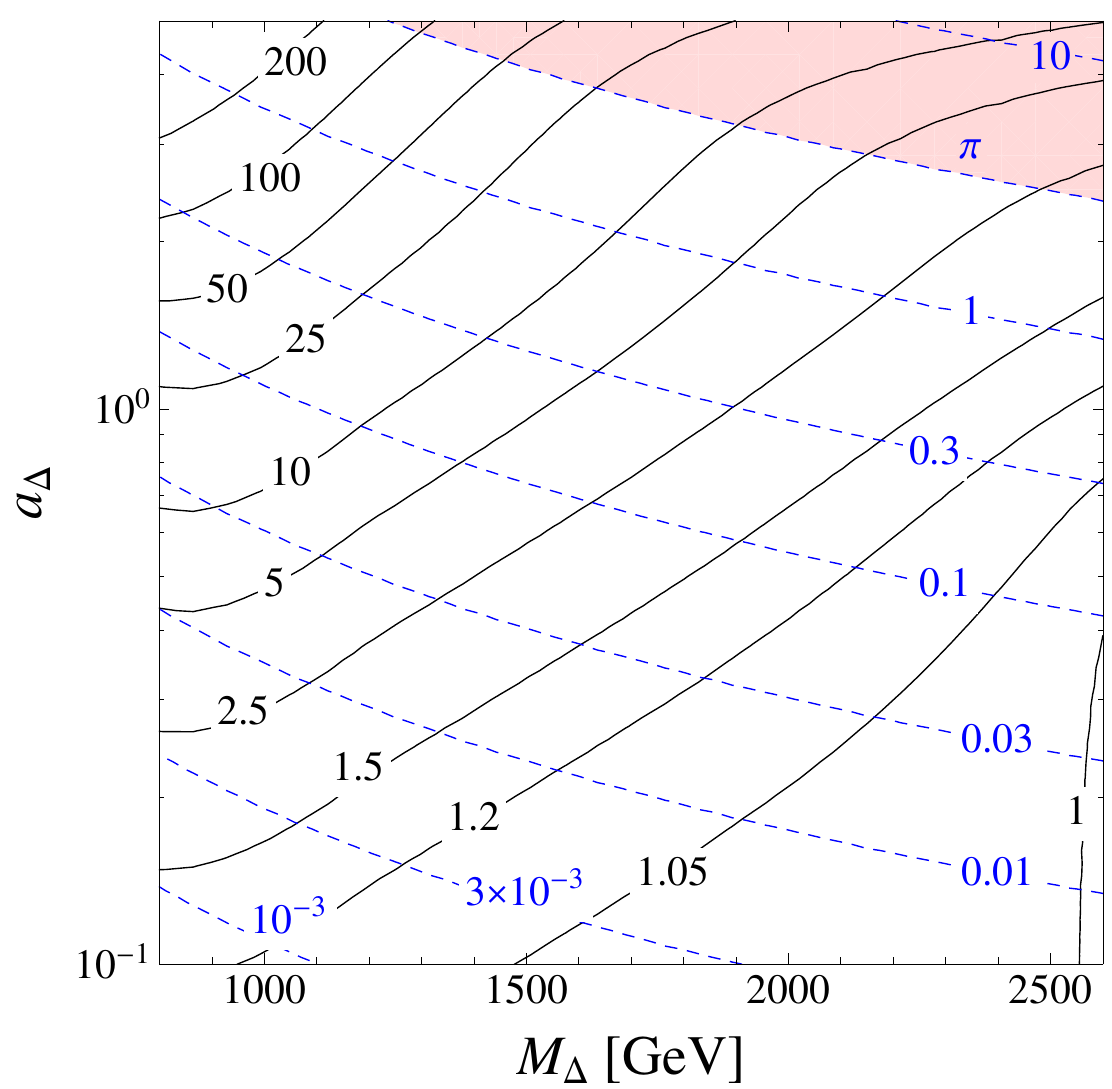}  
\end{center}
\end{minipage}
\end{center}
\caption{\label{fig:deltacontours} 
\small
Contours of constant $R(\Delta,\xi,m_\text{cut})$ (continuous black lines) and constant $\Gamma_\Delta/m_\Delta$ (dashed blue lines)
in the plane ($a_\Delta, m_\Delta$) for $\xi =0.5$ and $m_\text{cut} = 800\,$GeV.  In the  pink (darker)  region  $\Gamma_\Delta/m_\Delta > \pi$,
and our perturbative calculation cannot be trusted, see text.
}
\end{figure}

\vspace{0.6cm}
Finally, the results for the resonance $\Delta$ are shown in Fig.~\ref{fig:deltacontours}.
In this case, given the quantum numbers of the $\Delta$, the exchange of the resonance tends to enhance all the channels.
As for the $\eta$, our  calculation cannot be trusted in the pink (darker) area of the $(m_\Delta, a_\Delta)$ region shown 
in  Fig.~\ref{fig:deltacontours}, where the $\pi\pi\Delta$ coupling is non-perturbative ($\Gamma_\Delta/m_\Delta > \pi$).
For reference values $m_{\Delta} = 1.5\,$TeV, $a_{\Delta} = 1$, we predict an enhancement in all channels as follows: 
$R = 2.3$ in  $W^+_L W_L^-$; $R= 1.7$ in $hh$; $R = 6.6$ in $W^+_L W_L^+$; $R=5.3$ in $W^+_L Z_L$.
An increase of $\xi$ implies a suppression of the ratio $R$ in all the channels. This is illustrated 
by the  following tables, which report the value of $R$ in the channels $hh$ (left) and $W^+_L W^+_L$ (right) for $a_{\Delta} = 1$:
\vspace{0.1cm}
\begin{center}
\begin{tabular}{rc|ccc}
$hh$ & & \multicolumn{3}{c}{$m_{\Delta}\,$[GeV]} \\[0.15cm]
& $R$ & $1500$ & $2000$ & $2500$  \\[0.1cm]
\hline
&&&& \\[-0.3cm]
\multirow{3}{*}{$\xi$}\hspace{0.1cm}  & 0.1 & 3.2 & 1.7 & 1.2 \\[0.1cm]
                                                           & 0.5 & 1.7 & 1.3 & 1.1 \\[0.1cm]
                                                           & 0.8 & 1.6 & 1.2 & 1.1
\end{tabular}
\hspace{1.2cm}
\begin{tabular}{rc|ccc}
$W^+_L W^+_L$ & & \multicolumn{3}{c}{$m_{\Delta}\,$[GeV]} \\[0.15cm]
& $R$ & $1500$ & $2000$ & $2500$  \\[0.1cm]
\hline
&&&& \\[-0.3cm]
\multirow{3}{*}{$\xi$}\hspace{0.1cm}  & 0.1 & 33 & 11 & 4.3 \\[0.1cm]
                                                           & 0.5 & 6.6 & 2.4 & 1.2 \\[0.1cm]
                                                           & 0.8 & 4.3 & 1.7 & 1.0
\end{tabular}
\end{center}
\vspace{0.25cm}

\subsection{Check of the analytic approximation}
\label{sec:checkanalytic}

At this point we would like to discuss the precision of our approximate analytic calculation
of the cross sections. We will do so by comparing  with a full calculation
performed by  using  a Montecarlo simulation.
We consider the models MCHM4 and MCHM5 defined in Refs.~\cite{Agashe:2004rs} and \cite{Contino:2006qr},
and their  linearized versions, which we will denote respectively as LMCHM4 and LMCHM5 in the following. 
The LMCHM4, in particular,  has been already   considered in Ref.~\cite{Barbieri:2007bh}.
In both linearized models a scalar resonance $\eta$ is added to the original $SO(5)/SO(4)$
chiral lagrangian to  form a linear representation of $SO(5)$ together with the four NG bosons.
The lagrangian which describes the derivative interactions between $\eta$ and the NG bosons is that
of eq.(\ref{eq:etaLag}) with $a_\eta = b_\eta = 1$. 
The two linearized models thus differ only by (non-derivative) potential terms, as well the original MCHM4 and MCHM5
differ only by the Higgs trilinear coupling.
The reader can find all the details and the relevant formulas in Appendix~\ref{sec:LMCHM}. 

\begin{figure}[!t]
\begin{center}
\includegraphics[width=110mm]{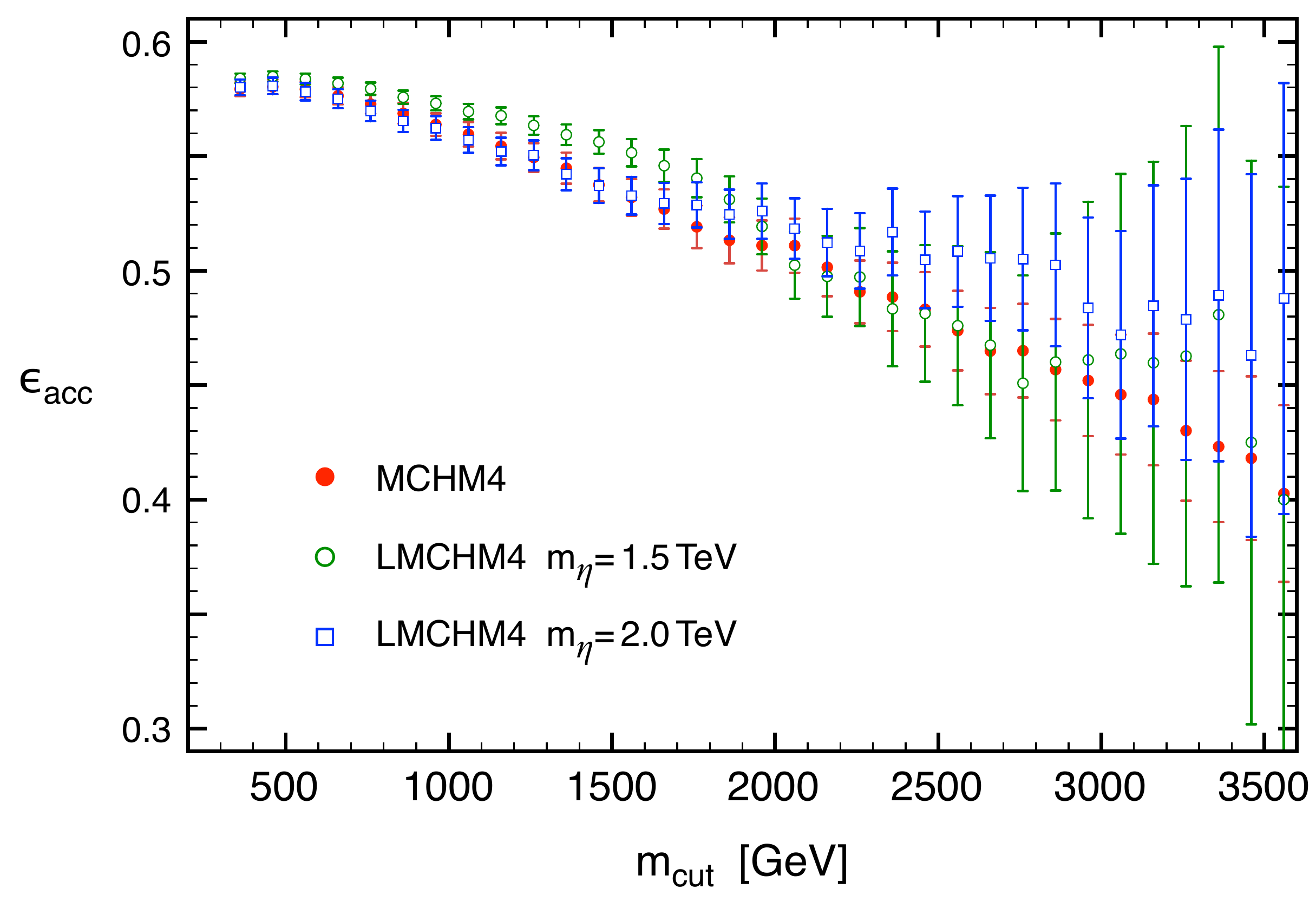}   
\end{center}
\caption{\label{fig:effvsmhhcut} 
\small
Efficiency of the acceptance cuts of eq.(\ref{eq:AC})  as a function of $m_\text{cut}$ for  the process $pp\to hhjj$  (with $m(hh) \geq m_\text{cut}$).
Filled red circles, empty green circles and empty blue squares respectively correspond to the MCHM4, 
the LMCHM4 with $m_\eta = 1.5\,$TeV, the LMCHM4 with $m_\eta = 2.0\,$TeV.
In all the models  $\xi=0.5$. The vertical bars report the statistical theoretical error in each point.
}
\end{figure}

We first checked that the efficiency of the acceptance cuts defined in eq.(\ref{eq:AC}) is universal, \textit{i.e.} it does not
depend on the model, as prescribed by the EWA.
For simplicity, we focused on the process $pp\to hhjj$ and set $m_h=180\,$GeV,~\footnote{For this value of the Higgs mass,
both in the MCHM4 and MCHM5 as well as in their linearized versions, the region at small $\xi$ is excluded by 
the LHC results recently presented at the 2011 Lepton Photon Conference. As explained below in the text, 
in this section we choose $\xi=0.5$, which is  presently not excluded by any Higgs search.} 
so that our results partly extend  the analysis of Ref.~\cite{Contino:2010mh}.
We used  the Montecarlo program MadGraph~\cite{MG-ME}  to derive the cross sections in the models  MCHM4 and MCHM5
and in their linearized versions.~\footnote{Like in the previous section, we used the 
 set of PDFs CTEQ611 and fixed the factorization scale to $Q = m_W$.}
Figure~\ref{fig:effvsmhhcut} reports the efficiency of the acceptance cuts, $\epsilon_{acc}$, as a function of the value
of $m_\text{cut}$ in the MCHM4 and in the LMCHM4 for $m_\eta = 1.5\,$TeV and $m_\eta = 2.0\,$TeV. 
In all the models $\xi$ has been set to 0.5.
The efficiency is the same in the three models within the statistical error.
We find that the same conclusion holds in the MCHM5 and its linearized version, and when $\xi$ is varied.
The decrease of $\epsilon_{acc}$ with $m_\text{cut}$ is a simple kinematic effect: for fixed 
center of mass energy $\hat s$ of the quarks initiating the scattering, the outgoing jets tend to be softer for larger values of $m_\text{cut}$,
hence the efficiency of the acceptance cuts falls off.

\begin{figure}[!t]
\begin{center}
\begin{minipage}[t]{\textwidth}
\begin{center}
\includegraphics[width=100mm]{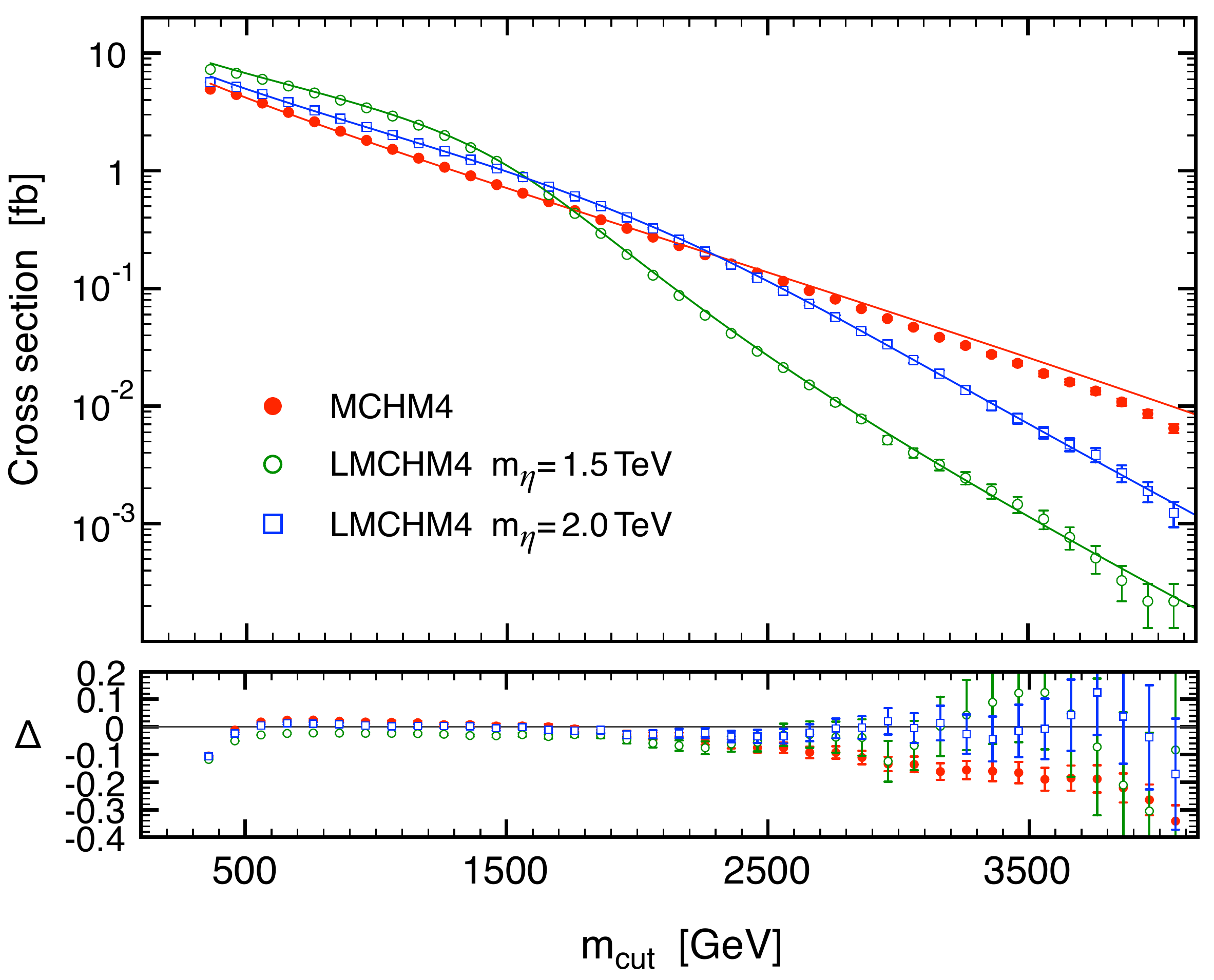} 
\caption{\label{fig:sigvsmhhcutA} 
\small
Upper panel: Cross section of $pp\to hhjj$ as a function of the cut on $m_{hh}$. 
Filled red circles, empty green circles and empty blue squares respectively correspond to the MCHM4, 
the LMCHM4 with $m_\eta = 1.5\,$TeV, the LMCHM4 with $m_\eta = 2.0\,$TeV. The continuous curves denote the corresponding
analytic results. 
Lower panel: relative difference between the Montecarlo and the analytic predictions, $\Delta = (\sigma(MC)-\sigma(analytic))/\sigma(analytic)$,
as a function of the cut on $m_{hh}$. 
In all the models  $\xi=0.5$. The vertical bars report the statistical theoretical error in each point.
}
\end{center}
\end{minipage}
 \\[0.75cm]
\begin{minipage}[t]{\textwidth}
\begin{center}
\includegraphics[width=100mm]{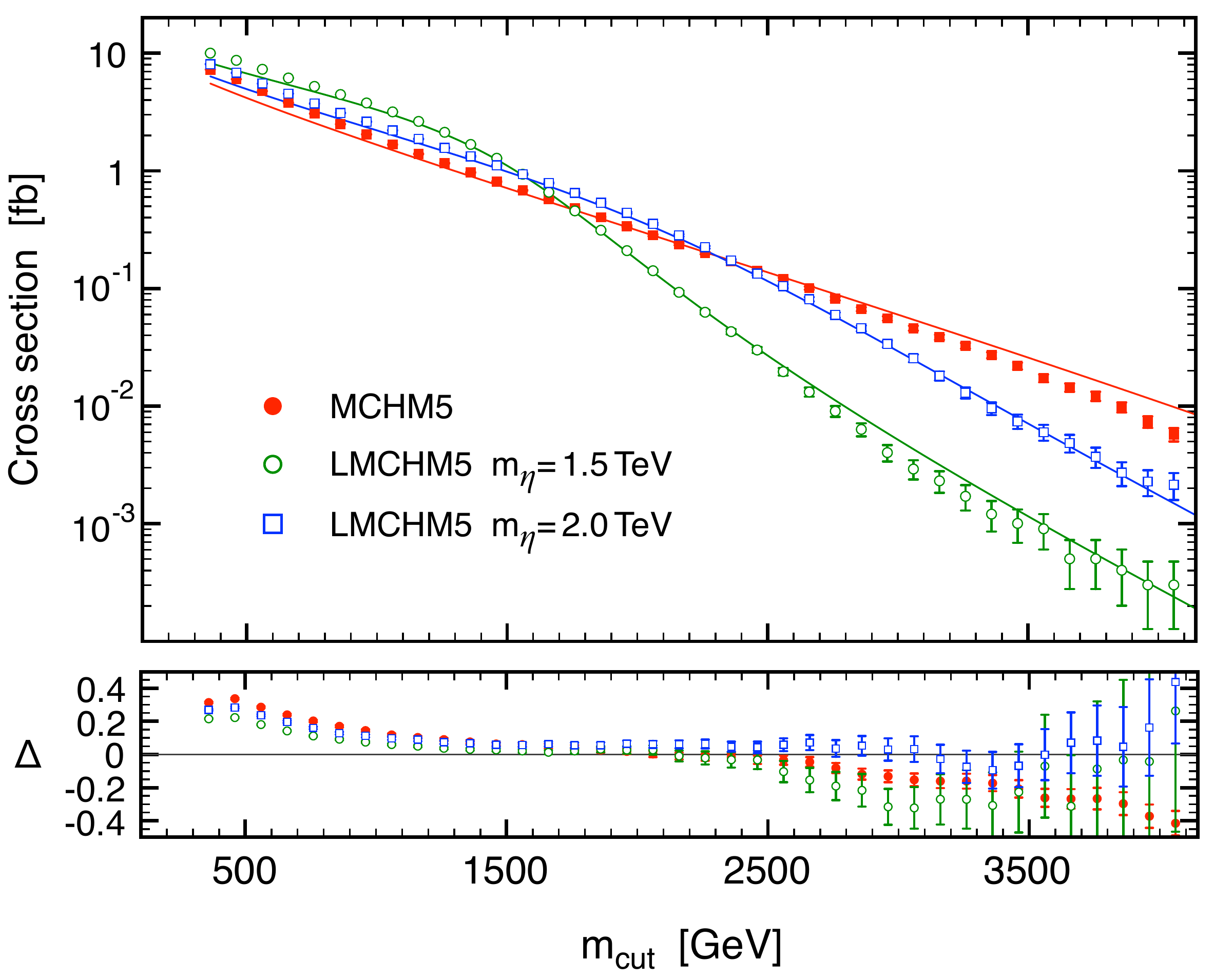} 
\caption{\label{fig:sigvsmhhcutB} 
\small
As in Fig.~\ref{fig:sigvsmhhcutA} but for the MCHM5 and its linearized version.
}
\end{center}
\end{minipage}
\end{center}
\end{figure}

\begin{figure}[!]
\begin{center}
\begin{minipage}[t]{\textwidth}
\begin{center}
\includegraphics[width=100mm]{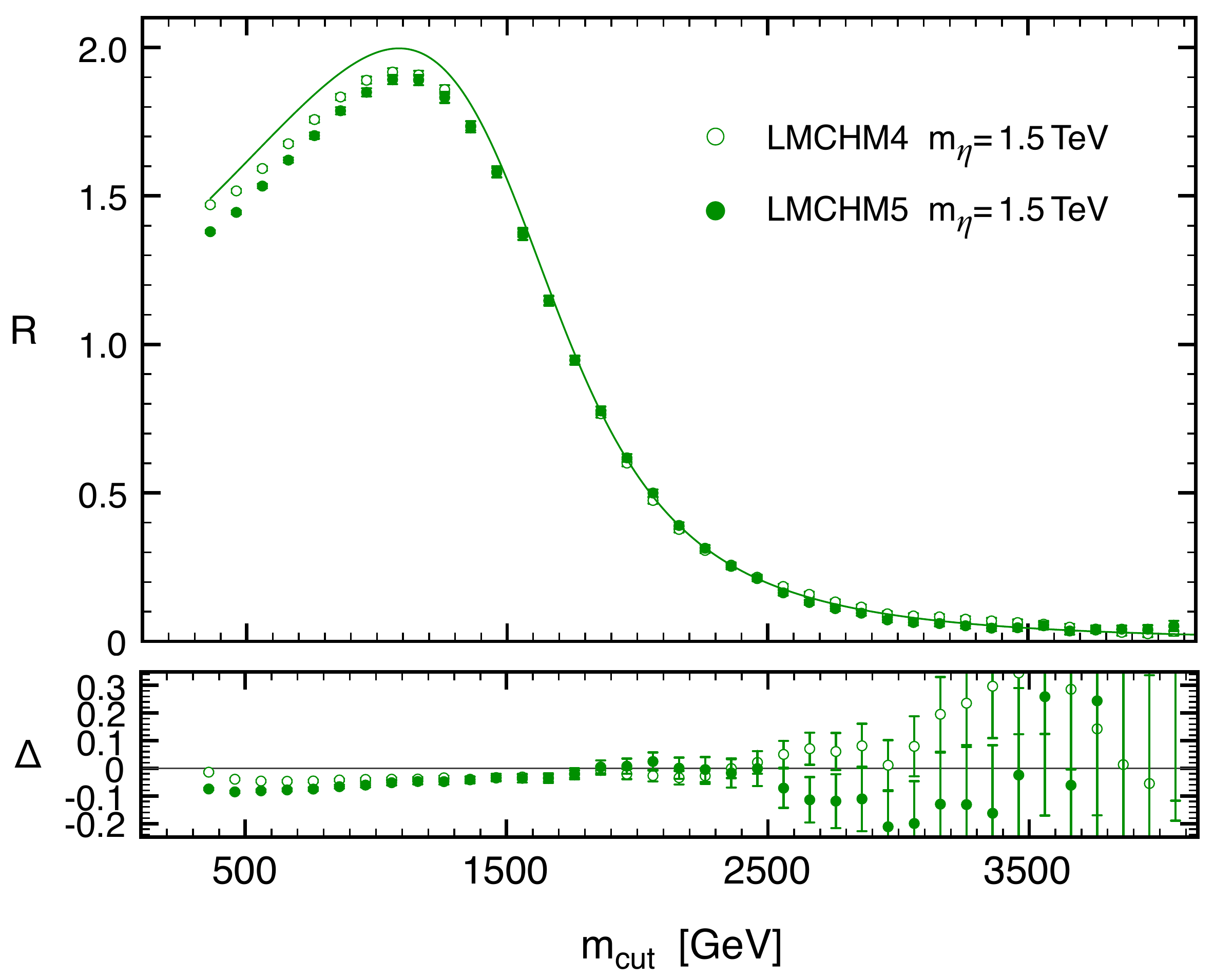} 
\caption{\label{fig:RvsmhhcutA} 
\small
Upper panel: Ratio $R(\eta,\xi=0.5,m_\text{cut})$  for the process $pp\to hhjj$ as a function of the cut on $m_{hh}$. 
Empty (filled) green circles  correspond to the LMCHM4 (LMCHM5) with $m_\eta = 1.5\,$TeV. 
The continuous curve denotes the analytic result. 
Lower panel: relative difference between the Montecarlo and the analytic prediction, $\Delta = (R(MC)-R(analytic))/R(analytic)$,
as a function of the cut on $m_{hh}$. The vertical bars report the statistical theoretical error in each point.
}
\end{center}
\end{minipage}
 \\[0.75cm]
\begin{minipage}[t]{\textwidth}
\begin{center}
\includegraphics[width=100mm]{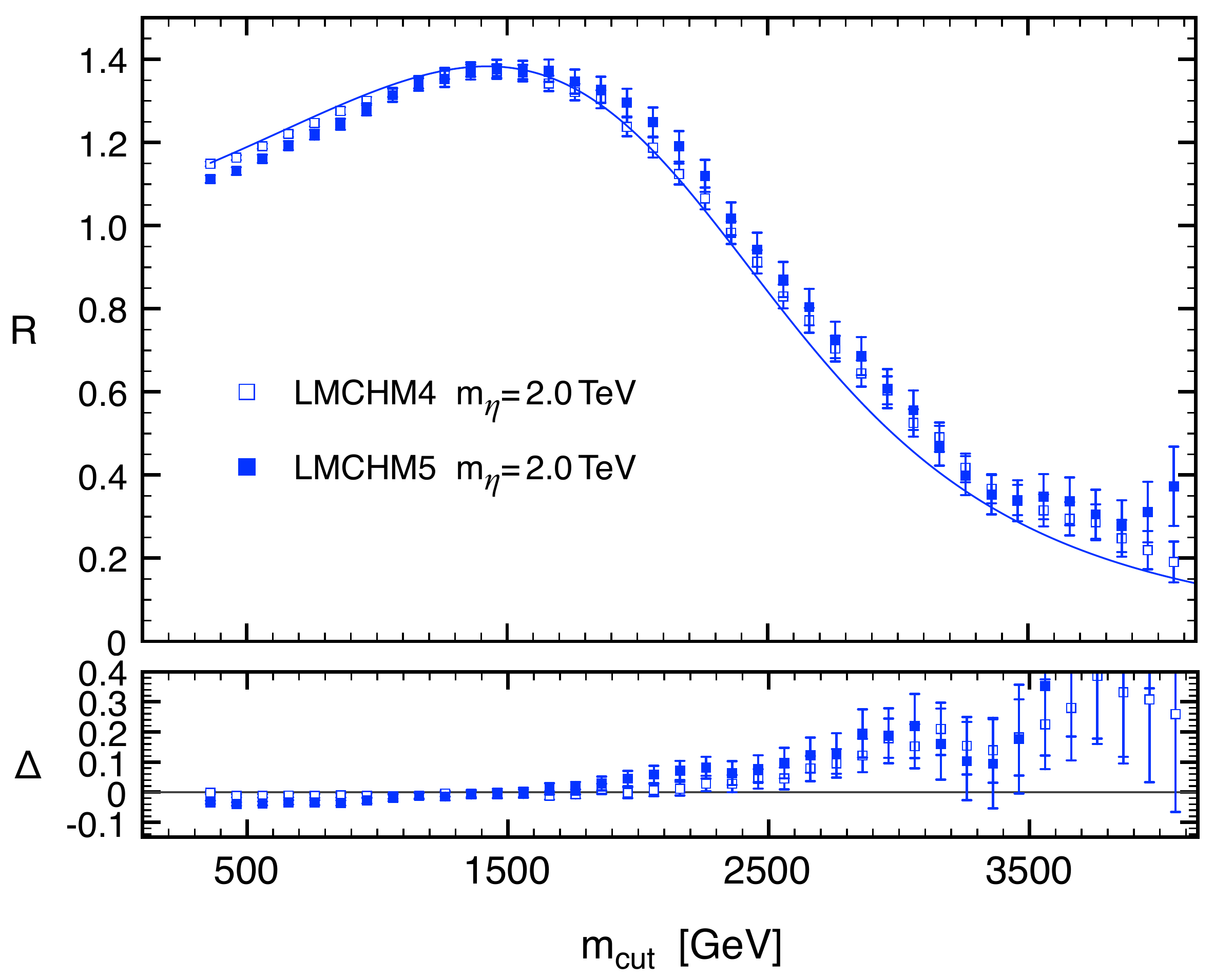} 
\caption{\label{fig:RvsmhhcutB} 
\small
Upper panel: Ratio $R(\eta,\xi=0.5,m_\text{cut})$  for the process $pp\to hhjj$ as a function of the cut on $m_{hh}$. 
Empty (filled) blue squares  correspond to the LMCHM4 (LMCHM5) with $m_\eta = 2.0\,$TeV. 
The continuous curve denotes the analytic result. 
Lower panel: relative difference between the Montecarlo and the analytic prediction, $\Delta = (R(MC)-R(analytic))/R(analytic)$,
as a function of the cut on $m_{hh}$. The vertical bars report the statistical theoretical error in each point.
}
\end{center}
\end{minipage}
\end{center}
\end{figure}

The second issue to address is how well our analytic calculation of the cross section, in which we neglected the effect of non-derivative
couplings and relied on the Equivalence Theorem,  reproduces the full result. As already discussed, we expect the agreement to be good
for large values of $m_\text{cut}$, whereas  the contribution from non-derivative couplings and the transverse polarizations
of the vector bosons should become important approaching the threshold.
This expectation is indeed confirmed by comparing the  Montecarlo and the analytic results for the cross section:
the plots of Figures~\ref{fig:sigvsmhhcutA} and~\ref{fig:sigvsmhhcutB} show that while in the MCHM4 and its linearized versions the agreement turns out
to be at the level of $5\%$ or better, except for points right at threshold, in the case of the (L)MCHM5 the analytic curve 
differs from the Montecarlo prediction   by up to $30-40\%$ for $m_\text{cut}\lesssim 1\,$TeV. In other words, we find that 
the model dependency due to non-derivative couplings starts to be important at energies $m_{hh} \sim 1\,$TeV, in agreement with
what found by Ref.~\cite{Contino:2010mh}.
Interestingly,  part of this model dependency cancels out in taking the ratio of cross sections, so that $R$ is 
better reproduced by the  analytic calculation. Figures~\ref{fig:RvsmhhcutA} and~\ref{fig:RvsmhhcutB}  show that 
the agreement is better that $10\%$, even for values of the cut near threshold, both in the LMCHM4 and the LMCHM5.
We thus conclude that $R$ is a quite robust quantity to monitor the importance of the resonance contribution.

\section{Summary}
\label{sec:conclusions}

In this paper we have extended the  study of the phenomenology of a general $SO(5)/SO(4)$ composite Higgs model beyond the leading $O(p^2)$ 
chiral lagrangian.  We  have done that along two different, but related, approaches.

Our first approach consisted in the study of the general lagrangian at $O(p^4)$. In doing that we have classified the relevant discrete symmetries, 
accidental or approximate, and discussed their possible phenomenological role.
First of all we have pointed out  $R$, the grading $Z_2$ symmetry of $SO(5)/SO(4)$ under which the broken generators are odd while the unbroken ones 
are even. Since $R\subset SO(4)$, it is  exact to all orders in the chiral lagrangian in the limit in which the SM couplings are neglected. One consequence 
is that no processes involving an odd number of Goldstones can be significantly enhanced by the strong interaction, even after  taking into account 
a non-vanishing Higgs expectation value. In particular, the process $W^+_LW^-_L\to hhh$ vanishes at leading order in the SM gauge couplings~\cite{us}. 
Another relevant symmetry  is $P_{LR}$, the exchange of the L and R components of $SU(2)_L\times SU(2)_R\sim SO(4)$. We have shown that $P_{LR}$ is an 
accidental symmetry up to $O(p^2)$ and is broken by several $O(p^4)$ operators. Moreover,  the leading effect of $P_{LR}$ breaking in $\pi\pi$ scattering 
can only  arise at $O(p^6)$ in the chiral lagrangian. Such a term could be induced at tree level by the exchange of a vector resonance either in the $(3,1)$ 
or in the $(1,3)$ of $SO(4)$, of course provided the spectrum breaks $P_{LR}$. Such a possibility is perfectly sensible although often disregarded in explicit 
5D constructions. However,  because of the $O(p^6)$ dependence,  the potentially  interesting $P_{LR}$-breaking process $W_L^+W_L^-\to Z_L h$ is very 
suppressed at low energy, and can only become relevant at the LHC if a $P_{LR}$-breaking vector resonance happens to be rather light.
Finally, another relevant symmetry is Higgs parity $P_h\equiv R\times P_{LR}$. The validity of $P_h$ at the chiral lagrangian level coincides with the 
validity of $P_{LR}$. However, for the specific case where $v=f$, $P_h$ survives in the bosonic sector even after gauging $SU(2)_L\times U(1)_Y$.
 In the limit where the SM couplings are turned off, $P_{h}$ is valid under the same conditions as $P_{LR}$, since $R\subset SO(4)$ is exact. 
 
In our second approach we modeled the lagrangian of possible low-lying resonances and studied their effects. Our picture is that there may be some 
resonances that are lighter than the ``bulk" of the spectrum and that their effect may be reasonably well described by a phenomenological lagrangian. 
In order to formalize that picture we have worked under the assumption that the bulk of the spectrum is broadly characterized by a coupling $g_*$, and 
a mass scale $\Lambda \sim g_* f$, while the low-lying resonance $\Phi$ has  mass $m_\Phi$ parametrically smaller than $\Lambda$.
We have introduced a principle, that we call {\it partial UV completion} (PUVC), to parametrize the couplings of such $\Phi$. The simplest way to phrase 
PUVC is that it  corresponds to the request that all couplings be of order $g_*$ at the scale $\Lambda$. The main practical consequence of PUVC is that 
the contribution of $\Phi$  in $\pi\pi$ scattering is well described by just $m_\Phi$ and a coupling $g_\Phi $ such that 
$a_\Phi \equiv m_\Phi /g_\Phi f\sim 1$. ~\footnote {One possible explicit realization of PUVC in the parametric limit $g_\Phi \ll g_*$ can be obtained in 5D 
constructions by introducing large kinetic terms on the IR brane. In that case $g_*$ corresponds to the KK coupling, while $g_\Phi$ is controlled by the 
boundary kinetic term.}
One way to see that PUVC is sensible is by noticing that it corresponds to the range of coupling and mass such that  the interaction among NG bosons 
is not made stronger by the presence of the light resonance. 
Moreover, within the range of parameters dictated by PUVC, one can accommodate models that reduce over a certain range the strength of the coupling 
among NG bosons. 
In this sense PUVC generalizes to a broader set of models the restrictive parameter choice of theories where the lighter resonance is required
to exactly unitarize $\pi\pi$ scattering.
Our description is strictly speaking  perturbative and consistent as long as $g_\Phi \ll g_*$. In particular, for the case $g_*\sim 4\pi$ we need 
$g_\Phi\ll 4\pi$. Nonetheless the hope is that our parametrization could be qualitatively correct in realistic and genuinely strongly coupled cases. 
That is in analogy with other situations, like large $N_c$ in QCD, where the insights of an approximation seemingly extend beyond its expected range of validity.

We have applied our parametrization to study  resonances of spin 0 and 1 that  can affect $\pi\pi$ scattering. For spin 0, Bose symmetry dictates that we 
can either have a $\eta\equiv (1,1)$ of $SO(4)$ or a $\Delta\equiv (3,3)$. In the spin-1 channel we can instead have either $\rho_L\equiv (3,1)$ or 
$\rho_R\equiv (1,3)$. Two broad properties should be remarked. The first, perhaps not obvious, is that according to PUVC vector resonances are narrower, 
relative to their mass, than the scalar ones. That is even more true when comparing $\rho_L$ or $\rho_R$ to the total singlet~$\eta$. This fact was already 
remarked before \cite{slava} and is compatible with the situation in QCD, where the $\sigma$ is relatively much broader than the $\rho$. The narrowness of the 
vectors makes them suitable for discovery in Drell-Yan production.
Indeed  $a_\rho$, $m_\rho$ and possibly $\alpha_2$ offer a simple, and theoretically robust, parametrization of the Drell-Yan
production and decay of  composite vectors. That seems  ideal to express the results of the searches at the LHC.
The second and more obvious property  is that each resonance enhances those processes where it is exchanged in the $s$-channel while it depletes those 
in which it is exchanged only in the $t$- and $u$-channels. Thus $\rho_{L,R}$ exchange enhances the cross section for $W^+W^-$ and $WZ$ final states 
and depletes the one in $hh$ and $W^+W^+$. Exchange of the total singlet $\eta$ enhances $W^+W^-$ and $hh$ and reduces the others. Finally, the 
tensor scalar $\Delta$ appears in the $s$-channel for all processes, and thus enhances them all. 

Aside these qualitative features it turns out that the effect of the scalars is quantitatively more significant at comparable value of the mass and of the 
corresponding coefficient $a_\Phi$. 
Indeed, as already noticed for the case of technicolor, the exchange of a vector resonance does not significantly
change the $pp\to VVjj$ cross section: less that a factor of 2 over a broad range of parameters, and this is even more true in the region which is favored 
by the bound on the $S$ parameter.  More precisely, focussing on the case where $\xi\gsim 0.3$ for which there is some chance to study $WW$ scattering
at the LHC, one can synthesize the results as follows (see Figs.~\ref{fig:rhoLcontours},~\ref{fig:etacontours},~\ref{fig:deltacontours}).
For $m_\Phi\gsim 2\,$TeV the cross section is modified by less than $\sim 50\%$ for $\rho$ and $\eta$ (with the exception of the $W^+W^-$ channel, 
where  the effect of the exchange of  $\eta$ is larger), while for $\Delta$ one can have an enhancement of order $2-3$. 
For $m_\Phi\sim 1.5\,$TeV and $a_\phi \sim 1$, the exchange of $\rho$ can enhance $WZ$ by a factor of $3$, while the exchange of $\eta$ can enhance $W^+W^-$ and $hh$ by up to a factor of $5$ and $3$ respectively. 
For the same mass and coupling, the exchange of $\Delta$ enhances $W^+W^+$  by a factor of $11$ and $WZ$ by a factor of $9$.

A consequence of these results is that only  for scalar resonances below  $\sim 1.5\,$TeV and $a_\Phi \gsim  1$
can one hope to successfully study $hh$ production before the luminosity upgrade of the LHC. Of course it all crucially depends on the 
value of the Higgs mass. But that should become known in a very short time.

\section*{Acknowledgments}

We would like to thank
R.~Barbieri,
J.~Galloway,
C.~Grojean,
G.~Isidori, 
M.~Moretti,
T.~Okui,
M.~Redi,
S.~Rychkov,
R.~Torre,
E.~Trincherini,
A.~Wulzer
for useful discussions and comments. 
We also thank C.~Grojean and M.~Salvarezza for pointing out a missing factor 2 in eq.(\ref{eq:Spar}) in the first version of the paper and other typos.
The work of D.P. and R.R. is supported by the Swiss National Science Foundation
under contracts No. 200021-116372 and No. 200021-125237.  
D.P. and R.R would also like to thank the Kavli Institute of Theoretical Physics at the University of California Santa Barbara 
for hospitality during the completion of this work. R.C. acknowledges the Aspen Center for Physics for hospitality during part of this work.

\appendix
\section*{Appendix}

\section{Generators of $SO(5)$}
\label{app:SO5generators}

We  report here the expression of the $SO(5)$ generators adopted in our work.  We define as follows the generators of
$SO(5)\to SO(4)^\prime$, where $SO(4)^\prime$ is the gauged subgroup:
\begin{equation}
\begin{split}
	T^\acap_{ij}(0) &= - \frac{i}{\sqrt{2}} \left( \delta^{\acap i} \delta^{5 j} - \delta^{\acap j} \delta^{5 i}   \right)\, , \\
	T^{a \: L / R}_{ij}(0) &= - \frac{i}{2} \left( \frac{1}{2} \epsilon^{abc} (\delta^{b i} \delta^{c j} - \delta^{b j} \delta^{c i} ) 
                                            \pm (\delta^{ai} \delta^{4j} - \delta^{aj} \delta^{4i} ) \right) \, .
\end{split}
\end{equation}
Here $\acap = 1,2,3,4$,  $a,b,c = 1,2,3$ and $i,j = 1, \dots, 5$. 
The generators of $SO(5)\to SO(4)$, denoted as $T^{\alpha}(\theta)$ in the text ($\alpha = a,\hat a$), 
can be obtained by rotating the $T^{\alpha}(0)$ by an angle~$\theta$:
\begin{equation} \label{eq:thetarotation}
T^{\alpha}(\theta) = r(\theta)\, T^{\alpha}(0) \, r^{-1}(\theta)\, ,  \qquad
r(\theta) = \begin{pmatrix} 
1 & 0 & 0 & 0 & 0 \\ 
0 & 1& 0 & 0 & 0 \\ 
0 & 0 & 1 & 0 & 0 \\
0 & 0 & 0 & \cos\theta & \sin\theta \\
0 & 0 & 0 & -\sin\theta & \cos\theta
 \end{pmatrix}\, .
\end{equation}
In particular, one has ($i,\hat i=1,2,3$):
\begin{equation}
\begin{split}
T^i_{L/R}(\theta) &= \frac{T^i_L(0)+T^i_R(0)}{2} \pm \cos\theta \, \frac{T^i_L(0)-T^i_R(0)}{2} \mp \frac{\sin\theta}{\sqrt{2}} \, T^{\hat i}(0) \\[0.1cm]
T^{\hat i}(\theta) &= \sin\theta \, \frac{T^i_L(0)-T^i_R(0)}{\sqrt{2}}  + \cos\theta \, T^{\hat i}(0) \\[0.1cm]
T^{\hat 4}(\theta) &= T^{\hat 4}(0)\, .
\end{split}
\end{equation}
With the above definitions, all the generators are normalized as $\trace[T^\alpha T^\beta] = \delta^{\alpha\beta} $. 
They satisfy the following commutation rules:
\begin{equation} \label{eq:SO5algebra}
\begin{gathered}
\left[T^{aL} , T^{bR} \right] = 0\, ,  \qquad \left[ T^{aL} , T^{bL} \right] =  i \epsilon^{abc} T^{cL} \, , 
 \qquad  \left[ T^{aR} , T^{bR} \right] = i \epsilon^{abc} T^{cR}  \\[0.5cm]
\left[ T^\capp{i} , T^\capp{4} \right] =\frac{i}{2  } \delta^{\capp{i} k} (T^{k_L} - T^{k_R})\, ,  \qquad
\left[ T^\capp{i} , T^\capp{j} \right] = \frac{i}{2 } \epsilon^{\capp{i} \capp{j} k} (T^{k_L} + T^{k_R}) \\[0.5cm]
\begin{aligned}
        \left[ T^\capp{i} , T^{a_L} \right] =& - \frac{i}{2} \delta^{\capp{i} a } T^\capp{4} + \frac{i}{2} \epsilon^{\capp{i} a \capp{j}} T^\capp{j}\, ,  & \quad
        \left[ T^\capp{i} , T^{a_R} \right] =& \frac{i}{2  } \delta^{\capp{i} a } T^\capp{4} + \frac{i}{2} \epsilon^{\capp{i} a \capp{j}} T^\capp{j} \\[0.15cm]
        \left[ T^{\capp{4}} , T^{a_L} \right] =& \frac{i}{2  } \delta^{a \capp{i}} T^\capp{i} \, , & 
	\left[ T^{\capp{4}} , T^{a_R} \right] =& - \frac{i}{2  } \delta^{a \capp{i}} T^\capp{i} \, .
\end{aligned}
\end{gathered}
\end{equation}

From the above relations one can extract the $SO(5)$ structure  constants, $C^{\alpha\beta\gamma}$,
which are defined by $[T^\alpha, T^\beta] = i \, C^{\alpha\beta\gamma} T^\gamma$. 
They satisfy the following identities (an implicit sum over repeated indices is understood):
\begin{equation}
\label{eq:costanti_struttura_parziale}
\begin{split}
C^{\acap \capp{b} a_L} C^{\capp{c} \capp{d} a_L} &= \frac{1}{2} C^{\acap \capp{b} e} C^{\capp{c} \capp{d} e} - \frac{1}{4}  \epsilon^{\capp{a} \capp{b} \capp{c} \capp{d} } \\[0.15cm]
C^{\acap \capp{b} a_R} C^{\capp{c} \capp{d} a_R} &= \frac{1}{2} C^{\acap \capp{b} e} C^{\capp{c} \capp{d} e} +  \frac{1}{4}  \epsilon^{\capp{a} \capp{b} \capp{c} \capp{d} }\, , 
\end{split}
\end{equation}
where
\begin{equation}
\label{eq:somma_cost_struttura_SO5}
C^{\acap \capp{b} e} C^{\capp{c} \capp{d} e} = - \frac{1}{2} \left( \delta^{ad} \delta^{cb} - \delta^{ac} \delta^{bd} \right)\, .
\end{equation}

Finally,  the following identities and Fiertz relations hold  for the $SO(5)$ generators
($T^{\hat a}(\theta)$, $T^{a}(\theta)$ respectively denote a broken and an unbroken generator):
\begin{equation} \label{eq:55identities}
\begin{split}
\big(T^{a}(0) T^{b}(0) \big)_{55} &= 0 \\[0.1cm]
\big(T^{a}(0) T^{\hat b}(0) \big)_{55} &= \big(T^{\hat a}(0) T^{b}(0) \big)_{55} = 0 \\[0.1cm]
\big(T^{\hat a}(0) T^{\hat b}(0) \big)_{55}  & = \frac{\delta^{\hat a\hat b}}{2} = \frac{1}{2}\, \Tr\big[T^{\hat a}(\theta) T^{\hat b}(\theta) \big]\, ,
\end{split}
\end{equation}
\begin{align}
& \Tr\!\left( T^{\hat a} T^{\hat b} T^{\hat c} T^{\hat d}  \right) = \frac{1}{4} \,\Tr\!\left( T^{\hat a} T^{\hat b}  \right) \Tr\!\left( T^{\hat c} T^{\hat d}  \right) +
   \frac{1}{4} \,\Tr\!\left( T^{\hat a} T^{\hat d}  \right) \Tr\!\left( T^{\hat c} T^{\hat b}  \right)  \\[0.25cm]
& \Tr\!\left(  T^a \{ T^{\hat a} , T^{\hat b} \} \right) = 0
\end{align}
for  $\hat a, \hat b, \hat c, \hat d = 1,2,3,4$ and $a=a_L,a_R$, $b=b_L,b_R$;
\begin{equation} 
\begin{split}
& \sum_{\alpha=\hat{a},a_L,a_R} \!\left( T^{\alpha} \right)_{ij} \left( T^{\alpha} \right)_{kl} = 
 \frac{1}{2} \left( \delta^{il}\delta^{jk} - \delta^{ik}\delta^{jl} \right), \\[0.1cm]
& \sum_{a_L} \left( T^{a_L} \right)_{ij} \left( T^{a_L} \right)_{kl} - \sum_{a_R} \left( T^{a_R} \right)_{ij} \left( T^{a_R} \right)_{kl} = 
 -\frac{1}{2} \epsilon^{ijkl5}, \\[0.1cm]
& \sum_{a_L} \left( T^{a_L} \right)_{ij} \left( T^{a_L} \right)_{kl} + \sum_{a_R} \left( T^{a_R} \right)_{ij} \left( T^{a_R} \right)_{kl} = 
 -\frac{1}{4} \epsilon^{ijmn5} \epsilon^{klmn5}\, ,
\end{split} 
\end{equation}
for $i,j,k,l,m,n = 1,\dots,5$.

\section{Chiral lagrangian and identities for $\calG/\calH$ symmetric}
\label{app:GoHsymm}

The quotient space $\calG/\calH$ is said to be symmetric if there exists an automorphism of
the algebra (grading), $R$, under which the broken generators change sign.
Notable examples are $SO(5)/SO(4)$, as discussed 
in section~\ref{sec:Op2chiralLag}, as well as $SO(4)/SO(3)$ and $SU(n)\times SU(n)/SU(n)$.
As originally noticed in~\cite{CCWZ}, in the case of a symmetric $\calG/\calH$
the field $\bar\Sigma \equiv U^2 = \exp(2i\Pi(x))$ transforms linearly under global
transformations $g\in \calG$. This can be easily shown by acting on eq.(\ref{eq:Utransf}) with the grading $R$  and taking the hermitian
conjugate; one finds: $U(\Pi) \to h(g,\Pi) U(\Pi) R(g)^\dagger$, where $R(g)$ is the element of $\calG$ obtained by acting on $g$ with $R$.
Combining this rule with that of eq.(\ref{eq:Utransf}) one finds:
\begin{equation}
\bar\Sigma \equiv U(\Pi)^2 \, , \qquad \bar\Sigma \to g \,\bar\Sigma\, R(g)^\dagger\, .
\end{equation}
Hence $\bar\Sigma$ transforms linearly under $\calG$, $R(g)$ being  a \textit{global} element of $\calG$ with no dependence upon 
the NG field $\Pi(x)$.

A  derivative of $\bar\Sigma$ covariant under local $\calG$ transformation is defined as
\begin{equation}
D_\mu \bar\Sigma \equiv \partial_\mu \bar\Sigma + i A_\mu \bar\Sigma - i \bar\Sigma A^{(R)}_\mu\, ,
\end{equation}
where $A_\mu = A_\mu^a T^a + A_\mu^{\hat a} T^{\hat a}$ is the external gauge field and $A^{(R)}_\mu$ is obtained by acting on $A_\mu$ 
with the grading: $A_\mu^{(R)} \equiv R[A_\mu] = A_\mu^a T^a - A_\mu^{\hat a} T^{\hat a}$.
Under a (local) transformation $g\in \calG$, the gauge field transforms as follows:
\begin{equation}
\begin{split}
A_\mu & \to g A_\mu g^\dagger + i (\partial_\mu g) g^\dagger \\[0.25cm]
A^{(R)}_\mu & \to R(g) A^{(R)}_\mu R(g)^\dagger + i (\partial_\mu R(g)) R(g)^\dagger\, .
\end{split}
\end{equation}
The following identities thus allow one to rewrite the CCWZ covariant variables $d_\mu$, $E_{\mu\nu}$ in terms of $\bar\Sigma$:
\begin{align}
d_\mu =& - \frac{i}{2} \, U^\dagger \!\left( D_\mu \bar\Sigma \right) U^\dagger =  
                + \frac{i}{2} \, U \left( D_\mu \bar\Sigma \right)^\dagger U \\[0.15cm]
E_{\mu\nu} =& -\frac{i}{4} U^\dagger \!\left( D_\mu \bar\Sigma D_\nu \bar\Sigma^\dagger - D_\nu \bar\Sigma D_\mu \bar\Sigma^\dagger \right) U
 + \frac{1}{2} \left( U^\dagger  A_{\mu\nu} U +   U A^{(R)}_{\mu\nu} U^\dagger \right)  
 = -i [ d_\mu , d_\nu] + f^+_{\mu\nu}\, .
\end{align}
In the last equality we used the fact that, for a symmetric $\calG/\calH$, $f_{\mu\nu}^{\pm}$ coincide  respectively with the
even and odd components of $f_{\mu\nu}$ under  the grading $R$:
\begin{equation} \label{eq:fpmforGoHsym}
f_{\mu\nu}^{\pm}  = \frac{1}{2} \left( f_{\mu\nu} \pm f^{(R)}_{\mu\nu} \right)\, ,
\end{equation}
where $f^{(R)}_{\mu\nu}\equiv R[f_{\mu\nu}] = U A^{(R)}_{\mu\nu} U^\dagger$. 
Finally, one can prove that:
\begin{equation} 
f_{\mu\nu}^- = \nabla_{[ \mu}  d_{\nu ]} \, .
\end{equation}

Use of the linear field $\bar\Sigma$ suggests an alternative approach to the formulation
of the chiral lagrangian, in which this latter is constructed in terms of operators manifestly
invariant under global $\calG$ transformations. 
Although this approach, historically adopted for the QCD pion lagrangian, allows one to 
easily  classify and construct all $P_{LR}$-even operators purely in terms of the field $\bar\Sigma$,
we find that this is not true in the case of the operators odd under $P_{LR}$, 
which are most efficiently and systematically constructed by means of the CCWZ technique.

\section{Proof of two identities used in the text}
\label{app:proofs}

We present here the proof of two identities used in the text.

First, we prove eq.(\ref{eq:standardOp2}) and its equivalence with eq.(\ref{eq:CCWZOp2}).
From the definition (\ref{eq:Phi}), eq.(\ref{eq:DefOfdEgauged}) and the identities (\ref{eq:55identities}), it follows:
\begin{equation} \label{eq:proofL2}
\begin{split}
(D_\mu \Phi)^T (D^\mu \Phi)  
 & = \Phi_0^T (D_\mu U)^\dagger (D^\mu U) \Phi_0 \\[0.15cm]
 & = \left( r(\theta)^{-1}  (D_\mu U)^\dagger (D^\mu U) \, r(\theta)\right)_{55} \\[0.15cm]
 & = \left( r(\theta)^{-1}  (d_\mu + E_\mu)^2 \, r(\theta)\right)_{55} \\[0.15cm]
& = \frac{1}{2}\Tr\!\left[ d_\mu d^\mu \right]\, ,
\end{split}
\end{equation}
where $r(\theta)$ is defined in eq.(\ref{eq:thetarotation}).
To prove eq.(\ref{eq:standardOp2})  it is convenient to first evaluate the kinetic term of $\Phi$ on the $SO(5)/SO(4)$ vacuum,
and use as before eqs.(\ref{eq:Phi}), (\ref{eq:DefOfdEgauged}) and (\ref{eq:55identities}):
\begin{equation}
\begin{split}
\langle (D_\mu \Phi)^T (D^\mu \Phi)  \rangle
 & = \Phi_0^T \big( A_\mu^{\hat a} T^{\hat a}(\theta) + A_\mu^{a} T^{a}(\theta) \big)^2 \Phi_0 \\[0.15cm]
 & =  \left( r(\theta)^{-1} \big( A_\mu^{\hat a} T^{\hat a}(\theta) + A_\mu^{a} T^{a}(\theta) \big)^2  r(\theta)\right)_{55} \\[0.15cm]
 & = \frac{1}{2} \, \Tr\!\left[  \big( A_\mu^{\hat a} T^{\hat a}(\theta)  \big)^2 \right] \\[0.15cm]
 & = \frac{1}{4} \sin^2\!\theta \left( 2 W_\mu^+W^{\mu\, -} + (W_\mu^3 - B_\mu)^2 \right)\, ,
\end{split}
\end{equation}
where the last identify follows from eqs.(\ref{eq:brokenA}) and (\ref{eq:unbrokenA}).
Then, eq.(\ref{eq:standardOp2}) is obtained by performing a local $SU(2)_L\times SU(2)_R$ rotation which 
reintroduces the eaten NG bosons $\chi$, and by allowing for ($SU(2)_L\times SU(2)_R$)-invariant fluctuations $h(x)$
around the vacuum.

A similar strategy can be adopted to prove eq.(\ref{eq:O5}). 
From eq.(\ref{eq:fpmforGoHsym}), in terms of the linear field $\bar\Sigma$, the operator $O_5^+$
can be rewritten as
\begin{equation}
O_5^+ = \frac{1}{2} \left( \Tr\!\left[ F_{\mu\nu}^2 \right] - \Tr\!\left[ \bar\Sigma^\dagger F_{\mu\nu} \bar\Sigma  F^{(R)}_{\mu\nu} \right] \right)\, .
\end{equation}
On the $SO(5)/SO(4)$ vacuum, for $F^{(R)}_{\mu\nu} = R^\dagger F_{\mu\nu}  R$ with $R$ defined in eq.(\ref{eq:gradingR}), 
one has
\begin{equation}
\begin{split}
\langle O_5^+ \rangle 
 & = \frac{1}{2} \left( \Tr\!\left[ F_{\mu\nu}^2 \right] - \Tr\!\left[ F_{\mu\nu} R^\dagger F_{\mu\nu} R \right] \right) \\[0.15cm]
 & = \frac{1}{2} \sin^2\!\theta \left( (W_\mu^{a})^2 + (B_{\mu\nu})^2 - W_{\mu\nu}^{3} B^{\mu\nu} \right)\, .
\end{split}
\end{equation}
Then, eq.(\ref{eq:O5}) follows by performing a local $SU(2)_L\times SU(2)_R$ rotation to reintroduce the NG bosons $\chi$, 
and by allowing for ($SU(2)_L\times SU(2)_R$)-invariant fluctuations $h(x)$ around the vacuum.

\section{$\sigma$-model for the $\rho_L$ vector}
\label{sec:cosetrho}

As explained in the text, the interactions of the $\rho$ vector at energies above its mass are more easily characterized in terms of the eaten NG bosons. 
The sigma-model governing their dynamics is studied by taking a limit in which the transverse polarizations of the $\rho$ decouple ($g_\rho\to 0$) 
while the strength of the interactions of its longitudinal polarizations is kept fixed ($m_\rho/g_\rho$ constant).
The resulting coset is $SO(5)\times SU(2)_H/SU(2)'_L\times SU(2)_R$ where
\begin{equation*}
SU(2)_L\times SU(2)_R\sim SO(4)\subset SO(5),\qquad SU(2)'_L=[SU(2)_L\times SU(2)_H]_{\rm diag}\, .
\end{equation*}
The coset is parametrized by 7 Goldstone bosons $\pi^{\hat i}\, (\hat i=1,\ldots, 4)$ and $\eta^a\,(a=1,\ldots,3)$ transforming under the unbroken 
$SU(2)_R\times SU(2)'_L$ respectively as a (2,2) and a (3,1). The $\rho$ vector gauges the extra $SU(2)_H$ factor and eats the three $\eta$s. 

A convenient parametrization for the NG fields is~\footnote{Here and in the following equations of this section we absorb the decay constants into 
the NG fields for simplicity.}
\begin{equation}
U=e^{i\pi}e^{i\eta},\qquad \pi=\pi^{\hat i}T^{\hat i},\quad \eta=\eta^aX^a\, ,
\end{equation}
where $T^{\hat i}$ are the four $SO(5)/SO(4)$ generators, and $X^a=T_L^a-T'^a$ are the three generators of $SU(2)_L\times SU(2)_H/SU(2)'_L$.
Under a transformation $g_5\in SO(5)$, $g_H\in SU(2)_H$, the NG fields transform as
\begin{equation}
U(\pi,\eta)\to U(\pi',\eta')= g_5 g_H \,U\, h'_L \!^\dagger(\pi,\eta,g_5,g_H) h_R^\dagger(\pi,g_5)\, .
\end{equation}
We thus consider the combination
\begin{equation}\label{UdU}
U^\dagger\partial_\mu U=e^{-i\eta}e^{-i\pi}\partial_\mu e^{i\pi}e^{i\eta}+e^{-i\eta}\partial_\mu e^{i\eta}=
 e^{-i\eta}\left(i d_\mu+iE_\mu\right)e^{i\eta}+e^{-i\eta}\partial_\mu e^{i\eta}.
\end{equation}
which transforms as
\begin{equation}
U^\dagger\partial_\mu U\to h_R h'_L \,U^\dagger\partial_\mu U\, h'_L\!^\dagger h_R^\dagger - i h_R h'_L \partial_\mu \left( h_R h'_L \right)^\dagger\, .
\end{equation}
Notice that $d_\mu$ and $E_\mu$ are the usual covariant variables 
for $SO(5)/SO(4)$. Since the broken generators $T^{\hat i}$ and $X^a$ transform under two different irreducible representations of the unbroken group, 
it follows that the coset will contain two independent $d$ symbols, one in the algebra spanned by the generators $T^{\hat i}$ and another in the 
algebra of the $X^a$.
It is clear from the parametrization we are using that 
the only term in eq.(\ref{UdU}) that is parallel to the $T^{\hat i}$ is the first one in parenthesis. 
In particular
\begin{equation}
e^{-i\eta}\,id_\mu\,e^{i\eta}=i d_\mu^{\hat i}\,R(\eta)^{\hat i \hat j}T^{\hat j},
\end{equation}
where $R(\eta)$ is just an $\eta$-dependent $SO(4)$ rotation. We can thus add to the lagrangian the term
\begin{equation}
(d_\mu^{\hat i}R(\eta)^{\hat i \hat j})^2=d^{\hat j}_\mu d^{\hat j\mu}
\end{equation}
which contains just the $\pi$ fields and is the analogous of eq.(\ref{eq:CCWZOp2}). It is also clear that in (\ref{UdU}) only the term
\begin{equation}\label{dterm1}
e^{-i\eta}\left(\partial_\mu+iE_\mu\right)e^{i\eta}
\end{equation}
has a component in the $X^a$ algebra.  We denote this with $E^{(L)a}_\mu T_L^a$.
We can systematically expand (\ref{dterm1}) in terms of the $\pi$ and $\eta$ fields and project it on the $X^a$ algebra. 
We show the first few terms
\begin{equation}
\begin{split}
\widetilde d_\mu^a =
&\, \partial_\mu \eta^a-\frac{1}{6}\left(\delta^{bd}\delta^{ca}-\delta^{ba}\delta^{cd}\right)\partial_\mu\eta^b\eta^c\eta^d+\mathcal O (\eta^5)\\
&+\frac{1}{2}E^{(L)a}_\mu-\frac{1}{2}\epsilon^{abc}E^{(L)b}_\mu\eta^c+\mathcal O(\pi^2\eta^2)\, ,
\end{split}
\end{equation}
where
\begin{equation}
E_\mu^{(L)a}=\frac{1}{4}\epsilon^{a b c}\pi^{b}\partial_\mu\pi^{c}+\frac{1}{4}\left(\pi^a\partial_\mu\pi^4-\pi^4\partial_\mu\pi^a\right)+\mathcal O(\pi^4)\, .
\end{equation}
%

\section{Scattering amplitudes}
\label{app:scattampl}

In this Appendix we collect the expressions of the scattering amplitudes among longitudinally polarized vector bosons used in our analysis.
By means of the Equivalence Theorem these can be derived, at leading order in $(m_W/E)$, from the scattering amplitudes among  $SO(5)/SO(4)$ 
NG bosons defined in eq.(\ref{eq:NGBscattampldef}):
\begin{equation}
\begin{split}
	\label{eq:lista_ampiezze_scattering}
	\ampl(W^+_L W^-_L \rightarrow Z_L Z_L) \simeq &\; A(s,t,u) \\[0.15cm]
	\ampl(W^+_L W^-_L \rightarrow W^+_L W^-_L) \simeq &\;  A(s,t,u) + A(t,s,u)  \\[0.15cm]
	\ampl(Z_L Z_L \rightarrow Z_L Z_L) \simeq &\;  A(s,t,u) + A(t,s,u) + A(u,t,s)  \\[0.15cm]
	\ampl(W^\pm_L Z_L \rightarrow W^\pm_L Z_L) \simeq &\;  A(t,s,u)  \\[0.15cm]
	\ampl(W^\pm_L W^\pm_L \rightarrow W^\pm_L W^\pm_L) \simeq&\;  A(t,s,u) + A(u,t,s)  \\[0.15cm]
	\ampl(W^+_L W^-_L \rightarrow h h) \simeq &\;  A(s,t,u)  \\[0.15cm]
	\ampl(Z_L Z_L \rightarrow h h) \simeq &\;  A(s,t,u)  \\[0.15cm]
	\ampl(W^+_L W^-_L \rightarrow Z h) \simeq&\;  B(s,t,u)\, .
\end{split}
\end{equation}
In terms of the $SO(3)$ isospin amplitudes
\be
\begin{split}
	T(0) =&\;  3 A(s,t,u) + A(t,s,u) + A(u,t,s) \\[0.15cm]
	T(1) =&\;  A(t,s,u) - A(u,t,s) \\[0.15cm]
	T(2) =&\;  A(t,s,u) + A(u,t,s) \\[0.15cm]
	T'(0) =&\;  \sqrt{3} A(s,t,u)\, ,
\end{split}	
\ee
one has:
\begin{equation}
	\label{eq:classificazione_ampl_isospin}
\begin{split}
	\ampl(W^+_L W^-_L \rightarrow Z_L Z_L)  =& \frac{1}{3} [T(0) - T(2) ] \\[0.15cm]
	\ampl(W^+_L W^-_L \rightarrow W^+_L W^-_L) =& \frac{1}{6} [2T(0) +3T(1) + T(2)]  \\[0.15cm]
	\ampl(Z_L Z_L \rightarrow Z_L Z_L)  =&  \frac{1}{3} [T(0) + 2T(2)]  \\[0.15cm]
	\ampl(W^\pm_L Z_L \rightarrow W^\pm_L Z_L)  =& \frac{1}{2} [T(1) + T(2)]  \\[0.15cm]
	\ampl(W^\pm_L W^\pm_L \rightarrow W^\pm_L W^\pm_L)  =& T(2)  \\[0.15cm]
	\ampl(W^+_L W^-_L \rightarrow h h)  =& \frac{1}{\sqrt{3}} T'(0)\, . 
\end{split}
\end{equation}
%

\section{Operators of the $\rho$ effective lagrangian}
\label{sec:operators}

We report here the full list of independent operators
that must be included in the effective lagrangian of the $\rho$ at leading order in $\partial/\Lambda$, 
that contribute to scattering processes involving up to 2$\,\rho$ vectors.
According to the criterion of PUVC,  they give a subleading contribution to $\pi\pi\to\pi\pi$ scattering at high energy.
In addition to those already defined in eq.(\ref{eq:Qi}) and to the mass and kinetic terms of eq.(\ref{eq:rhoLag}) we find:
\\[0.5cm]
{\sc Operators with 1 $\rho$}
\begin{subequations}
  \begin{alignat}{2}
    &Q_5=\Tr\left[\bar\rho^\mu[\nabla_\mu d_\nu,d^\nu]\right]     
   &&\qquad Q_6 =\epsilon^{\mu\nu\rho\sigma}\Tr\left[\bar\rho_\mu \nabla_\nu d_\rho d_\sigma\right]    
  \end{alignat}
  \end{subequations}
  \\
{\sc Operators with 2 $\rho$}
\begin{subequations}
  \begin{alignat}{2}
    &Q_{1(2)}=\Tr\left[(\nabla_\mu \bar\rho^\mu)^2\right]     && Q_{7(2)} =\Tr[\bar \rho_\mu\bar\rho^\mu d_\nu d^\nu]  \\[0.05cm]
    &Q_{2(2)}= \Tr\left[\bar \rho_\mu \bar \rho_\nu f_{\mu\nu}^+\right]&& Q_{8(2)} = \Tr[\bar \rho_\mu d_\nu \bar\rho^\mu d^\nu]\\[0.05cm]
    &Q_{3(2)} =\Tr [\bar\rho_\mu\bar\rho^\mu]\Tr[d_\mu d^\mu]   \qquad &&Q_{9(2)} =  \Tr[\bar \rho_\mu d^\mu \bar\rho_\nu d^\nu]\\[0.05cm]
    &Q_{4(2)} =\Tr [\bar\rho_\mu\bar\rho_\nu]\Tr[d^\mu d^\nu]&&Q_{10(2)} = \Tr[\bar \rho_\mu d_\nu \bar\rho^\nu d^\mu]\\[0.05cm]
    &Q_{5(2)} =\Tr \left[\bar\rho_\mu\bar\rho_\nu[d^\mu, d^\nu]\right]&&Q_{11(2)}=\epsilon^{\mu\nu\rho\sigma}\Tr[\bar\rho_\mu\bar\rho_\nu d_\rho d_\sigma]
    \\[0.05cm]
    &Q_{6(2)} =\Tr \left[\bar\rho_\mu\bar\rho_\nu\{d^\mu, d^\nu\} \right]&&Q_{12(2)} =\epsilon^{\mu\nu\rho\sigma}\Tr[\bar\rho_\mu d_\nu \bar\rho_\rho d_\sigma]
  \end{alignat}
  \\
\end{subequations}
{\sc Operators with 3 $\rho$}
\begin{subequations}
  \begin{alignat}{2}
    &Q_{1(3)}=\Tr\left[\bar\rho_\mu\bar\rho_\nu\rho^{\mu\nu}\right]     &&Q_{3(3)} =\epsilon^{\mu\nu\rho\sigma}\Tr[\bar \rho_\mu\bar\rho_\nu \rho_{\rho\sigma}]  
    \\[0.05cm]
    &Q_{2(3)}= \Tr\left[\nabla_\mu\rho_\nu\{\bar\rho^\mu,\bar\rho^\nu\}\right]&&\qquad
          \end{alignat}
\end{subequations}
In order to eliminate other linearly dependent operators
we made use of the relations reported in Appendix A and B 
as well as of the following identity:
\begin{equation}
\rho_{\mu\nu}=\nabla_{[\mu}\bar\rho_{\nu]}+E_{\mu\nu}+i[\bar\rho_\mu,\bar\rho_\nu]\, .
\end{equation}
Notice that since the grading of $SO(5)/SO(4)$ is an internal automorphism, $R\subset SO(4)$,
all operators involving an odd number of $d$ symbols vanish. 

Assuming the validity of the derivative expansion up to a scale $\Lambda=g_* f$ and requiring all interactions to remain weaker than $g_*$ at the scale 
$\Lambda$ (see section~\ref{subsec:rhoL}), the coefficient of each of the above operators is bounded to be $\ll 1/g_*^2$.  This implies in particular
that the contribution of $Q_{1(2)}$, $Q_5$, $Q_6$, the only operators that modify $\pi\pi$ scattering, is subdominant.
Finally, none of the above operators contribute to any low-energy observable
at $\mathcal O(p^4)$. This follows from the equation of motions of the vector resonance: $\bar \rho_\mu = \rho_\mu-E_\mu\propto \nabla_\nu E_{\nu\mu}$.

\section{$SO(5)/SO(4)$ linearized models}
\label{sec:LMCHM}

In this Appendix we define the models LMCHM4 and  LMCHM5, which were considered in section~\ref{sec:checkanalytic} to check
the validity of our analytic approximation. They respectively correspond to the linearized versions of the Minimal Composite Higgs Models
MCHM4 and MCHM5  of Refs.\cite{Agashe:2004rs},~\cite{Contino:2006qr}. 
The LMCHM4, in particular, has been previously introduced in Ref.\cite{Barbieri:2007bh}.
The two models are defined by the lagrangians
\be\label{linearized}
\mathcal L_{4(5)}=\frac{1}{2}(D_\mu \varphi)^TD^\mu \varphi-\lambda( \varphi^T \varphi-f_0^2)^2-V_{4(5)}(\vec\varphi,\varphi_5)
\ee
in terms of a scalar multiplet $\varphi$  transforming in the fundamental representation of $SO(5)$.
The potential term $V_{4(5)}$, the only one which differentiates the LMCHM4 from the LMCHM5, 
breaks explicitly the $SO(5)$ symmetry and triggers the EWSB.
In the original models it arises from the one-loop contribution of the SM fermion and gauge fields.
Here, for simplicity, we take $V_{4(5)}$ to match the leading term in the expansion of the logarithm that appears in the 
Coleman-Weinberg potential of Refs.~\cite{Agashe:2004rs, Contino:2006qr}.~\footnote{As noticed by Refs.~\cite{Agashe:2004rs, Contino:2006qr}, 
this approximation is quite good thanks to the fast convergence of the one-loop integral.}

The field $\varphi$ can be redefined as
\begin{equation}
 \varphi(x)=\bar\eta(x) \Phi(x) = \bar\eta(x) U(x) \Phi_0,
\end{equation}
where $U(x)$ is defined as in eq.(\ref{eq:Phi}) upon identifying $f\equiv \langle \bar\eta\rangle$.
From eq.(\ref{eq:proofL2}) it thus follows
\begin{equation}
{\mathcal L}_{4(5)}
 = \frac{1}{2}\partial_\mu \bar\eta\partial^\mu\bar\eta+\frac{\bar\eta^2}{4}\Tr [d_\mu d^\mu]-\lambda(\bar\eta^2-f_0^2)^2-V_{4(5)}(\bar\eta, h).
\end{equation}
The derivative couplings that follow from the above lagrangian, after redefining $\bar\eta= \eta+\langle \bar\eta \rangle$, are of the
form of those in eq.(\ref{eq:etaLag}) with $a_\eta = 1 = b_\eta$.
The expression of the potential term $V_{4(5)}$ is explicitly reported in the following for each of the two models.

\vspace{0.5cm}
\noindent {\rm\bf LMCHM4}\\[0.2cm]
In the LMCHM4 the symmetry breaking potential is given by
\begin{equation}
V_4 =\alpha f_0^3 \, \varphi_5-\beta f_0^2 \, \vec\varphi^2\, ,
\end{equation}
where $\alpha$ and $\beta$ are two dimensionless free parameters.
Using eq.(\ref{eq:Phi2}), one can rewrite $V_4$ as:
\begin{equation}
V_4 =\alpha f_0^3 \,\bar\eta(x) \,\cos\!\left(\theta + \frac{h(x)}{f}\right)  -\beta f_0^2 \, \bar\eta^2 \, \sin^2\!\left(\theta + \frac{h(x)}{f}\right)\, .
\end{equation}
The expectation value $\langle\bar\eta\rangle = f$ and $\theta$ are thus given by
\begin{equation}
\langle\eta\rangle^2\equiv f^2=f_0^2\left(1+\frac{\beta}{2\lambda}\right)\equiv f_0^2(1+z)\, , \qquad  
\cos\theta=-\frac{\alpha}{2\beta}\frac{1}{\sqrt{1+z}}\equiv\sqrt{1-\xi}, , 
\end{equation}
with $z\equiv\beta/2\lambda$.
By expanding the potential around the vacuum one gets the scalar mass matrix
\be
\mathcal M=
\begin{pmatrix}
m_{\eta\eta}^2 & m_{h\eta}^2\\[0.15cm]
m_{h\eta}^2 & m_{hh}^2
\end{pmatrix}\, ,
\ee
with
\begin{equation}
\begin{split}
m^2_{\eta\eta} =& 8\lambda f^2 \frac{1+ 3/2z}{1+ z}-m_{hh}^2,\\[0.15cm]
m^2_{\eta h}    =&-m^2_{hh}\frac{\sqrt{1-\xi}}{\sqrt\xi},\\[0.15cm]
m^2_{hh}         =&\frac{2\beta v^2}{1+ z}. 
\end{split}
\end{equation}
Working in the small mixing limit, $m_h\ll m_\eta$, one can easily derive from the potential 
the non-derivative $hhh$ and $hh\eta$ couplings, the only ones relevant to our analysis:
\begin{equation}
\begin{split}
V_4(h,\eta) \supset & \, \frac{g_{hhh}}{6}\, h^3+\frac{g_{h h\eta}}{2}\, h^2\eta,\\[0.25cm]
g_{hhh} \approx & \, 3 \, \frac{m_h^2}{v}\sqrt{1-\xi},\\[0.15cm]
g_{hh\eta} \approx& \,-\frac{m_h^2}{v} \, \frac{1-3\xi}{\sqrt \xi}\, .
\end{split}
\end{equation}

\vspace{0.5cm}
\noindent {\rm\bf LMCHM5}\\[0.2cm]
In the case of the LMCHM5 the potential has the form
\be
V_5 =\alpha f_0^2 \, \vec\varphi^2-\beta \, \varphi_5^2 \, \vec\varphi^2\, ,
\ee
which can be rewritten as:
\be
V_5 =\alpha f_0^2 \, \bar\eta^2 \sin^2\!\left(\theta + \frac{h(x)}{f}\right) 
 -\beta \, \bar\eta^4\sin^2\!\left(\theta + \frac{h(x)}{f}\right)  \cos^2\!\left(\theta + \frac{h(x)}{f}\right) \, .
\ee
By minimizing the potential, one obtains:
\be
\langle\eta\rangle^2\equiv f^2=f_0^2\left(\frac{\alpha-4\lambda}{\beta-4\lambda}\right)\equiv f_0^2(1+z)\, , \qquad
 \sin^2\theta=\frac{2 \lambda}{\beta}\frac{\alpha-\beta}{\alpha-4\lambda}\equiv \xi\, ,
\ee
with $z\equiv(\alpha-\beta)/(\beta-4\lambda)$.
The scalar masses are given by
\begin{equation}
\begin{split}
m^2_{\eta\eta} =&\, 8\lambda f^2 \frac{1-z}{1+ z}-m_{hh}^2\frac{\xi}{1-\xi},\\[0.15cm]
m^2_{\eta h} =&\, -m^2_{hh}\frac{1-2\xi}{\sqrt{\xi(1-\xi)}},\\[0.15cm]
m^2_{hh} =& \, 8\beta v^2(1-\xi)\, .
\end{split}
\end{equation}
In the small mixing limit the relevant scalar couplings are
\begin{equation}
\begin{split}
g_{hhh} \simeq & \, 3\, \frac{m_h^2}{v} \, \frac{\sqrt{1-2\xi}}{\sqrt{1-\xi}},\\[0.2cm]
g_{hh\eta} \simeq & \,-\frac{m_h^2}{v} \, \frac{1-12(1-\xi)\xi}{2(1-\xi)\sqrt \xi}\, .
\end{split}
\end{equation}


\end{document}